\documentclass{aa}  

\usepackage{graphicx}
\usepackage{txfonts}
\usepackage{lipsum}
\usepackage{subcaption}         
\usepackage{lscape}             
\usepackage{placeins}           

\usepackage{amsmath}
\usepackage{xcolor}

\usepackage{rotating}

\usepackage{multirow}

\newcommand{\kms}{km~s$^{-1}$}
\newcommand{\ergs}{erg~s$^{-1}$}
\newcommand{\ergcms}{erg~cm$^{-2}$~s$^{-1}$}
\newcommand{\Msun}{M$_\sun$}

\newcommand{\nii}{[\ion{N}{ii}]}
\newcommand{\hii}{\ion{H}{ii}}
\newcommand{\sii}{[\ion{S}{ii}]}
\newcommand{\oi}{[\ion{O}{i}]}
\newcommand{\oiii}{[\ion{O}{iii}]}
\newcommand{\oii}{[\ion{O}{ii}]}
\newcommand{\nai}{\ion{Na}{i}}
\newcommand{\heii}{\ion{He}{ii}}
\newcommand{\niii}{\ion{N}{iii}}
\newcommand{\nv}{\ion{N}{v}}
\newcommand{\ciii}{\ion{C}{iii}}
\newcommand{\civ}{\ion{C}{iv}}

\usepackage[normalem]{ulem}

\usepackage{makecell} 

\usepackage[colorlinks=true, allcolors=blue]{hyperref}

\usepackage{comment}

\usepackage{titlesec}
\titleformat{\paragraph}
  {} 
  {\theparagraph}         
  {1em}                   
  {}

\begin{document}

\title{Multiwavelength study of NGC 2993 and its candidate ultraluminous X-ray sources}

\author{A. Kilina\inst{1}, O. Godet\inst{1}, B. Epinat\inst{2, 3}, C. Morisset\inst{4}, A. Gúrpide\inst{5}
    }
\authorrunning{A. Kilina}    
\institute{Université de Toulouse, CNRS, CNES, IRAP, 9 avenue du colonel Roche, 30128 Toulouse Cedex 4, France\\
         \email{anastasia.kilina@utoulouse.fr}
    \and
        Aix-Marseille Univ., CNRS, CNES, LAM, 38 Rue Frédéric Joliot Curie, 13338 Marseille, France
    \and
        French-Chilean Laboratory for Astronomy, IRL 3386, CNRS and Universidad de Concepción, Departamento de Astronomía, Barrio Universitario s/n, Concepción, Chile
    \and 
        Universidad Nacional Autónoma de México,  Instituto de Ciencias Físicas, Av. Universidad s/n, 62210 Cuernavaca, Mor., México
    \and 
       Anton Pannekoek Institute for Astronomy,  University of Amsterdam, Science Park 904, 1098 XH Amsterdam, Netherlands
     }

\date{Received 17 June, 2026; Accepted 26 November 2026}

\abstract
{Ultraluminous X-ray sources (ULXs) are accreting stellar-mass compact objects with X-ray luminosities above 10$^{39}$ \ergs. Some ULXs are found to ionize their environment by UV/X-ray radiation and/or winds from their supercritical accretion disk. Studying the influence of ULXs on their surroundings may help to better understand physics behind the supercritical accretion onto compact objects.
}
{We study the nature of two ULX candidates (ULX1 and ULX2) and of two central sources (X1 and X2) in the NGC 2993 star-forming galaxy interacting with NGC 2992. We aim to understand the local variations in the environment where ULXs are formed, and how their presence influences their surrounding interstellar medium (ISM).
} 
{We use archival X-ray observations from the \textit{Chandra} and \textit{XMM-Newton} X-ray observatories to search for spectral variability and to measure their X-ray fluxes. We use archival data from the MUSE instrument on the \textit{Very Large Telescope} to study spatially resolved properties of the NGC 2993 ISM. We also perform optical photometry on the counterparts of ULX1 and ULX2 using archival \textit{HST} images. 
}
{
We found an \textit{HST} point-like counterpart to ULX2, while ULX1 appears to be located in a \textit{HST} resolved stellar cluster with a radius of 5.1$\pm$1.1 pc and a mass below $3.50$$\times10^6$ \Msun. We do not detect any significant spectroscopic evidence of the ULX1 impact on its surroundings through either hard X-ray photoionization or shocks, although it may be hidden by the insufficient spatial resolution and/or strong stellar photoionization.
In addition, we found a possible local nitrogen enrichment at the ULX1 position, unresolved by MUSE. Its origin is unclear, but it might be associated with the evolution of the binary system leading to ULX1 or with Wolf-Rayet stars present in its host cluster. Finally, we did not find any indication of AGN activity in the galaxy. 
}
{}


\maketitle
\nolinenumbers

\section{Introduction}

Ultraluminous X-ray sources (ULXs) are accreting non-nuclear sources that sustain X-ray luminosities above 10$^{39}$ \ergs, which is the Eddington limit for a 10 M$_{\sun}$ black hole (\citealt{kaaret_ultraluminous_2017}, \citealt{king_ultraluminous_2023}). Therefore, these sources were originally seen as good candidates to host very massive or even intermediate mass black holes (IMBHs) \citep{colbert_nature_1999} accreting at Sub-Eddington rates.
However, several spectral features consistent with super-Eddington accretion were later observed for some ULXs. Some of them presented a high-energy cutoff in X-rays, indicating super-Eddington accretion onto a stellar-mass compact object (\citealt{gladstone_ultraluminous_2009}, \citealt{bachetti_ultraluminous_2013}, \citealt{gurpide_long-term_2021}). In addition, some ULXs were discovered to exhibit X-ray pulsations (\citealt{bachetti_ultraluminous_2014}, \citealt{furst_discovery_2016}, \citealt{israel_accreting_2017}, \citealt{carpano_discovery_2018}, \citealt{quintin_new_2021}), establishing their accreting compact objects as neutron stars. 
Some ULXs have been found to exhibit absorption features in their X-ray spectra interpreted as cyclotron resonance scattering features \citep{staubert_cyclotron_2019}, indicating the presence of a magnetic field and further supporting the accretors of some ULXs as being neutron stars (\citealt{brightman_magnetic_2018}, \citealt{kong_insight-hxmt_2022}, \citealt{allak_candidate_2026}, \citealt{jayasurya_discovery_2026}).
Furthermore, studies discovered powerful relativistic outflows launched from supercritical ULX disks using atomic emission and absorption lines  (\citealt{pinto_resolved_2016}, \citealt{kosec_evidence_2018},  \citealt{robba_broadband_2021}).
Nowadays, it is largely accepted that most ULXs are either stellar mass black holes or neutron stars accreting at super-Eddington rates, with a few exceptions that are still considered robust IMBH candidates (\citealt{godet_first_2009}, \citealt{amato_population_2025}).

In the super-Eddington regime, the accretion disk becomes optically thick, emitting strong mildly relativistic and optically thick outflows (for a black hole or low-magnetized neutron star accretor), creating a funnel geometry around the rotational axis of the compact object \citep{poutanen_supercritically_2007}. 
These outflows are thought to carry away a large amount of angular momentum and kinetic energy. However, the geometries of supercritical outflows are still poorly understood.  
The angular size of this funnel decreases with the increasing accretion rate (\citealt{king_masses_2009}, \citealt{kawashima_comptonized_2012}). Therefore, the X-ray photons from the inner edges of the accretion disk are thought to only escape through this funnel, which may result in a highly anisotropic emission. 
However, current X-ray telescopes only allow us to see ULXs as unresolved sources, with no information about their inclination and the opening angle of the funnel. Therefore, there are still many uncertainties related to the geometries of the ULX accretion disks and their outflows, and it remains challenging to estimate the real energy output of ULXs from the X-ray observations only.

To overcome this limitation, one can use multiwavelength observations to study the impacts of supercritical accretion on the ULX surroundings. Indeed, the strong UV/X-ray radiation of ULXs and/or shocks caused by their winds ionize their surrounding environment. These processes inflate bubbles of ionized gas around some ULXs, reaching up to 200 pc in radius (\citealt{pakull_optical_2002}, \citealt{berghea_detection_2020}, \citealt{soria_ultraluminous_2021}). Studying these bubbles can provide us with an independent measure of the ULX energy output, giving us insights into the physics of super-Eddington accretion flows. 

A few bubbles blown by photoionization and/or by shocks from ULXs have been identified using optical imaging \citep{kaaret_high-resolution_2004} or spectroscopy \citep{moon_large_2011} or both \citep{urquhart_multiband_2018}.
But not so many have been studied using integral field units (IFUs). IFUs are very powerful instruments to study ionized nebulae, as they provide us with 3D datacubes with two dimensions for imaging and another one for spectra. 
The  Multi-Unit Spectroscopic Explorer \citep[MUSE,][]{bacon_muse_2010} IFU instrument on the \textit{Very Large Telescope (VLT)} telescope has been especially useful to explore ULX bubbles. Indeed, it was used to study for example NGC 1313 ULX-1 \citep{gurpide_muse_2022}, NGC 1313 ULX-2 \citep{zhou_very_2022}, and NGC 55 ULX-1 \citep{zhou_identification_2023}. In this paper, we use archival data from the \textit{VLT}/MUSE instrument to study the environment of NGC 2993 ULX1. 

Optical and/or infrared instruments with high spatial resolution such as the \textit{Hubble Space Telescope} (\textit{HST}, \citealt{leckrone_hubble_2003}) or the \textit{James Webb Space Telescope} (\textit{JWST}, \citealt{gardner_james_2023}) can be used to look for possible ULX counterparts. The observed emission may be dominated by the accretion disk (\citealt{tao_compact_2011}, \citealt{allak_unveiling_2025}) or by the donor stars \citep{gladstone_optical_2013}. Near infrared \citep{heida_near-infrared_2014, heida_discovery_2015, heida_keckmosfire_2016, allak_first_2024} and optical (\citealt{vinokurov_optical_2018}, \citealt{allak_optical_2022}) observations of ULX donor star counterparts show that many of them are supergiants. In this paper, we study the optical counterparts of ULX1 and ULX2 using \textit{HST} observations.

\medskip

NGC 2993 together with NGC 2992 form the interacting system Arp 245. It is located at the distance of 30.5 Mpc from us \citep{tully_catalog_1988}. 
The two galaxies are in strong tidal interaction, with a tidal bridge between them and tidal tails escaping from them. The clear presence of the bridge favors a prograde encounter \citep{toomre_galactic_1972}. The Arp 245 system is in its early stage of interaction, 100 Myr after the first encounter and 700 Myr before the merger \citep{duc_formation_2000}.

\begin{figure}[ht!]
    \centering
    \includegraphics[width=\linewidth]{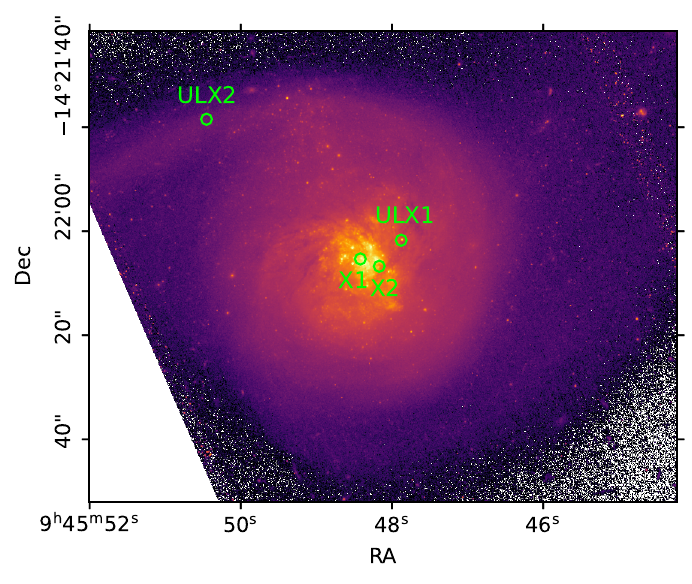}
    \caption{\textit{HST} ACS/WFC/F606W image of the NGC 2993 galaxy. The positions of the \textit{Chandra} X-ray sources in the galaxy are identified by the lime circles with arbitrary sizes (see Fig. \ref{fig:chandra} for the \textit{Chandra} broadband X-ray image). There is a tidal bridge to the North-West of the galaxy in the direction of NGC 2992 and a tidal tail where ULX2 is found.}
    \label{fig:ngc2993}
\end{figure}

There are four X-ray sources within NGC 2993 (see Figs.~\ref{fig:ngc2993} and \ref{fig:chandra}): two in the center of the galaxy \citep[sources 10 and 11 in][that we hereafter call X1 and X2]{colbert_extranuclear_2005}, and two ULX candidates: one to the North-West from the center \citep[source 9 in][that we hereafter call ULX1]{colbert_extranuclear_2005} and one in the spiral arm \citep[source 13 in][that we hereafter call ULX2]{colbert_extranuclear_2005}. Those sources are only mentioned in \cite{colbert_extranuclear_2005} without detailed discussion on their nature. They found observed X-ray luminosities of $\sim5\times10^{39}$ \ergs\ in the 0.3$-$8.0 keV band for all four sources using data from the \textit{Chandra X-ray observatory} \citep{weisskopf_chandra_2000}.
The galaxy imaged by \textit{HST} is shown in Fig. \ref{fig:ngc2993} with source positions from \textit{Chandra}.

In this paper, we aim to confirm the ULX nature of the ULX1 and ULX2 candidates as well as to study the possible impact of ULX1 on its surrounding medium. We also discuss the nature of the two central X-ray sources and whether NGC 2993 hosts an AGN. We discuss the link between the local environment of the X-ray sources and the galaxy scale processes. In particular, we focus on the NGC 2993 ULX1 source. In Sect. \ref{sec:obs}, we present the archival data that were used for this study. In Sect. \ref{sec:data_analysis}, we discuss the multi-wavelength data analysis of the observations of the galaxy and X-ray sources. In Sect. \ref{sec:discussion}, we discuss our results and their implications on the possible origins of ULX1 and other X-ray sources in NGC 2993. Finally, Sect. \ref{sec:conclusion} serves as a conclusion to our work. Throughout the paper, we use a distance of 30.5 Mpc for NGC 2993.

\section{Observations and data reduction}
\label{sec:obs}

\begin{table*}[h!]
    \centering
    \caption{Observations of the NGC 2993 galaxy used in this paper.}
    \renewcommand{\arraystretch}{1.3}
    \begin{tabular}{cccccccc}
    \hline\hline
        Telescope & \makecell{Instrument \\ and Filter} & Band & Obs Id & Date & \makecell{Net exposure \\ Time (ks)} \\ \hline
        \textit{Chandra} & ACIS-S & 0.5$-$7.0 (keV) & 3956 & 2003-02-16 & 	49.55 \\ 
        \textit{Chandra} & ACIS-S & 0.5$-$7.0 (keV) & 11858 &  2010-02-09 & 94.40 \\ 
        \textit{XMM-Newton} & MOS1, MOS2, pn & 0.3$-$12 (keV) & 0147920301 & 2003-05-19 & 27.04, 28.00, 28.09 \\ \hline
        \textit{HST} & ACS/WFC/F606W & 4632$-$7179 (\AA) & - & 2018-05-22 & 0.78 \\
        \textit{HST} & ACS/WFC/F435W & 3599$-$4884 (\AA) & - & 2018-11-27 & 2.10 \\ 
        \textit{VLT} & MUSE & 4750$-$9352 (\AA) & 3114238, 3115072 & 2021-11-24 & 4.80  \\ \hline
    \end{tabular}
    \label{tab:observations}
\end{table*}     

We use archival X-ray observations from the \textit{Chandra X-ray observatory} and from the \textit{XMM X-ray observatory} \citep{jansen_xmm-newton_2001} to study the X-ray sources in the NGC 2993 galaxy. We use optical IFU archival data from the \textit{VLT}/MUSE instrument in order to study the properties of the galaxy, which gives us information about the environment of the X-ray sources. Finally, we use archival optical data from \textit{HST} to study the galaxy with a better spatial resolution and to look for potential optical counterparts of the X-ray sources. In this section, we describe the observations used in this paper. Their characteristics are summarized in Table \ref{tab:observations}. We applied an astrometric correction to all the data as described in Appendix \ref{app:astrometry}.

\subsection{\textit{Chandra}}
\label{subsec:chandra_data}
\begin{figure}[!ht]
    \centering
    \includegraphics[width=1\linewidth]{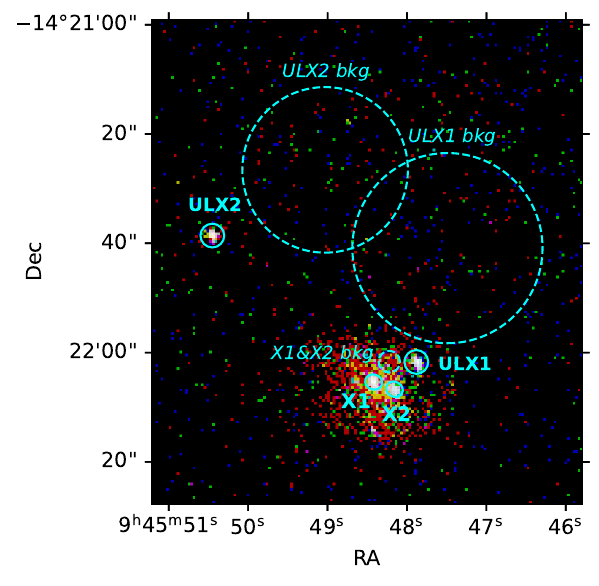}
    \caption{\textit{Chandra} RGB image of the NGC 2993 galaxy. Red: soft band (0.5$-$1.2 keV), green: medium band (1.2$-$2.0 keV), blue: hard band (2.0$-$7.0 keV). Source/background extraction regions are shown with solid/dashed circles and ellipses. The radius of the extraction circle of ULX candidates is 2.17\arcsec. The semi-minor radius of the elliptical extraction regions for X1 and X2 is 1.37\arcsec\ and the semi-major radius is 1.58\arcsec. The radii of the background extraction circles are 2.00\arcsec\ for X1 and X2, 17.4\arcsec\ for ULX1 and 15.1\arcsec\ for ULX2.}
    \label{fig:chandra}
\end{figure}

We considered two \textit{Chandra} observations of the NGC 2993 galaxy\footnote{Two other observations (ObsID \textit{12103} and \textit{12104}) were discarded as their observing time is too short to extract useful information from them.}. The \textit{Chandra} data has been reprocessed using the \texttt{chandra\_repro} function from Chandra Interactive Analysis of Observations (\texttt{CIAO}) 4.16 \citep{fruscione_ciao_2006} using the \texttt{v4.11.0 CALDB} files. Although the \textit{11858} observation was made with the High-Energy Transmission Grating (HETG) instrument \citep{canizares_chandra_2005}, we used the zeroth-order image in this work. The spectra used in this work were extracted using \texttt{specextract} from \texttt{CIAO 4.16} and grouped using the optimal grouping \citep{kaastra_optimal_2016} from the \texttt{ftgrouppha} routine from the \texttt{HEAsoft} 6.31 software (NASA High Energy Astrophysics Science Archive Research Center (Heasarc) \citeyear{nasa_high_energy_astrophysics_science_archive_research_center_heasarc_heasoft_2014}). 
The source and background extraction regions are shown in Fig. \ref{fig:chandra}. Using the \texttt{src\_psffrac} routine from \texttt{CIAO}, we computed the encircled energy fraction in the source extraction regions at the \textit{Chandra} broad band effective energy (2.3 keV), and obtained 93\% for ULX1 and ULX2, 89\% for X1 and X2. 
There is soft (mostly below 2.0 keV) diffuse X-ray emission in the central part of the galaxy, likely due to hot gas. 

\subsection{\textit{XMM-Newton}}
\label{subsec:xmm_data}
The NGC 2992 galaxy has been observed by \textit{XMM-Newton} multiple times. However, there is only one observation in which NGC 2993 is also observed. We have reprocessed the \textit{EPIC-pn} \citep{struder_european_2001} data files with \texttt{epproc} and the \textit{EPIC-MOS} \citep{turner_european_2001} data files with \texttt{emproc} (both version 2.25.4) from the Science Analysis Software \citep[\texttt{SAS},][]{gabriel_xmm-newton_2017} version 22.1.0. We used the latest calibration files as of March 2025. We then filtered the events files in order to get rid of time intervals with high background flaring activity using \texttt{evselect} and \texttt{tabgtigen} functions from \texttt{SAS}.

Contrarily to \textit{Chandra}, the spatial resolution of \textit{XMM-Newton} is insufficient to distinguish ULX1, X1 and X2. Therefore we only used \textit{XMM-Newton} data to extract the spectrum of ULX2. The source exposure time was of 19.4 ks for \textit{pn} and 23.4 ks for both \textit{MOS1} and \textit{MOS2}. To extract the spectrum, we used the \texttt{evselect} function from \texttt{SAS} by defining a source and a background region (with radii of 15\arcsec\ and 66.5\arcsec\ respectively). We kept patterns $\leq$4 for the \textit{pn} instrument and $\leq$12 for the \textit{MOS} instruments. We used the function \texttt{rmfgen} to create redistribution matrices and the function \texttt{arfgen} to create auxiliary response files. Finally, we grouped the spectra using the optimal grouping from \texttt{ftgrouppha}.

\subsection{\textit{HST}}
All four X-ray sources are covered in the \textit{HST} ACS/WFC/F606W \citep{ford_overview_2003} filter image of the NGC 2993 galaxy. ULX2 is also observed by \textit{HST} in the ACS/WFC/F435W filter. The \texttt{drc.fits} files that we used are already corrected for charge transfer efficiency and the data is given in $\mathrm{e}^-\,\mathrm{s}^{-1}$. We multiplied it by the inverse sensitivity to convert it into units of erg s$^{-1}$ cm$^{-2}$ \AA$^{-1}$ given by the \texttt{PHOTFLAM} in the data headers. To ensure that we analyze the emission that comes from the NGC 2993 galaxy itself, we performed a sky residual correction on the \textit{HST} images. To do so, we selected a region far from the galaxy where there was no emission from stars or gas, and computed its median flux value. Then, we subtracted this value from the flux in each pixel of the image, so that the flux in regions with no sources fluctuates around zero.

\subsection{\textit{MUSE}}
\label{subsec:muse}
There are two \textit{VLT}/MUSE observations of the NGC 2993 galaxy: one taken on 2021-11-24 and another on 2021-12-17, both with exposures of 2400 seconds.
The resolving power goes from 1770 at 480 nm to 3590 at 930 nm with a mean value at 3000 \citep[see e.g.][]{bacon_muse_2017}. We use a "MUSE-DEEP" cube available in the ESO archive\footnote{\href{https://archive.eso.org/scienceportal/home}{https://archive.eso.org/scienceportal/home}}. It was created by stacking these two observations, thus doubling their individual exposure times. The data has been reduced by the ESO data processing pipeline for the MUSE instrument \citep{weilbacher_data_2020} version 2.8.6. This pipeline takes care of corrections such as subtraction of sky lines. The FWHM effective spatial resolution of the deep cube's whitelight image measured by the pipeline is 0.867\arcsec. We replaced pixels with NaN values by interpolating them with their surrounding pixels using the \texttt{nanclean} routine from the \texttt{zap} 2.1 python package \citep{soto_zap_2016}. This led to the replacement of $\sim$0.1\% of pixels in the cube.

\section{Data analysis}
\label{sec:data_analysis}

\subsection{X-ray spectra}
\label{subsec:xrays}

To calculate the X-ray fluxes of our sources, we modeled their spectra with an absorbed powerlaw model and an absorbed multi-blackbody accretion disk model using \texttt{jaxspec 0.3.3} \citep{dupourque_jaxspec_2024}. We used the solar abundance table from \cite{anders_abundances_1989} and the photoionization cross-section table from \cite{verner_atomic_1996}. 
For X1 and ULX2, the absorption column $n_{\mathrm{H}}$ was found to be compatible with the galactic value ($5.26 \times 10^{20}$ cm$^{-2}$, \citealt{dickey_h_1990}), so we fit their spectra using a \texttt{tbabs*powerlaw} \citep{wilms_absorption_2000} and \texttt{tbabs*diskbb} \citep{mitsuda_energy_1984} models. For X2 and ULX1 we found an additional absorption column, therefore we used the \texttt{tbabs*tbabs*powerlaw} and \texttt{tbabs*tbabs*diskbb} models with the second absorption parameter set free to vary. The data quality does not allow for more complex models.
The spectral fitting was carried out in the energy bands in which the instruments were well calibrated, i.e. 0.5$-$7 keV for \textit{Chandra} and 0.3$-$10 keV for \textit{XMM}.

Since X1, X2 and ULX1 are situated in a region where X-ray diffuse emission is present, we considered different background extraction regions around the sources. We did not find any significant difference in the resulting fits. We also estimated the contribution from the diffuse background emission to each source by fitting the optimally grouped spectra of the X1 and X2 background region using \texttt{tbabs*(mekal+mekal)} \citep{mewe1995update} in \texttt{XSPEC} \citep{arnaud_xspec_1996}. We used data from the 3956 observation. The corresponding region in the 11858 observation does not have enough counts for spectral analysis. As for X1 and ULX2, no additional absorption was found and we fixed $n_{\mathrm{H}}$ at the galactic value. We obtained $kT_1=0.54\pm0.18$ keV and $kT_2=2.70\pm0.80$ keV. The unabsorbed flux in the 0.5$-$7.0 keV band was found to be ($1.03\pm0.15$)$\times$10$^{-14}$ erg cm$^{-2}$ s$^{-1}$. By rescaling this flux to the area of the X1, X2 and ULX1 extraction regions, we found that the flux of the diffuse background emission is at maximum 12\% of the sources fluxes.

The best-fit parameters for X1, X2, ULX1 and ULX2 can be found in Table \ref{tab:xray}. The provided parameter values correspond to the mean values of the posterior distribution, and their 1$\sigma$ uncertainties are the corresponding standard deviations.
For each source and both models, we provide the parameters and fluxes obtained for the 3956 and 11859 \textit{Chandra} observations (for X2, only 3956 results are provided since the quality of the 11858 spectrum was too poor for any meaningful modeling). For X1, ULX1 and ULX2, the best-fit parameters and X-ray fluxes are consistent between the 3956 and 11858 observations at the 3$\sigma$ level.
We also provide the best-fit parameters for the \textit{XMM/EPIC-pn} observation for ULX2. \textit{MOS1} and \textit{MOS2} results are consistent at 1$\sigma$ with the \textit{pn} ones. We observe a flux variability by 60\% between the \textit{Chandra} 3956 and the \textit{XMM} observations of ULX2.
 
\begin{figure}[h!]
    \centering
    \begin{subfigure}[c]{0.85\linewidth}
        \includegraphics[width=\linewidth]{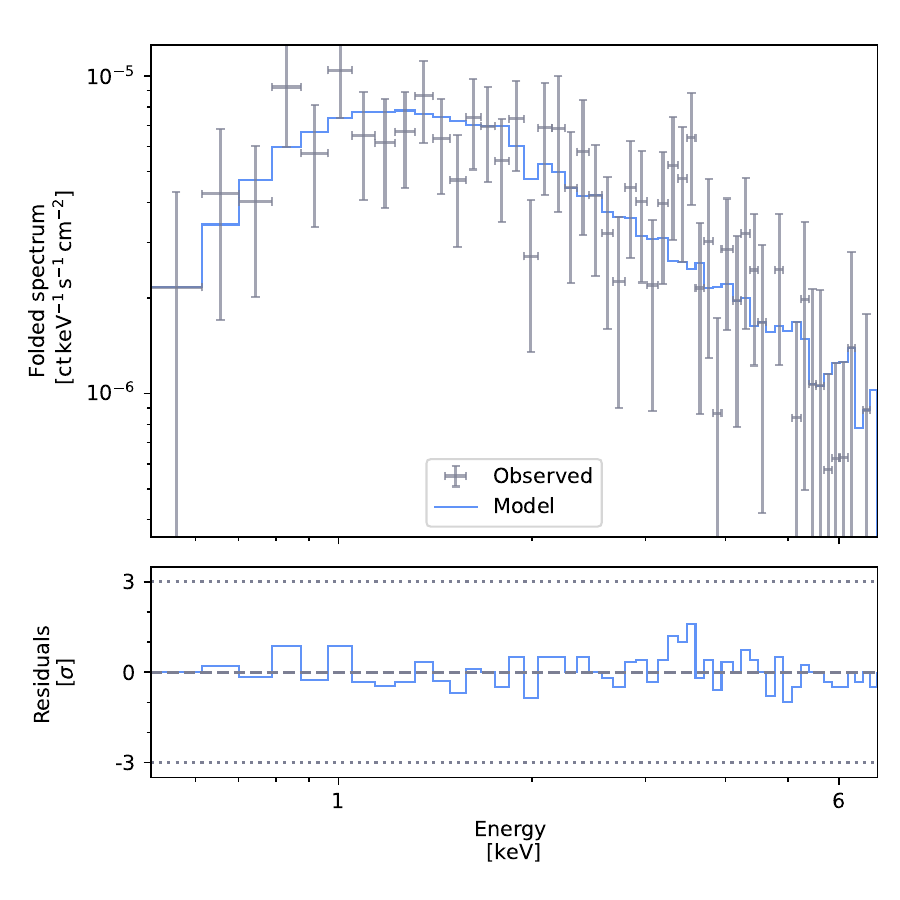}
    \end{subfigure}
    \begin{subfigure}[c]{0.8\linewidth}
        \includegraphics[width=\linewidth]{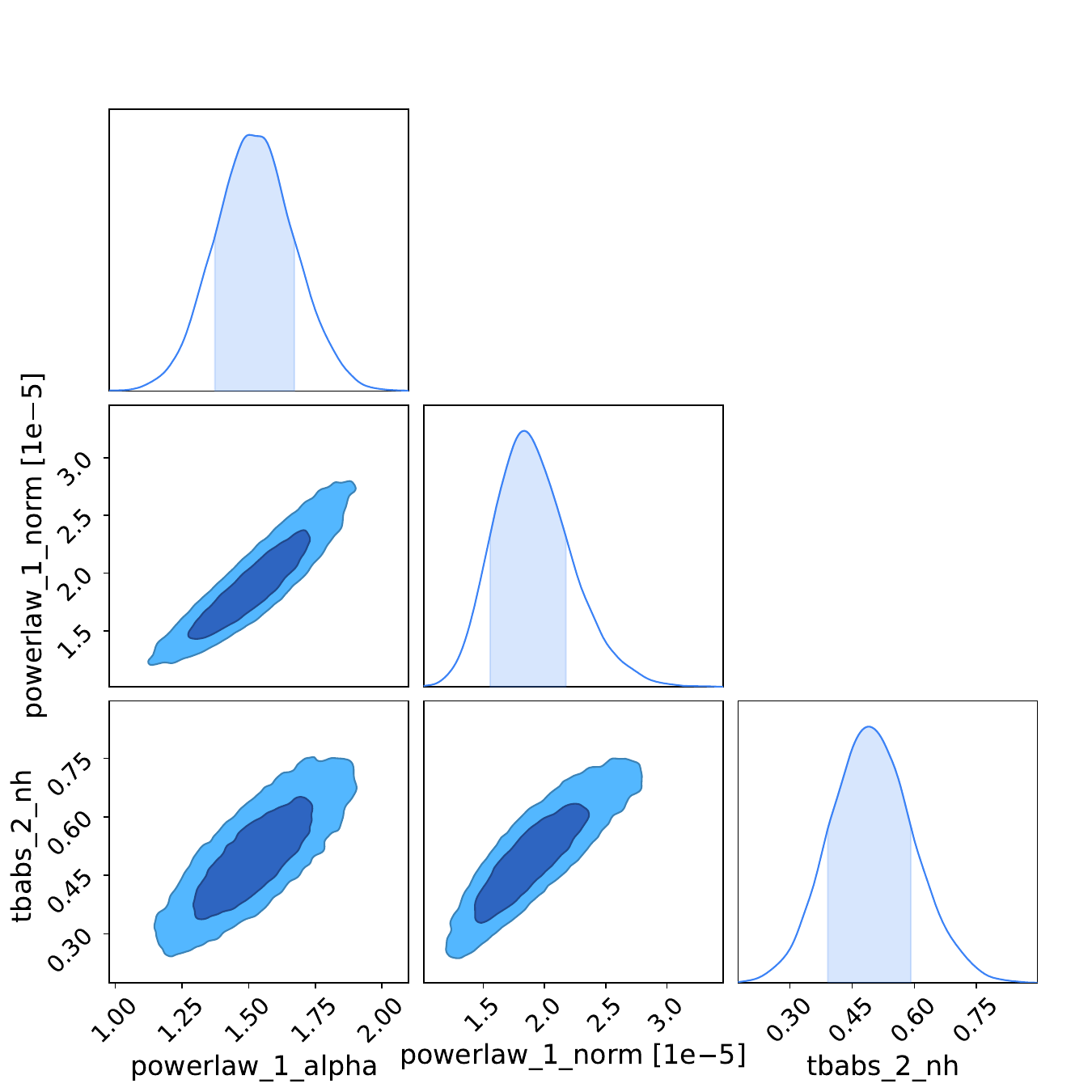}
    \end{subfigure}
    \caption{Top panel: ULX1 best-fit \texttt{tbabs*tbabs*powerlaw} model for the 3956 \textit{Chandra} observation and the corresponding residuals. Bottom panels: corner plots displaying the 1D (with the shaded region showing the 1$\sigma$ value range) and the 2D (with the shaded regions showing the 1$\sigma$ and 2$\sigma$ value ranges) projections of the parameter posterior distribution. $n_{\mathrm{H}}$ is in units of 10$^{22}$ cm$^{-2}$.}
    \label{fig:ulx1_spectrum}
\end{figure}

Figure \ref{fig:ulx1_spectrum} shows the data and the corresponding fit using the \texttt{tbabs*tbabs*powerlaw} model for ULX1 for the 3956 \textit{Chandra} observation. The fit for ULX2 using both \textit{Chandra} and \textit{XMM-Newton} observations for both \texttt{tbabs*powerlaw} and \texttt{tbabs*diskbb} models are shown in Appendix \ref{app:x_spectra}. The residuals show that the \texttt{tbabs*diskbb} model does not fit the ULX2 spectrum well. To evaluate what model described best the spectra for each source, we computed their expected log pointwise predictive densities (ELPD) using Pareto smoothed importance sampling leave-one-out cross-validation (LOO) (\citealt{vehtari_practical_2017}, \citealt{vehtari_pareto_2024}). The ELPD values can only be used in relative terms \citep{sivula_uncertainty_2025}, but for ULX1, X1, and X2 $\Delta$ELPD was $<4$, which is not significant enough to discriminate between the \texttt{powerlaw} and \texttt{diskbb} models. As expected from Fig. \ref{fig:app_ulx2_spectrum}, the \texttt{powerlaw} model is preferred for ULX2 for both \textit{Chandra} and \textit{XMM-Newton} observations ($\Delta$ELPD>19). 
Using the information from the \texttt{powerlaw} fit, we found that the photon index $\Gamma$ for the two ULXs is in the range usually found for ULXs \citep[1.6<$\Gamma$<3.3,][]{gladstone_ultraluminous_2009}. As for the ULX1 \texttt{diskbb} fit, the inner disk temperature $T_{in}$ is also in the expected range \citep[1.06<$T_{in} $ (keV)<1.7,][]{gladstone_ultraluminous_2009}.

\begin{table*}[h]
    \centering
    \caption{Best-fit parameters using an absorbed powerlaw (top) or multi-blackbody (bottom) accretion disk fit and their 1$\sigma$ uncertainties.}
    \renewcommand{\arraystretch}{1.3}
    \begin{tabular}{cccccccccc}
    \hline\hline
        Source & Obs. & \makecell{$n_{\mathrm{H}}$\\ (10$^{22}$ cm$^{-2}$)} & $\Gamma$ & cstat & dof & \makecell{Flux\\ ($10^{-14}$~\ergcms)} & \makecell{Luminosity\\ ($10^{39}$~\ergs)} \\ \hline
        \multirow{2}{*}{ULX1} &  \textit{Chandra} 3956 & 0.50$\pm$0.10 & 1.52$\pm$0.15 & 64 & 50 & 11.8$\pm$2.55 & 13.1$\pm$2.84\\ 
        & \textcolor{gray}{\textit{Chandra} 11858} & \textcolor{gray}{0.67$\pm$0.23} & \textcolor{gray}{2.08$\pm$0.32} & \textcolor{gray}{32} & \textcolor{gray}{42} & \textcolor{gray}{6.46$\pm$2.82} & \textcolor{gray}{7.19$\pm$3.14} \\ 
        \multirow{3}{*}{ULX2} & \textit{Chandra} 3956 & - & 2.11$\pm$0.11 & 44 & 47 & 3.74$\pm$0.31 & 4.16$\pm$0.35 \\
        & \textcolor{gray}{\textit{Chandra} 11858} & \textcolor{gray}{-} & \textcolor{gray}{1.41$\pm$0.22} & \textcolor{gray}{33} & \textcolor{gray}{42} & \textcolor{gray}{4.07$\pm$1.37} & \textcolor{gray}{4.53$\pm$1.52} \\
        & \textit{XMM} & - & 2.03$\pm$0.08 & 63 & 61 & 6.14$\pm$0.46 & 6.83$\pm$0.51\\
        \multirow{2}{*}{X1} & \textit{Chandra} 3956 & - & 1.65$\pm$0.09 & 61 & 51 & 7.29$\pm$0.69 & 8.11$\pm$0.77\\
        & \textcolor{gray}{\textit{Chandra} 11858} & \textcolor{gray}{-} & \textcolor{gray}{1.94$\pm$0.51} & \textcolor{gray}{15} & \textcolor{gray}{18} & \textcolor{gray}{3.79$\pm$2.16} & \textcolor{gray}{4.22$\pm$2.40} \\
        X2 & \textit{Chandra} 3956 & 0.46$\pm$0.19 & 1.28$\pm$0.20 & 55 & 49 & 8.77$\pm$4.19 & 9.76$\pm$4.66\\
        \hline
        Source & Obs. & \makecell{$n_{\mathrm{H}}$\\ ($10^{22}$ cm$^{-2}$)} & \makecell{$T_{in}$\\ (keV)} & cstat & dof & \makecell{Flux\\ ($10^{-14}$~\ergcms)} & \makecell{Luminosity\\ ($10^{39}$~\ergs)} \\ \hline
        \multirow{2}{*}{ULX1} & \textit{Chandra} 3956 & 0.28$\pm$0.07 & 1.88$\pm$0.25 & 62 & 50 & 9.85$\pm$0.62 & 11.0$\pm$0.69 \\ 
        & \textcolor{gray}{\textit{Chandra} 11858} & \textcolor{gray}{0.33$\pm$0.24} & \textcolor{gray}{1.64$\pm$0.47} & \textcolor{gray}{32} & \textcolor{gray}{42} & \textcolor{gray}{4.09$\pm$0.62} & \textcolor{gray}{4.55$\pm$0.69} \\ 
        \multirow{3}{*}{ULX2} & \textit{Chandra} 3956 & - & 0.83$\pm$0.07 & 52 & 47 & 3.35$\pm$0.24 & 3.73$\pm$0.27 \\
        & \textcolor{gray}{\textit{Chandra} 11858} & - & \textcolor{gray}{1.97$\pm$0.55} & \textcolor{gray}{32} & \textcolor{gray}{42} & \textcolor{gray}{3.69$\pm$0.49} & \textcolor{gray}{4.12$\pm$0.55} \\ 
        & \textit{XMM} & - & 0.57$\pm$0.05 & 74 & 61 & 4.73$\pm$0.36 & 6.00$\pm$0.40 \\
        \multirow{2}{*}{X1} & \textit{Chandra} 3956 & - & 1.88$\pm$0.26 & 60 & 51 & 6.49$\pm$0.42 & 7.22$\pm$0.47 \\
        & \textcolor{gray}{\textit{Chandra} 11858} & \textcolor{gray}{-} & \textcolor{gray}{1.64$\pm$0.47} & \textcolor{gray}{15} & \textcolor{gray}{18} & \textcolor{gray}{3.15$\pm$1.08} & \textcolor{gray}{3.51$\pm$1.20} \\
        X2 & \textit{Chandra} 3956 & 0.30$\pm$0.13 & 2.44$\pm$0.54 & 53 & 49 & 6.80$\pm$0.58 & 7.57$\pm$0.65 \\ 
        \hline
    \end{tabular}
    \tablefoot{We fit the spectra with either \texttt{tbabs*powerlaw}/\texttt{tbabs*diskbb} model or \texttt{tbabs*tbabs*powerlaw}/\texttt{tbabs*tbabs*diskbb} model if an additional absorption column $n_{\mathrm{H}}$ is found. For X1 and ULX2, no additional $n_{\mathrm{H}}$ was found. The galactic $n_{\mathrm{H}}$ is fixed at $0.052 \times 10^{22}$ cm$^{-2}$ for all sources. For the \texttt{powerlaw} model, $\Gamma$ is the dimensionless photon index. For the \texttt{diskbb} model, $T_{in}$ is the inner-disk temperature. 
    To assess the goodness of fit, we provide the C-statistics (cstat) \citep{cash_parameter_1979} and the number of degrees of freedom of the fit (dof).
    Flux and luminosity are the unabsorbed values in the 0.5$-$7.0 keV band.}
    \label{tab:xray}
\end{table*}

\subsection{HST imaging}
\label{subsec:hst_imaging}

\begin{figure*}[h!]
    \centering
    \begin{subfigure}[b]{0.19\linewidth}
        \centering
        \includegraphics[width=\linewidth]{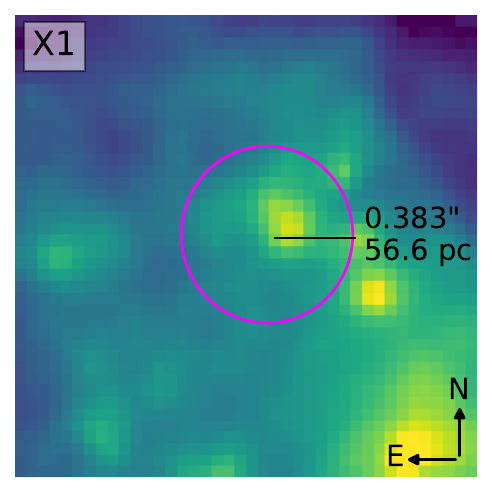}
        \label{fig:x1_f606W}
    \end{subfigure}
    \begin{subfigure}[b]{0.19\linewidth}
        \centering
        \includegraphics[width=\linewidth]{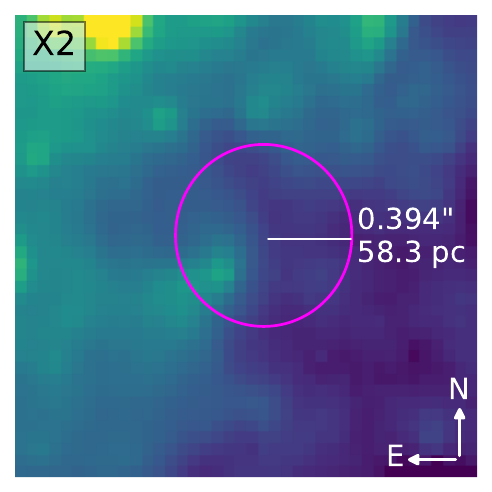}
        \label{fig:x2_f606W}
    \end{subfigure}
    \begin{subfigure}[b]{0.19\linewidth}
        \centering
        \includegraphics[width=\linewidth]{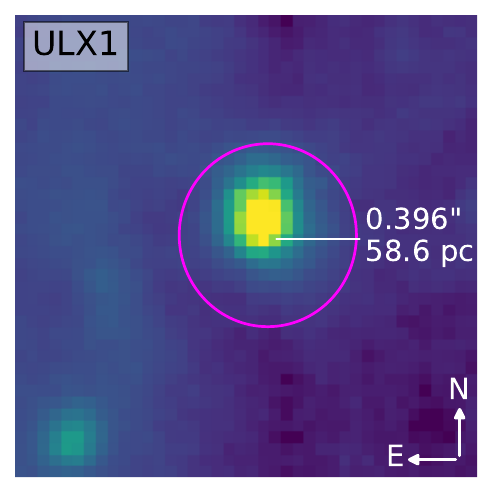}
        \label{fig:ulx1_f606W}
     \end{subfigure}
    \begin{subfigure}[b]{0.19\linewidth}
        \centering
        \includegraphics[width=\linewidth]{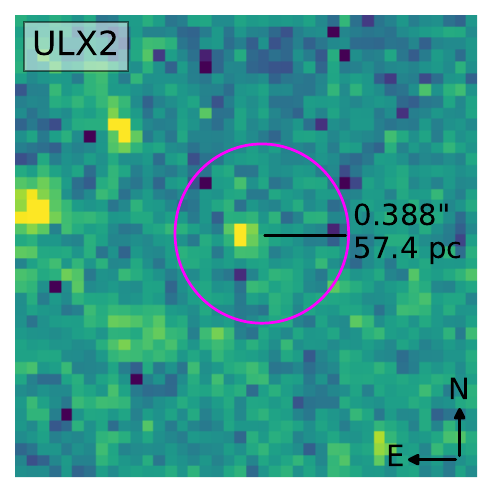}
        \label{fig:ulx2_f606W}
    \end{subfigure}
    \begin{subfigure}[b]{0.19\linewidth}
        \centering
        \includegraphics[width=\linewidth]{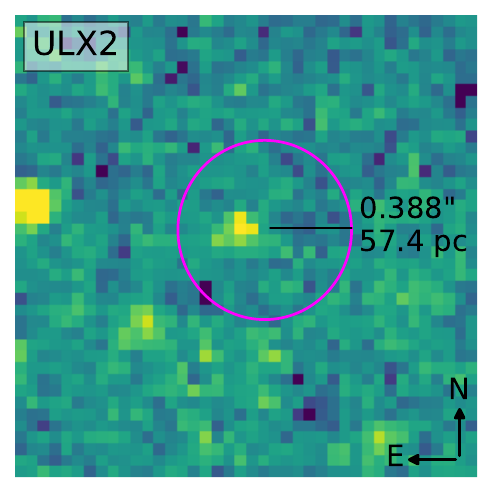}
        \label{fig:ulx2_f435W}
    \end{subfigure}
    \caption{Optical counterparts of the X-ray sources. From left to right: X1, X2, ULX1 and ULX2 in the HST/ACS/F606W filter, ULX2 in the HST/ACS/F435W filter. The magenta circles show the \textit{Chandra} source positions and their uncertainties (see Appendix~\ref{app:astrometry}).
    The corresponding radii are indicated in each image.}
    \label{fig:opt_counterparts}
\end{figure*}

The positions of the four X-ray sources in NGC 2993 and their corresponding error circle radii are given in Appendix \ref{app:astrometry}. Unique optical counterparts in the \textit{Chandra} error circles have been identified for the ULX candidates, as shown in Fig. \ref{fig:opt_counterparts}. The extinction corrected Vega magnitudes of ULX2 are 26.38$\pm$0.28 mag and 26.52$\pm$0.32 mag in the F606W and F435W filters respectively (see Sect.~\ref{subsubsec:photometry}). We determined the 3$\sigma$ Vega limiting magnitudes from 2000 empty apertures around the ULX2 position. We found the extinction corrected values of 27.60 mag and 27.77 mag for the ACS/WFC/F606W and ACS/WFC/F435W filters respectively. Therefore, although the counterpart of ULX2 seems faint, it is detected at the 4.4$\sigma$ level for F606W and 3.9$\sigma$ level for F435W.
We note that the X1 position lies in a star-forming region. There is no evident optical counterpart associated with X2.

\subsubsection{Size of the counterparts}
\label{subsubsec:size_counter}
 
In order to check whether the optical counterparts of the two ULXs are resolved or not in the \textit{HST} data, we created flux radial profiles for several isolated bright stars located in different parts of the images. To do so, we calculated the flux contained in apertures of different sizes centered at the positions of the stars taking into account the background. We used stars that showed similar point spread function (PSF) shapes. For the ACS/WFC/F606W filter, we used eight sources that were classified by Gaia as stars. For the ACS/WFC/F435W filter, we used five sources, three of which were classified by Gaia as stars and two that follow similar profiles. The profiles showed that the counterpart of ULX1 seems extended (see Fig. \ref{fig:profiles}), while the counterpart of ULX2 seems point-like (see Appendix \ref{app:ulx2_profile}). 

\begin{figure}[h!]
    \centering
    \includegraphics[width=0.95\linewidth]{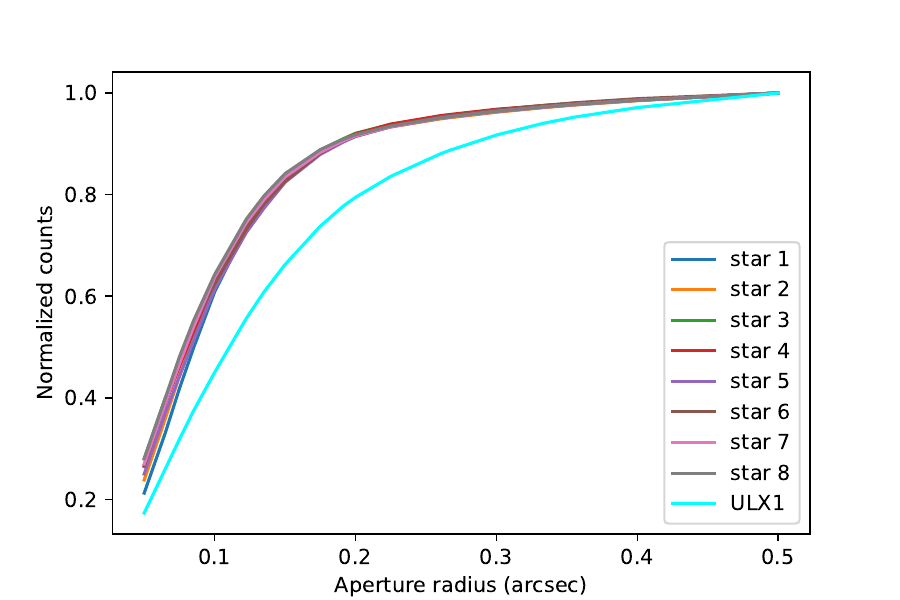}
    \caption{Radial profiles of stars in the \textit{HST} ACS/WFC/F606W observation compared to the radial profile of the counterpart of ULX1.}
    \label{fig:profiles}
\end{figure}

To get the size of the ULX1 counterpart, we first modeled the PSF of the \textit{HST} ACS/WFC/F606W image. To do so, we cut out a 1\arcsec $\times$1\arcsec\ image around each star, to which we fitted a background using a polynomial of degree two and a 2D circular Moffat profile using the \texttt{curve\_fit} function from the python package \texttt{scipy} 1.14.1. 
Then, we modeled a 1\arcsec $\times$1\arcsec\ image around the ULX1 counterpart using the corresponding Moffat functions convolved with a Gaussian using \texttt{fftconvolve} from \texttt{scipy}. We then used the \texttt{least\_squares} function from \texttt{scipy} to find the best parameters for this Gaussian. We obtained the size of the ULX1 counterpart of $0.068\pm0.015$\arcsec\ or 10.1$\pm$2.3 pc in diameter.
We note that a simpler fit where the PSF is also modeled as a 2D Gaussian provides results that are consistent at the 1$\sigma$ level.

For the ULX2 counterpart, we applied the same method to the observations in both ACS/WFC/F435W and ACS/WFC/F606W filters. We found a size compatible with zero. This confirms that the ULX2 counterpart is not resolved in the \textit{HST} images.

\subsubsection{Optical photometry}
\label{subsubsec:photometry}

We used the \texttt{photutils} 2.2.0 package to perform photometry on the optical counterparts of the ULXs. We started by calculating the net counts from the counterparts in a given circular aperture. For the counterpart of ULX2, we used an aperture radius of 0.1\arcsec, which we found to yield the best compromise between maximizing the signal and minimizing the noise. The background was subtracted by taking the median value in an annulus centered on the sources. We set its inner radius to 0.75\arcsec\ and outer radius to 1.25\arcsec. Flux obtained from the aperture photometry needs to be corrected for the finite aperture size. We followed the recommendations from the HST User Documentation\footnote{\url{https://hst-docs.stsci.edu/acsdhb/chapter-5-acs-data-analysis/5-1-photometry}} to do this correction in two steps. First, we corrected the finite aperture flux to 0.5\arcsec\ using the isolated bright stars in the images that are mentioned in Sect. \ref{subsubsec:size_counter}. Then, we applied the correction factor (0.915 for F606W; 0.907 for F435W) provided in the same documentation to transform the flux at 0.5\arcsec\ to the flux at infinity. For the counterpart of ULX1, we calculated the flux in the 0.5\arcsec\ aperture directly, since this source is slightly extended, bright, and no other sources appear in this aperture. We then corrected the flux to infinity as for ULX2. The uncertainties were calculated by propagating the data uncertainties from the source and from the background assuming Poisson statistics.

\begin{table*}[h!]
    \centering
    \caption{Optical photometry results for the ULX counterparts.}
    \renewcommand{\arraystretch}{1.3}
    \begin{tabular}{ccccccc}
    \hline\hline
        Source & Filter & \makecell{Apparent magnitude\\ (mag)} & \makecell{Flux density\\ ($10^{-20}$~erg~s$^{-1}$~cm$^{-2}$~\AA$^{-1}$)} & \makecell{Luminosity\\ ($10^{39}$ erg s$^{-1}$)} & log(f$_X$/f$_V$) & $\xi$ \\ \hline
        ULX1 & F606W & 19.90$\pm$0.01 & 3204$\pm$16
        & 3.31$\pm$0.02
        & 0.24$\pm$0.07 & -- \\
        ULX2 & F606W & 26.38$\pm$0.28
        & 8.1$\pm$2.1
        & (8.42$\pm$2.14)
        $\times$ 10$^{-3}$ & 2.69$\pm$0.11
        & -- \\
        ULX2 & F435W & 26.52$\pm$0.32
        & 15.7$\pm$4.6
        & (5.57$\pm$1.62) $\times$ 10$^{-3}$
        & -- & 19.7$\pm$0.35
        \\ \hline 
    \end{tabular}
    \tablefoot{The filters correspond to the \textit{HST} ACS/WFC instrument. The uncertainties are given at 1$\sigma$. The apparent magnitude is provided in the Vega system.
    log(f$_X$/f$_V$) is the X-ray-to-optical ratio as defined in \cite{maccacaro_medium_1982}. $\xi$ is the X-ray to optical color index. For ULX1, we use the X-ray flux from the 3956 \textit{Chandra} observation to compute the ratios, and for ULX2 from the \textit{pn} instrument on \textit{XMM-Newton}. Using the 3956 \textit{Chandra} observation for ULX2, we obtain log(f$_X$/f$_V$) = 2.5$\pm$0.1 and $\xi$ = 19.1$\pm$0.4.}
    \label{tab:optical_fluxes}
\end{table*}

The fluxes and the apparent magnitudes are given in Table \ref{tab:optical_fluxes}. They have been corrected for the galactic extinction using the extinction of the pivot wavelength of the filter $A_{filt}$. To calculate this value, we used the extinction law $R_V = A_V/E(B-V)$, where $R_V=3.1$ is the parameter of relative visibility from \cite{cardelli_relationship_1989}. $E(B-V)$ at the position of ULXs was taken from \cite{schlafly_measuring_2011} (0.0518 for ULX1, 0.0520 for ULX2). We then obtained the corresponding extinction factor $A_{filt}$ for each filter from the $A_{filt}/A_V$ ratio (0.924 for F606W, 1.32 for F435W) which is calculated using the extinction law by \cite{fitzpatrick_correcting_1999} and where $A_V$ is the visual extinction.
To calculate the luminosities, we integrated the flux in units of erg s$^{-1}$ cm$^{-2}$ \AA$^{-1}$ into units of erg s$^{-1}$ cm$^{-2}$ by multiplying it by the rectangular width of the filters given in the \texttt{stsynphot 1.5.1} \citep{stsci_development_team_synphot_2018} package.

We computed two X-ray over optical flux ratios. First, we used the definition of the X-ray over optical flux ratio from \cite{maccacaro_medium_1982}: $\log(f_X/f_V) = \log(f_X) + m_V/2.5 + 5.37$, where $f_X$ is the unabsorbed X-ray flux in the $0.3-3.5$ keV band and $m_V$ is the visual magnitude. We use the ACS/WFC/F606W magnitude as the visual magnitude (although the ACS/WFC/F606W band is slightly larger). We also computed the X-ray to optical color index, which is defined as $\xi = B_0 + 2.5 \log(F_X)$ \citep{lewin_x-ray_1995}, where $B_0$ is the dereddened B-band magnitude (we use the F435W filter as B-band) and $F_X$ (in $\mathrm{\mu}$Jy) is the unabsorbed flux in the 2$-$10 keV band. The obtained results are given in Table \ref{tab:optical_fluxes}.

For ULX1, the $\log(f_X/f_V)$ ratio is unusual for a ULX, as values around 0 usually indicate AGN \citep{stocke_einstein_1991}. However, this is not surprising in our case, since the ULX1 counterpart appears to be extended, therefore the optical emission does not seem to come from the ULX alone.

For ULX2, the $\log(f_X/f_V)$ ratio is as expected for ULXs \citep{tao_compact_2011}. Its color index falls in the range of low-mass X-ray Binaries, and is consistent with that of other ULXs \citep{tao_compact_2011}. This might indicate that the optical emission from the possible counterpart is dominated by the disk emission instead of the companion star. However, since observations in only two optical filters are available and the X-ray and optical observations are not simultaneous, we cannot draw any firm conclusions on the nature of ULX2 from this color index.


\subsection{MUSE IFU data analysis}

We study the spatially-resolved properties of NGC 2993 in order to look for the influence of the X-ray sources on their host galaxy. MUSE data are well suited to infer stellar and ionized gas kinematics as well as the stellar population mass distribution, age, and metallicity.

\subsubsection{Stellar population properties}
\label{subsubsec:stellar_kinem}

In order to study the stellar kinematics of the NGC 2993 galaxy, we applied the \texttt{PPXF} 9.4.1 code \citep{cappellari_full_2023} to the \textit{VLT}/MUSE observation of the galaxy. This code allows us to model the stellar continuum part of the spectrum, and thus helps to constrain the kinematics of stars in the galaxy. First of all, we considered a wavelength range of 5099$-$5487 \AA, where the spectrum is free of emission lines. The surface brightness map in this range is shown in the top panel of Fig. \ref{fig:PPXF}. It shows an excess of flux at the ULX1 position (see Appendix \ref{app:zooms} for a zoom). X1 seems to be situated in a central region of the galaxy with an intense flux level.
Then, we divided the cube into spatial bins keeping the S/N values above 20 using Voronoi binning \citep{cappellari_adaptive_2003}. \texttt{PPXF} works by fitting a given spectrum to existing stellar population templates. We used templates from the \texttt{E-MILES} \citep{vazdekis_uv-extended_2016} stellar library because it includes the wavelength range of the MUSE instrument with a similar spectral resolution. In order to compare our spectrum to these templates, we de-redshifted the cube.
\texttt{PPXF} fitting was done in two steps\footnote{\url{https://github.com/micappe/PPXF\_examples}}. First, we applied \texttt{PPXF} to the binned cube to find and mask all the emission lines. We then reapplied the code to the cube masking the emission lines. We used multiplicative polynomials of order 4 in the fitting of the continuum. This allowed us to produce stellar kinematics maps of the NGC 2993 galaxy. 

\begin{figure}[h!]
    \centering
    \begin{subfigure}[b]{0.9\linewidth}
        \centering
        \includegraphics[width=\linewidth]{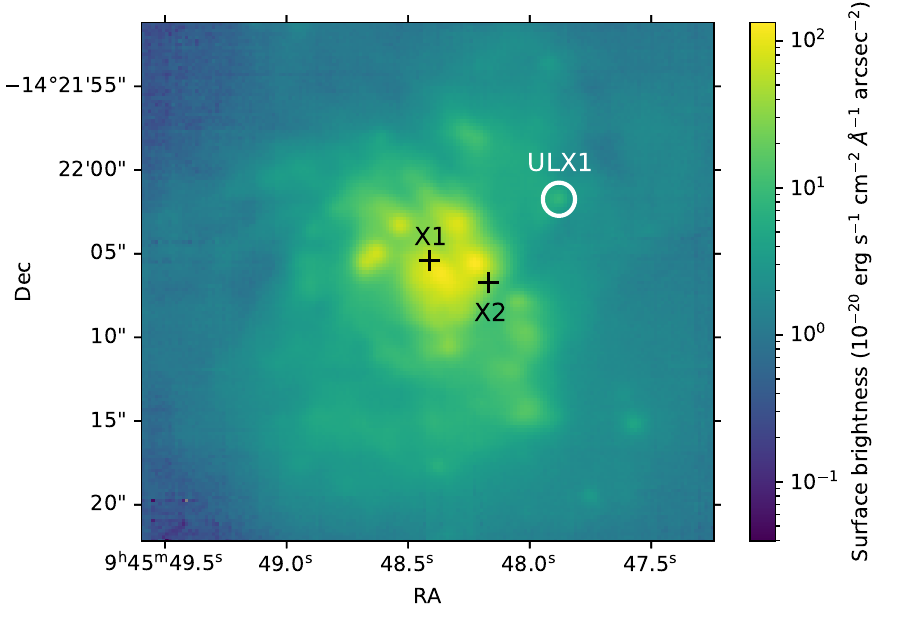}
        \phantomsubcaption
        \label{fig:PPXF_flux}
    \end{subfigure}
    \begin{subfigure}[b]{0.9\linewidth}
        \centering
        \includegraphics[width=\linewidth]{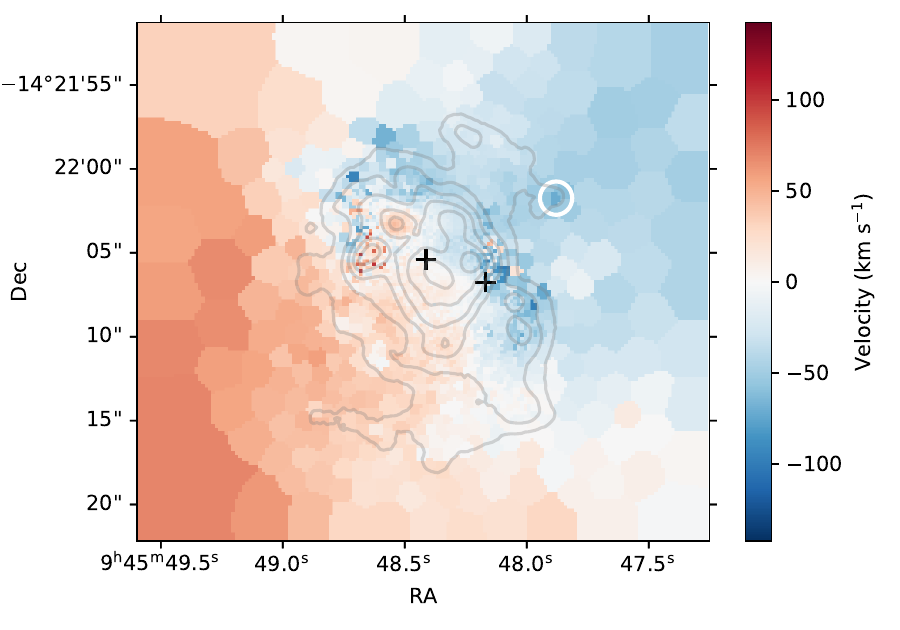}
        \phantomsubcaption
        \label{fig:PPXF_vel}
    \end{subfigure}
    \begin{subfigure}[b]{0.9\linewidth}
        \centering
        \includegraphics[width=\linewidth]{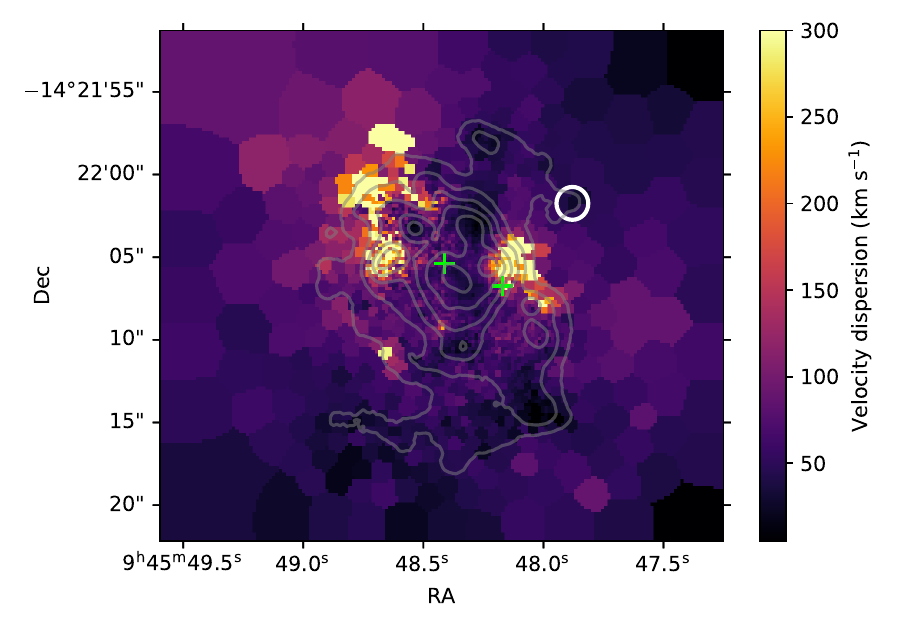}
        \phantomsubcaption
        \label{fig:PPXF_sig}
    \end{subfigure}
    \caption{Stellar kinematics of the NGC 2993 galaxy. The two crosses near the galaxy center correspond to the \textit{Chandra} positions of X1 and X2 sources, while the white circle shows the ULX1 position. Top: Continuum surface brightness from a wavelength range free of emission lines (5099$-$5487 \AA), used as contours in the other two images. Middle: De-redshifted LoS velocity map. 
    Bottom: Velocity dispersion map corrected for the instrumental broadening.}
    \label{fig:PPXF}
\end{figure}

To refine the systemic redshift of NGC 2993, we measured the median stellar velocity (middle panel in Fig. \ref{fig:PPXF}) in an aperture of 0.82\arcsec\ in radius centered on the peak of the continuum flux map (top panel in Fig. \ref{fig:PPXF}). We measured $z=0.007854$, and used this value hereafter as zero point in all line-of-sight (LoS) velocity maps, so that the stellar velocity at the center of the galaxy is null.
The smooth velocity gradient suggests that circular motions constitute the dominant stellar velocity component.
There is a clump of stars located North-East of X1 that has positive velocities of about 40 \kms, while its surroundings have velocities around $-20$~\kms . Its location along the minor axis of the galaxy indicates that it may move towards/outwards the center, or that it rotates in a different plane than the bulk of stars.

\begin{figure}[h!]
    \centering
     \begin{subfigure}[b]{0.5\textwidth}
        \centering
        \includegraphics[width=0.9\linewidth]{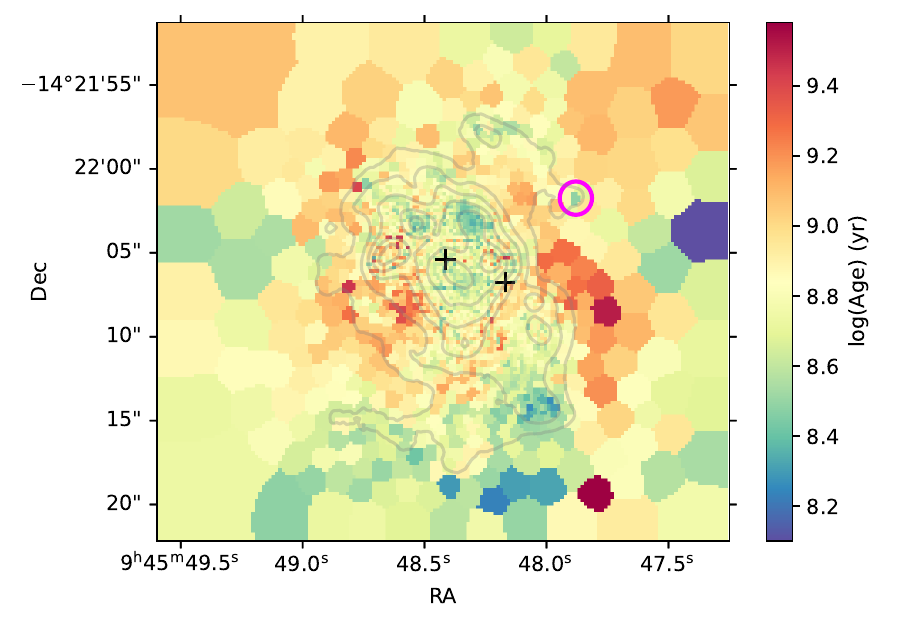}
        \phantomsubcaption
        \label{fig:PPXF_age}
    \end{subfigure}
    \begin{subfigure}[b]{0.5\textwidth}
        \centering
        \includegraphics[width=0.9\linewidth]{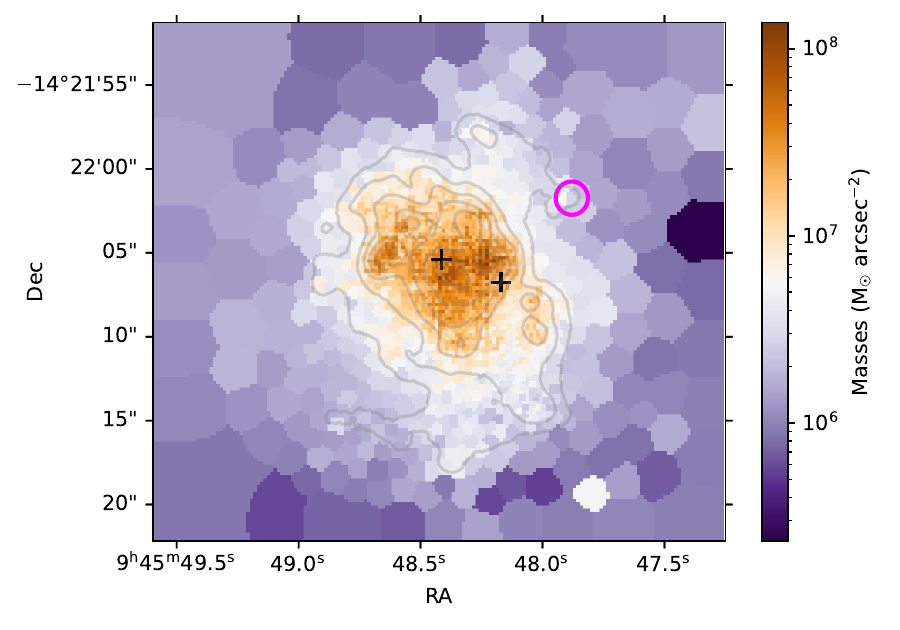}
        \phantomsubcaption
        \label{fig:PPXF_mass}
    \end{subfigure}
    \caption{Age (top) and mass (bottom) distribution maps of NGC 2993 from the \texttt{PPXF} modeling. Crosses and circle are like in Fig. \ref{fig:PPXF}. Contours display the stellar surface brightness (top panel in Fig. \ref{fig:PPXF}).
    }
    \label{fig:PPXF_bis}
\end{figure}

The velocity at the ULX1 position slightly stands out with a value of approximately $-$70 \kms compared to its surroundings at approximately $-$55 \kms. For a zoom-in on the ULX1 position for this and all of the following maps see Appendix \ref{app:zooms}.
The velocity dispersion (bottom panel of Fig. \ref{fig:PPXF}) was corrected for the instrumental broadening using the MUSE line spread function as a function of wavelength from \cite{bacon_muse_2017}. The Voronoi bins with velocity dispersion above 250 \kms\ may be bins where the \texttt{PPXF} fit was not as good as in other bins. If the spectra for these bins are fitted using additive polynomials, which are better suited for the kinematic analysis of galaxies, we find less bins with exceptionally high velocity dispersion. However, we decided to use multiplicative polynomials, as they are better suited for stellar population analysis and for the emission line modeling (see Sect. \ref{subsubsec:gas_kinem}).
The velocity dispersion is lower in the regions corresponding to higher surface brightness. Similar results have been previously observed in MUSE data, indicating that the \hii\ regions with high star formation rate inherit the lower velocity dispersion of the cold molecular gas \citep{breda_large-scale_2024}. This idea is confirmed by looking at the stellar age map of the galaxy (top panel of Fig. \ref{fig:PPXF_bis}).
Indeed, higher surface brightness regions (center and the spiral arms of the galaxy) correspond to younger stellar populations. At the position of ULX1, the stellar age is around $10^{8.5}$ yr, which is lower than the age of its surroundings, where it is around $10^{8.8}$ yr.

To create the mass distribution map shown in the bottom panel of Fig. \ref{fig:PPXF}, we used \texttt{PPXF} to obtain mass-to-light ratios (M/L) for each bin considering the \textit{HST} ACS/WFC/F606W band.
We then computed the observed luminosity in the same band using the \textit{VLT}/MUSE observed flux and the filter's response curve. The mass per bin is then the product of these two quantities. Most of the galaxy mass is concentrated in its center. The mass distribution is positively correlated with the surface brightness distribution in the galaxy. Although there is a surface brightness excess at the ULX1 position, no mass excess is observed, likely due to younger stellar populations with lower M/L values.
We note that since the Voronoi bins show the average value for their pixels, we cannot say that the mass of the bin at the position of ULX1 corresponds exactly to the mass of the ULX1 counterpart. Since the spatial binning decreases the contrast, the M/L ratio from the bin is probably higher than the real M/L at the ULX position (see Sect. \ref{subsubsec:stellar_counter}).
We find the total mass of the galaxy to be $3.61 \times 10^{9}$ M$_\sun$, consistent with the value of $(5.8\pm2.6) \times 10^{9}$ M$_\sun$ from the GLADE+ catalog \citep{dalya_glade_2022}. We expect our value to be slightly lower, since we did not consider the outer parts of the galaxy. 
We also computed the galaxy mass using the average mass-to-light ratio in the galaxy and the luminosity from the \textit{HST} ACS/WFC/F606W image. Using the same field of view (FoV) as in Fig. \ref{fig:PPXF} ($34.1$\arcsec$\times31.0$\arcsec), we obtain a mass of $5.41 \times 10^9$ M$_\sun$. Using a FoV equivalent to the full size of the MUSE image ($62.9$\arcsec$\times62.2$\arcsec), we obtain a mass of $6.82 \times 10^9$ M$_\sun$. These values also fall within the uncertainties of the galaxy mass from GLADE+.

\subsubsection{Gas flux and kinematics}
\label{subsubsec:gas_kinem}

We also studied the gas kinematics of the galaxy traced by emission lines. To do so, we first subtracted the stellar influence from the spectrum. From the \texttt{PPXF} fitting, we get a best-fit model of the stellar continuum, as well as of the H$\alpha$ and H$\beta$ absorption lines for each Voronoi bin. However, we would like to study emission lines of the galaxy pixel by pixel. Therefore, simply subtracting the Voronoi binned \texttt{PPXF} model is not enough, since it does not account for fluctuations that can arise from one pixel to another. Keeping the fit obtained for a Voronoi bin from \texttt{PPXF} and scaling it for each pixel of the bin does not suffice, since the shape of the spectrum changes from one pixel to another. Therefore, we reapplied \texttt{PPXF} to the cube for each individual pixel freezing the velocity and velocity dispersion values as found for their respective Voronoi bins. The resulting \texttt{PPXF} maps follow the same tendencies as the Voronoi binned ones, even though they are noisier. The model of the continuum obtained in this manner was then subtracted from the cube. 
Without this correction, the flux of the H$\alpha$ and H$\beta$ emission lines would be underestimated. However, \texttt{PPXF} may overestimate the intensity of the stellar absorption lines, resulting in the overestimation of the emission lines in the continuum-subtracted cube. It can also lead to the overestimation of the width of the H$\beta$ emission line. Fitting the spectrum using multiplicative polynomials instead of additive polynomials helps to reduce this risk. We view our derived fluxes for H$\alpha$ and H$\beta$ as upper limits.

\begin{figure}[h!]
    \centering
     \begin{subfigure}[b]{\linewidth}
        \centering
        \includegraphics[width=0.9\linewidth]{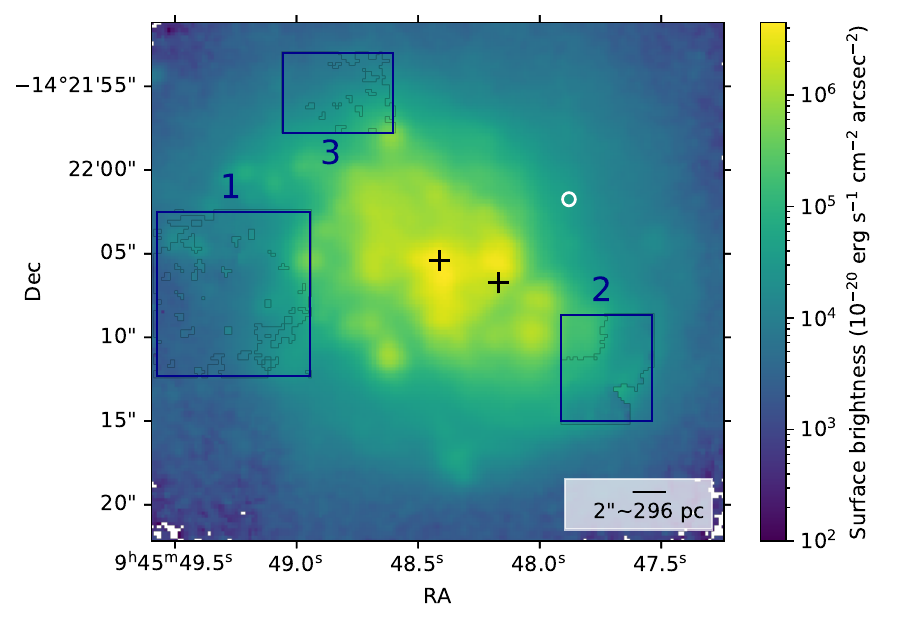}
        \phantomsubcaption
        \label{fig:camel_flux}
    \end{subfigure}
    \begin{subfigure}[b]{\linewidth}
        \centering
        \includegraphics[width=0.9\linewidth]{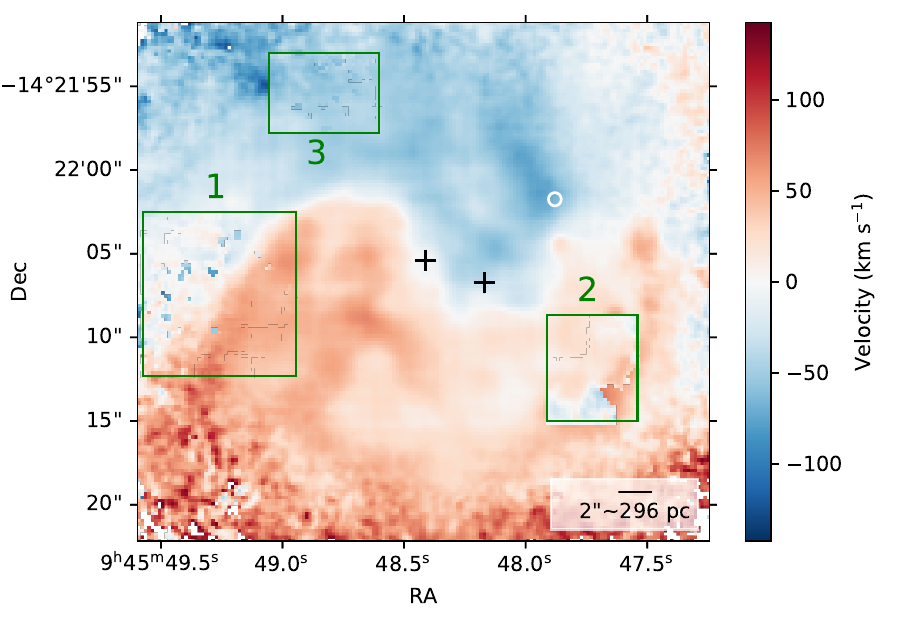}
        \phantomsubcaption
        \label{fig:camel_vel}
    \end{subfigure}
    \begin{subfigure}[b]{0.5\textwidth}
        \centering
        \includegraphics[width=0.9\linewidth]{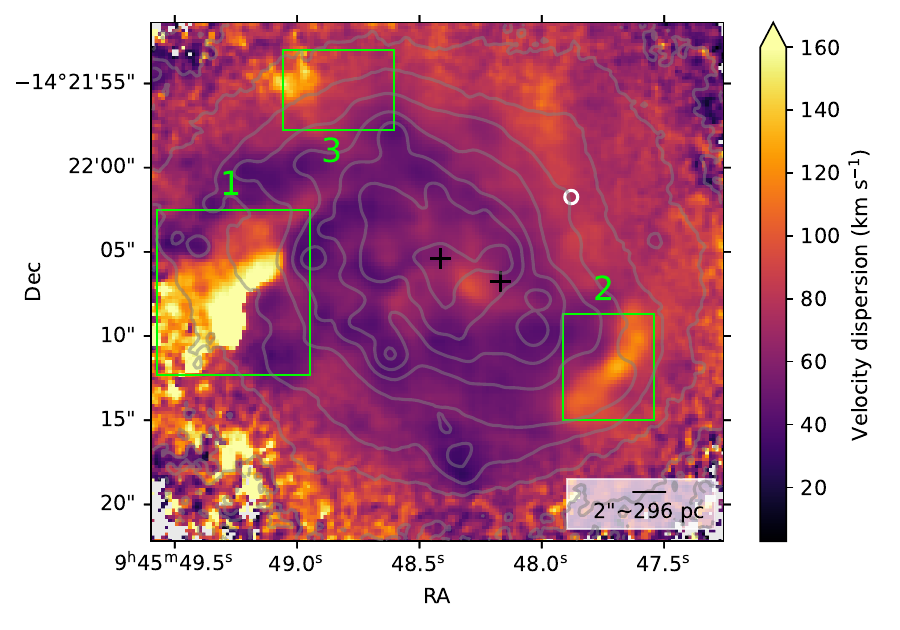}
        \phantomsubcaption
        \label{fig:camel_disp}
    \end{subfigure}
    \caption{H$\alpha$ emission line surface brightness and kinematics for the NGC 2993 galaxy. Crosses are like in Fig. \ref{fig:PPXF}. Here and in the following maps produced using \texttt{CAMEL} the circle shows the \textit{Chandra} ULX1 position with its uncertainty (see Appendix~\ref{app:astrometry}). White pixels correspond to masked areas where the S/N is lower than 4 and/or relative errors are higher than 1.5. The maps in the three rectangular regions correspond to those from the first component (see text). Top: Surface brightness of the galaxy. Middle: LoS velocity map. Bottom: Velocity dispersion map corrected for the instrumental broadening. The contours are from the surface brightness map.
    }
    \label{fig:camel}
\end{figure}

In order to study the gas kinematics of the galaxy, we chose the emission lines usually used for diagnostics in the BPT \citep*{baldwin_classification_1981} diagrams \citep{veilleux_spectral_1987}: H$\beta$, \oiii $\lambda$5007, [OI]$\lambda$6300, \nii $\lambda$6583, H$\alpha$, and the \sii  ($\lambda\lambda$6716, 6731) doublet. We modeled these emission lines by fitting a Gaussian to them with the code \texttt{CAMEL} \citep{epinat_integral_2009}. Best-fit parameters are derived using an iterative minimization technique based on the $\chi^2$-statistics. To get better S/N for each pixel, we applied a 2D Gaussian spatial smoothing with a FWHM of 2 pixels to the cube. We then fit a Gaussian to the emission lines of interest. We fit the \nii\ doublet together with the H$\alpha$ line, setting their velocities and width to be the same. The lines in the \oiii\ doublet and in the \sii\ doublet were also fitted together. We then extracted the flux, LoS velocity shift and velocity dispersion with their corresponding 1$\sigma$ errors for each pixel. Figure \ref{fig:camel} shows the resulting maps for the H$\alpha$ emission line. Maps for other emission lines are shown in Appendix \ref{app:add_camel}.

The gas velocities have been computed from a de-redshifted cube using the systemic redshift found in Sect. \ref{subsubsec:stellar_kinem}. The gas velocity map looks more perturbed than the stellar continuum one.
The orientation of the velocity gradient in the inner region is similar for ionized gas and stars, which suggests that rotation dominates in the central part of the galaxy. However, at larger radii (e.g. in the region 1), the gradient orientation rotates by about 90\degr , which may indicate either a warped gaseous disk or a transition from circular to radial motions at larger radii.
We corrected the velocity dispersion map for the instrumental broadening at the emission line wavelength. The velocity dispersion is lower in the regions with a higher surface brightness, i.e. in the spiral arms of the galaxy, as it is the case with the stellar velocity dispersion. A local anti-correlation between the surface brightness and the gas velocity dispersion in galaxies is expected \citep{denbrok_muse_2020}. There is nothing special in the kinematics maps at the position of X1, X2 and ULX1. There is an intense flux level at the position of X1. \\

\begin{figure}[h!]
    \centering
    \includegraphics[width=\linewidth]{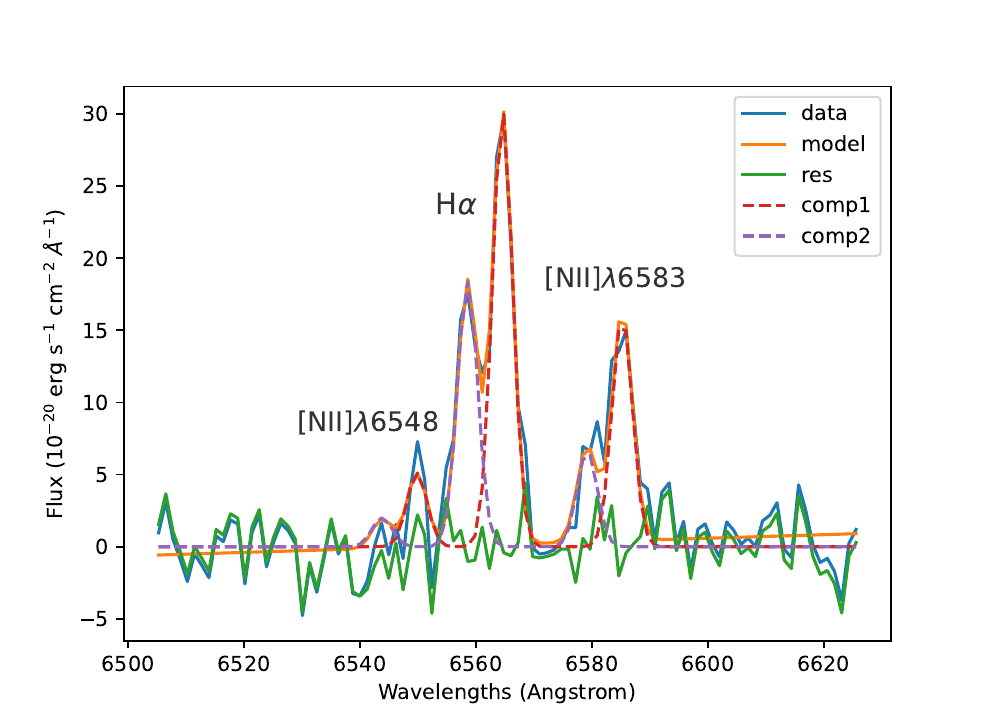}
    \caption{H$\alpha$ line and the \nii\  doublet double-peaked emission lines modeled by two Gaussian profiles using \texttt{CAMEL} for one pixel in the region 1. Common width for the \nii and H$\alpha$ lines were fixed independently for each component.}
    \label{fig:double}
\end{figure}

\begin{figure}[h!]
    \centering
    \begin{subfigure}[b]{0.9\linewidth}
        \centering
        \includegraphics[width=\linewidth]{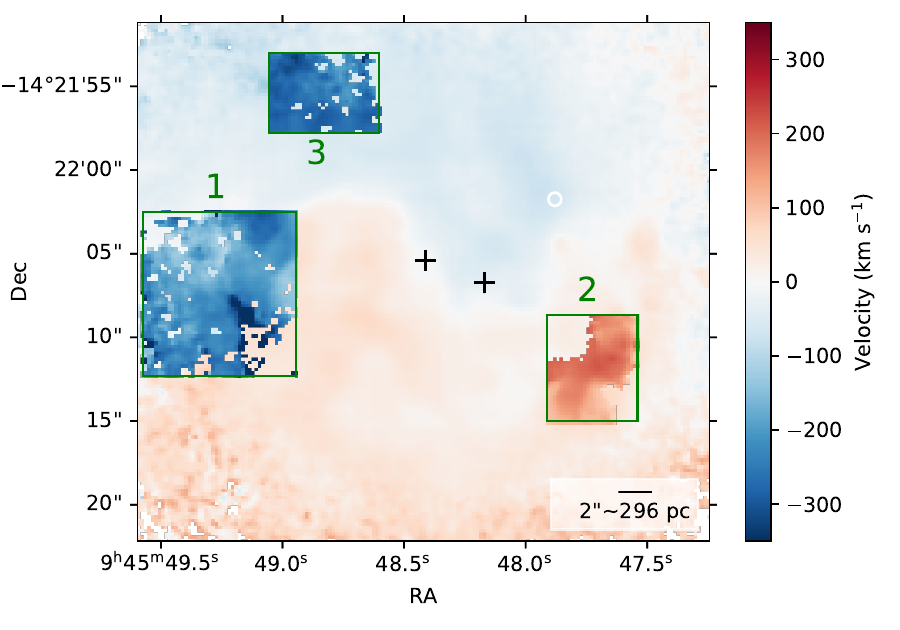}
    \end{subfigure}
    \begin{subfigure}[b]{\linewidth}
        \centering
        \includegraphics[width=1\linewidth]{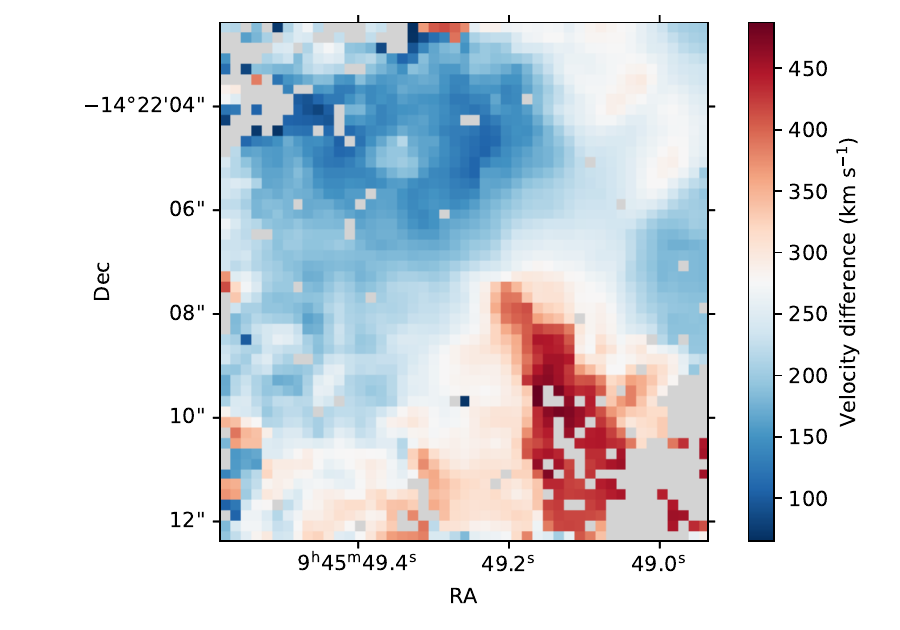}
    \end{subfigure}
    \begin{subfigure}[b]{\linewidth}
        \centering
        \includegraphics[width=1\linewidth]{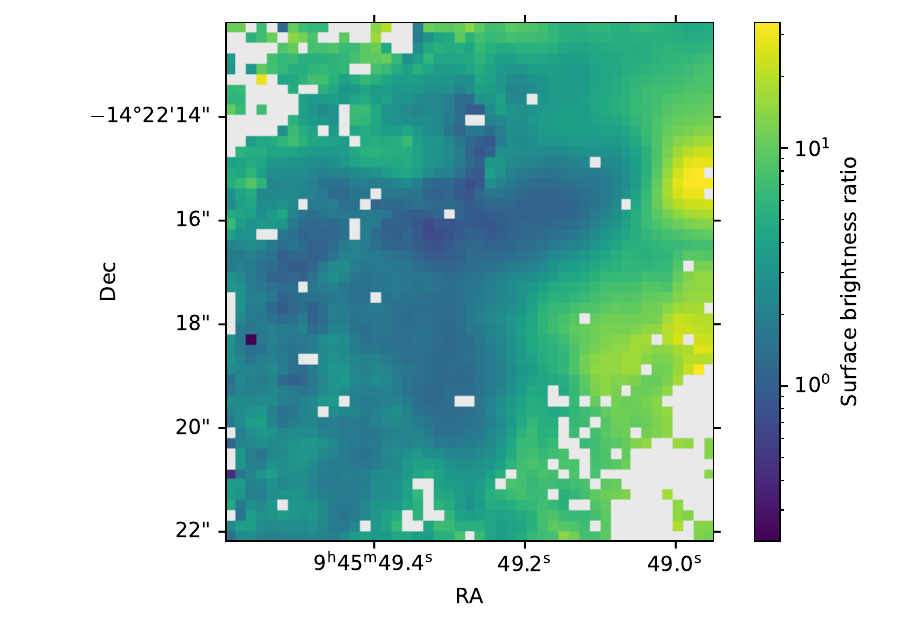}
    \end{subfigure}
    \caption{Maps for the second component. Top: same map as in the top panel of Fig. \ref{fig:camel} with the maps for the second component inserted in the rectangular regions. Middle: velocity difference between the two components for the region 1. Bottom: surface brightness ratio between the two components for the region 1. Grey pixels have S/N<4.}
    \label{fig:double_camel}
\end{figure}

The bottom panel of Fig. \ref{fig:camel} shows that there are three regions with velocity dispersion larger than 120 \kms\ in the galaxy. Gas dynamics appear to be different from their surroundings in these regions. \texttt{CAMEL} originally fits broader lines in these regions, but in reality the emission lines are double-peaked, as illustrated in Fig. \ref{fig:double} for the H$\alpha$+\nii\ complex. We defined rectangular regions near three regions where the velocity dispersion obtained from one component fit is larger than 120 \kms. We note that these regions were defined as to obtain the best S/N (>4) for the double components. Other regions exhibiting a double component line emission may be present in the MUSE cube. We use these three regions as representative of the gas motion in the galaxy. 
Then, we modeled the emission lines in these regions using two Gaussian components. The double components fit was best for H$\alpha$, \nii\ and \sii. Although other emission lines have lower S/N, similar results were found for H$\beta$, \oiii\ and \oi. We separate the two components by their velocities. The velocity of the first component (on the right in Fig. \ref{fig:double}) follows that of the surrounding regions of the galaxy. We inserted the maps for the first component into the top and middle panels of Fig. \ref{fig:camel}. For each region, we only replaced the original surface brightness and velocity values with those obtained from the first component if the global S/N ratio was >\,4. This corresponds to 1880, 639 and 694 spaxels in regions 1, 2 and 3 respectively.
The second component reaches a difference in velocity of around 400 \kms\ with respect to the first one, as shown in Fig. \ref{fig:double_camel}. The flux of the first component is usually stronger than that of the second component, with some exceptions such as the bottom part of the region 2, where the second component is stronger than the first one. In the regions 1 and 3 the second component indicates gas approaching at a velocity above that of the rest of the galaxy (around $-$250 \kms), while in the region 2 it is receding at a velocity around 200 \kms. This suggests the presence of extra-planar gas (see Sect. \ref{subsec:ion_field}).\\
 
\begin{figure}
    \centering
    \begin{subfigure}[b]{\linewidth}
        \centering
        \includegraphics[width=\linewidth]{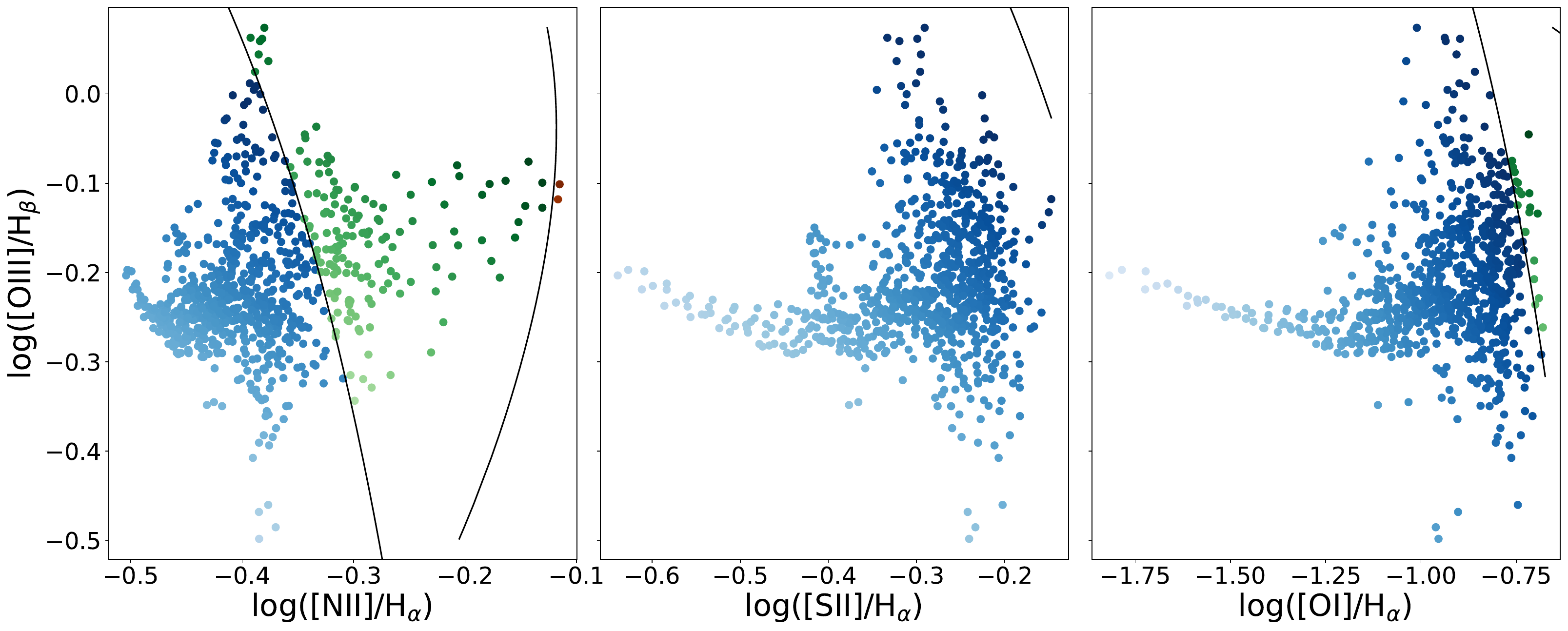}
        \phantomcaption
        \label{fig:bpt_zoom_diag}
    \end{subfigure}
    \begin{subfigure}[b]{\linewidth}
        \centering
        \includegraphics[width=\linewidth]{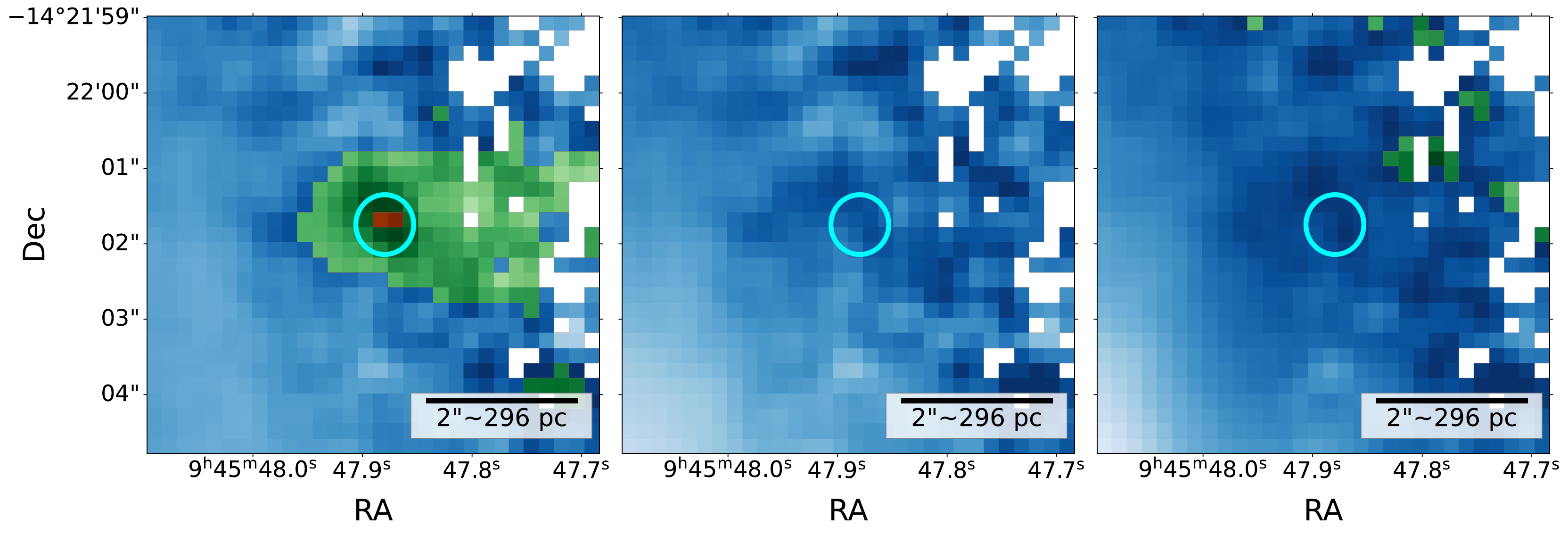}
        \phantomcaption
        \label{fig:bpt_zoom_map}
    \end{subfigure}
    \caption{BPT diagrams (top panel) and their corresponding maps (bottom panel): zoom-in on ULX1 (see Appendix~\ref{app:bpt} for the diagrams and maps of the full galaxy and for more details). The black lines in the BPT diagrams show separations \citep{law_sdss-iv_2021} between the \hii\ region (blue points), intermediate region (green points) and LINER (red points) according to the corresponding emission line flux ratios (given in the x and y axes). Darker colors show points that are close to the separation lines. The maps corresponding to each diagram show the spatial position of each point with the corresponding color.
    White pixels have S/N < 4. The circle is as in Fig. \ref{fig:camel}.}
    \label{fig:bpt_zoom}
\end{figure}

We used the emission line flux ratios to find the main sources of ionization of the interstellar medium (ISM) in the galaxy and in particular near ULX1. 
The corresponding BPT diagrams \citep{baldwin_classification_1981} are provided in Appendix \ref{app:bpt}. 
A zoom-in on ULX1 is shown in Fig. \ref{fig:bpt_zoom}.
We found that the ISM in the galaxy seems to be mainly ionized by stellar radiation, which is to be expected in a starburst galaxy. However, at the position of ULX1 (S/N = 78), the \nii$\lambda$6583/H$\alpha$ ratio values are unexpected. Indeed, this ratio reaches values near 0.8, which are not found elsewhere in the galaxy (except for the pixels on the outer parts of the galaxy with S/N < 15), as seen in Fig. \ref{fig:n2ha} in the Appendix.  According to the BPT classification, such a high \nii$\lambda$6583/H$\alpha$ ratio value may indicate that the surrounding gas is mainly ionized by shocks. Such line ratio variations are not observed for \sii/H$\alpha$, \oi/H$\alpha$ and \oiii/H$\beta$.

Although the H$\alpha$ emission does not significantly change near the ULX1 position, the flux from the \nii\ emission lines grows significantly for several pixels around this position. 
In order to check whether this \nii\ counterpart is resolved or not, we extracted a narrow-band ($6624-6648$ \AA) image around the \nii$\lambda$6583 emission line from the cube. We then compared the ULX counterpart's radial profile to that of a point-like star\footnote{Gaia DR3 5687298253440424320} found in our MUSE observation. To do so, we made 4.8\arcsec $\times$ 4.8\arcsec\ cutout images around both sources. We then fitted a circular 2D Gaussian and a second degree 2D polynomial to the cutout images to extract the FWHM of the star and of the ULX1 counterpart (as described in Sect. \ref{subsubsec:size_counter}).
We find a FWHM of $0.835\pm 0.014$\arcsec\ for the star and of $0.859\pm 0.028$\arcsec\ for the ULX1 \nii\ counterpart, which is compatible with an unresolved source in the MUSE data. 
We also estimated the minimum PSF-corrected FWHM that can be measured with a 1$\sigma$ precision. To do so, we determined the minimum value of the intrinsic source FWHM at which it becomes bigger than its 1$\sigma$ uncertainty computed by taking into account both the PSF and source uncertainties. We found that the source's diameter needs to be at least 0.163\arcsec\ or 24 pc to be resolved at the 1$\sigma$ level. This is larger than the 10.1~pc of the optical counterpart of ULX1 that we inferred from the \textit{HST} data in Sect. \ref{subsubsec:size_counter}. This confirms that the optical counterpart of ULX1 is unresolved by MUSE.


\subsection{Metallicity}
\label{subsec:metal}

\begin{figure}[ht!]
    \centering
    \begin{subfigure}[b]{0.9\linewidth}
        \centering
        \includegraphics[width=\linewidth]{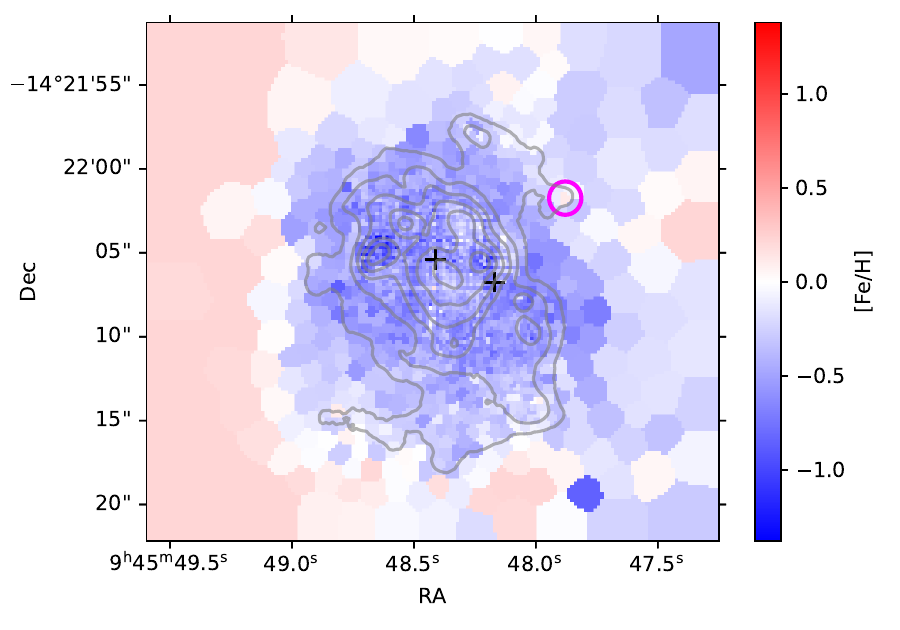}
        \phantomsubcaption
        \label{fig:PPXF_metal}
    \end{subfigure}
    \begin{subfigure}[b]{0.9\linewidth}
        \centering
        \includegraphics[width=\linewidth]{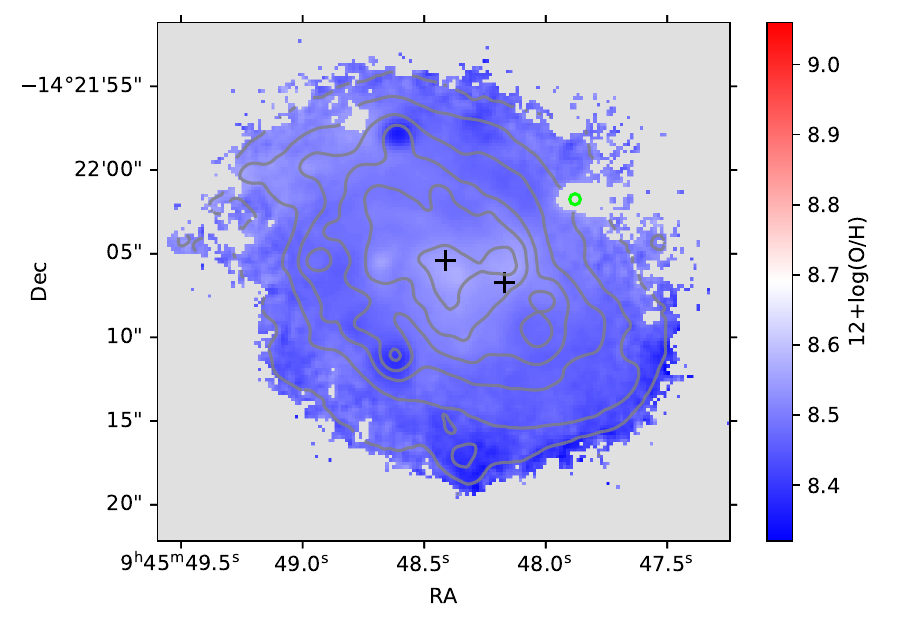}
        \phantomsubcaption
        \label{fig:camel_metal}
    \end{subfigure}
    \caption{Metallicity maps. Crosses and circles are like in Figs. \ref{fig:PPXF} and \ref{fig:camel}. Contours are from the surface brightness map in the top panel of Fig. \ref{fig:camel}. Top: metallicity map obtained by modeling the stellar continuum with \texttt{PPXF} making use of the Voronoi binning (Sect. \ref{subsubsec:stellar_kinem}). This map is in units of [Fe/H], with solar metallicity being equal to 0. Bottom: gas metallicity map obtained from fluxes extracted from nebular emission lines using \texttt{CAMEL} for each individual pixel of the MUSE image from the S2-R3 strong line method. The values are in units of 12+log(O/H), with solar metallicity being 8.69. Gray pixels correspond to masked areas which have S/N<4 or are not \hii\ regions.}
    \label{fig:metal}
\end{figure}

Fig. \ref{fig:metal} shows metallicity maps derived from the stellar populations and from the gas emission lines. The metallicities given in the \texttt{E-MILES} stellar library are in units of [Fe/H], i.e. log((Fe/H)/(Fe/H)$_{\odot}$). So negative values in the top panel in Fig. \ref{fig:metal} show regions with sub-solar metallicity, while positive values show super-solar metallicity.

The bottom panel in Fig. \ref{fig:metal} shows the metallicity map created using emission line fluxes and the S2-R3 calibration from \cite{pilyugin_new_2016}, which uses the \sii$\lambda\lambda6717,6731$ and \oiii$\lambda\lambda4959,5007$ emission line doublets. \cite{pilyugin_calibration-based_2022} showed that this calibration can be applied to IFU data, although it becomes less robust at metallicities higher than 12+log(O/H)=8.55, leading to underestimation of the oxygen abundances. This calibration is used over the R2 one that uses \oii$\lambda\lambda$3727,3729 lines because they are outside of the wavelength range of MUSE. To calculate the metallicity, we use the following flux ratios: \nii($\lambda$6548 + $\lambda$6584)/H$\beta$, \sii($\lambda$6717 + $\lambda$6731)/H$\beta$ and \oiii($\lambda$4959 + $\lambda$5007)/H$\beta$ with line fluxes extracted using \texttt{CAMEL} as described in Sect. \ref{subsubsec:gas_kinem}. 
This method should only be applied to \hii\ regions. Therefore, we have masked the pixels that do not correspond to \hii\ regions. To do so, we used the separation between the main ionization regions from the BPT diagram based on the \nii\ emission line (see Appendix \ref{app:bpt} for the full BPT diagrams and maps on the galaxy; see \cite{gurpide_muse_2022} for a similar approach). This map is in units of 12+log(O/H), with solar metallicity being 8.69.

It is tricky to compare the metallicity map obtained by modeling the stellar continuum to that obtained by modeling emission lines, since the former uses iron abundances and the latter oxygen abundances. To go from one metallicity parameter to another, one needs to know the relation between [Fe/H] and [O/Fe] which can vary depending on the metallicity \citep{mcwilliam_abundance_1997}. However, we can still compare the global tendencies. The galaxy has an overall sub-solar metallicity in both maps. Bigger Voronoi bins where the signal is not as good show slightly super-solar stellar metallicities in some regions, however we should not rely on those values. The gas metallicity approaches solar values in the central regions consistent with higher flux regions and with the position of X1. 
The stellar metallicity map shows a rise at the ULX1 position, with a slightly super-solar metallicity in the corresponding Voronoi bin compared to sub-solar metallicities around it (see Appendix \ref{app:zooms} for a zoom-in).


\subsection{Electron density}
\label{subsec:ne}

\begin{figure}[h!]
    \centering
    \includegraphics[width=\linewidth]{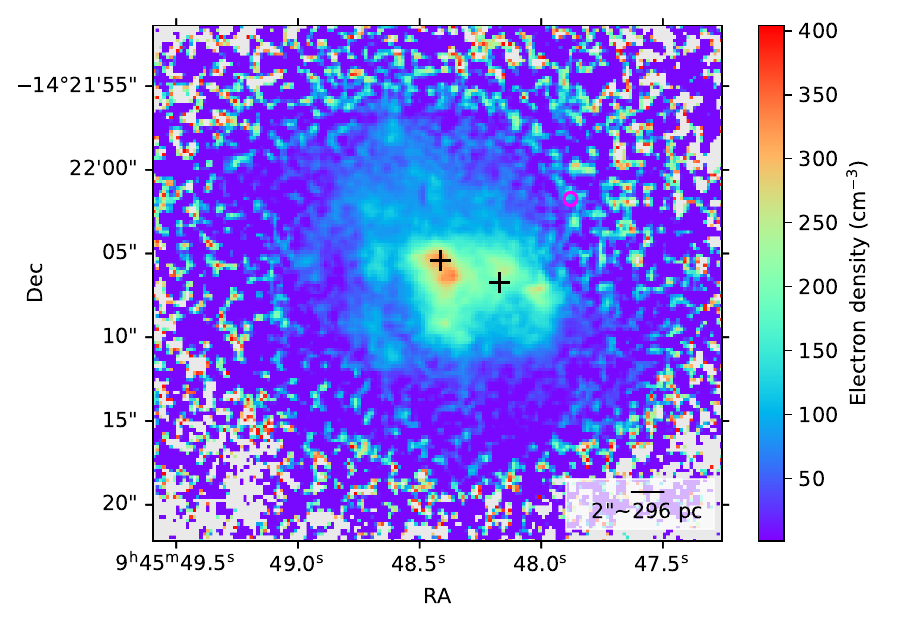}
    \caption{Electron density map from the \sii$\lambda$6716/$\lambda$6731 line ratio. Crosses and circle as in Fig. \ref{fig:camel}.}
    \label{fig:el_dens}
\end{figure}

The \sii$\lambda$6716/$\lambda$6731 ratio traces the electron density of the medium. Using the python package \texttt{PyNeb} \citep{luridiana_pyneb_2015}, we computed the corresponding electron density for each pixel in the galaxy. We assumed the electron temperature of $10^4$ K, since this is a classic value for \hii\ regions and the electron density diagnostic does not have a high dependence on the temperature \citep{osterbrock_astrophysics_2006}. The electron density map is shown in Fig. \ref{fig:el_dens}. The density is higher near the center of the galaxy, with a larger value at the position of X1. There is no significant excess at the position of ULX1.

\subsection{\texorpdfstring{\nai}{NaI} absorption}
\label{subsec:na1}

\begin{figure}[!ht]
    \centering
    \includegraphics[width=\linewidth]{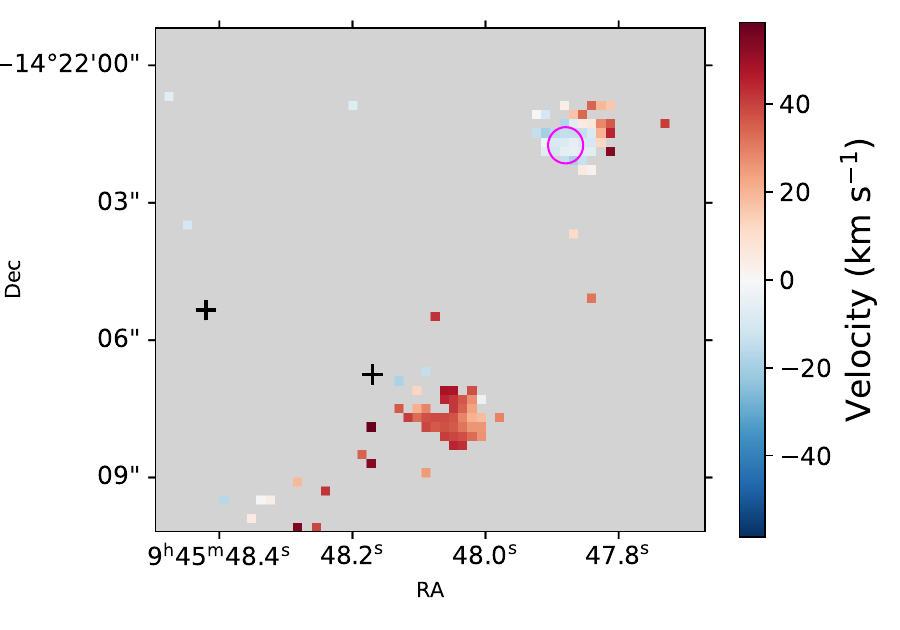}
    \caption{LoS Velocity of the \nai\ absorption lines. Crosses and circle as in Fig. \ref{fig:camel}.}
    \label{fig:na1}
\end{figure}

\nai\ lines are unambiguously observed in absorption at the ULX1 position. We thus modeled the \nai $\lambda$5890 and \nai $\lambda$5895 absorption lines in NGC 2993 using the code \texttt{LiMe} \citep{fernandez_lime_2024}, setting common kinematics for the doublet. We found that the \nai\ absorption doublet is present in only two regions of the galaxy: at the position of ULX1 and in a region to the South-West of X2, as shown in Fig. \ref{fig:na1}. The latter region is associated with elevated surface brightness in both continuum and emission lines (see top panels of Figs. \ref{fig:PPXF} and \ref{fig:camel}). 
The difference between the velocity of the \nai\ lines and the stellar velocity is around 63~\kms\ at the ULX1 position and around 70~\kms\ in the other region. The difference with the gas velocity is around 58~\kms\ at the ULX1 position and 62~\kms\ in the other region. Absorption lines trace the gas that is situated between the galaxy and the observer. Since the LoS velocity from the absorbing gas is higher than that of the gas and stars in the galaxy, this means that it is mostly moving in the direction of the galaxy.
The two regions showing \nai\ absorption are at different azimuth angles but show a similar velocity difference with respect to the ionized gas within the disk of NGC 2993. This, taken together with the fact that the disk has a relatively low inclination ($45$\degr) suggests that the gas in absorption is likely infalling onto the galactic disk.

An alternative interpretation of the velocity offset between the stars and \nai\ lines could be streaming motions within the galaxy disk due to the gravitational effects, for example due to its spiral structure \citep{rupke_spatially_2021}. However, such streaming motions are expected to be more prominent along the minor axis, where radial motions dominate the LoS velocity. Although one of the regions with observed \nai\ emission is indeed located along the minor axis, the other one is not. Importantly, the velocity offset between \nai\ and stars is very similar to the one found between \nai\ and the gas measured using emission lines. There is no strong reason for the streaming motions in the disk to differ in emission and absorption. \cite{shih_complex_2010} observed a decoupling between the velocities of \nai\ absorption and H$\alpha$ emission lines in a late-stage merger galaxy, with \nai\ lines being more blueshifted than the H$\alpha$ line. However, in our case the \nai\ lines are redshifted with respect to the stars and the emitting gas, indicating inflows instead of outflows.

\section{Discussion}
\label{sec:discussion}

\subsection{Galactic kinematics and strong photoionization field}
\label{subsec:ion_field}

The kinematic maps of both stars and gas reveal that the galaxy is perturbed. 
This perturbation may be the result of the occurring interaction of NGC 2993 with NGC 2992. Indeed, from the middle panels of Figs.~\ref{fig:PPXF} and ~\ref{fig:camel} we can see that the kinematic major axis of the stellar component is misaligned with respect to that of the gas component. This is a hallmark of galaxies that underwent a merger in the past \citep{raimundo_increase_2023}.
Since we clearly see a tidal bridge between the two galaxies and a tidal tail on the opposite side of the galaxy (see Fig. \ref{fig:ngc2993}), it is not surprising that some gas is either expelled outside of the disk of the galaxy or is falling back onto it. The velocity of the \nai\ absorption doublet seems to indicate the latter. We observe double emission lines in the galaxy, especially in the three regions marked in Fig. \ref{fig:camel}. One of the components has the same velocity as the surrounding part of the galaxy, meanwhile the other one has an average difference of approximately 250 \kms\ with respect to the first one (see Fig. \ref{fig:double_camel}). The second component may come from the gas that is located outside of the disk of the galaxy and is falling onto the disk. The fact that we are able to observe this gas means that it is ionized. Although the galaxy is kinematically disturbed, BPT diagrams show no evidence for shocks, falling in the purely star-forming regions. This is also the case for the second component in the three studied regions.  The source of the ionization is likely the stars in the galaxy, since this galaxy is in a starburst phase
and could have a strong photoionization field able to ionize gas outside of its disk.


\subsection{Optical counterparts of ULX1}

In this section, we discuss the possible explanations for the counterparts that we observe at the position of ULX1 in the \textit{HST} image (Sect. \ref{subsubsec:stellar_counter}) and in the continuum and the \nii\ lines emission in the MUSE data (Sect. \ref{subsubsec:integ_spec}). We note that our current optical
and X-ray data (see Sect. \ref{subsec:xrays}) is not enough to discriminate between some of them. Deeper observations, or detailed modeling will be necessary to do so.

\subsubsection{Stellar counterpart}
\label{subsubsec:stellar_counter}

Since the ULX1 counterpart is resolved in the \textit{HST} image, we assume that the ULX is located in a stellar cluster. From the \textit{HST} data, the radius of the cluster is 5.1$\pm$1.1 pc (see Sect. \ref{subsubsec:size_counter}). The stellar age at the position of ULX1 is around 0.32 Gyr, with stars there being younger than in its surroundings (see top panel of Fig. \ref{fig:PPXF_bis}). The top panel of Fig. \ref{fig:PPXF} shows that there is a flux excess at this position. 
Using the luminosity obtained in Sect. \ref{subsubsec:photometry} from the \textit{HST} ACS/WFC/F606W image and the M/L-value of 0.47 derived from \texttt{PPXF} for the corresponding Voronoi bin, we computed a mass of $(3.50\pm0.03) \times 10^6$ \Msun\ for the counterpart of ULX1. However, with no uncertainty on the M/L-value provided by \texttt{PPXF}, the quoted uncertainty on the mass only comes from the uncertainty on the luminosity. Since the M/L of the Voronoi bin corresponding to the ULX is lower than that of the surroundings, and the emission in this bin does not come from the cluster alone, the actual M/L in the cluster may be even lower, if its stars are not very massive but very luminous. Therefore, we consider this mass value an upper limit.

The presence of young and bright stars may indicate that we observe a young open cluster. According to \cite{bergh_diameters_2006}, the average age of an open cluster is 0.1 Gyr. In the sample with ages between 0.15 Gyr and 1.5 Gyr where our cluster would be situated, the average radius of an open cluster is 1.5 pc. However, some open clusters in their sample reach radii above 7.5 pc. Therefore, the optical counterpart of ULX1 may be an open cluster situated in the higher bracket of sizes. In addition, a known open cluster Messier 11 has a tidal radius of 29 pc \citep{santos_structure_2005}. However, Messier 11 has a mass of $1.1 \times 10^5$ solar masses, which is an order of magnitude below our value. 
Instead, we may be observing a young globular cluster. According to \cite{mandushev_dynamical_1991}, Galactic globular clusters have a mean mass of $1.9 \times 10^5$ \Msun\ and a maximum mass of $2.4 \times 10^6$~\Msun, but these masses should be considered as lower limits.
In other galaxies, some globular clusters are even more massive, such as Mayall II=G1 in the Andromeda galaxy, with a lower estimate of mass at $14 \times 10^6$ \Msun\ \citep{meylan_mayall_2001}. However, its age is estimated to be around 15 Gyr, which is much older than our estimated age for the ULX1 counterpart. Nevertheless, young globular clusters have been discovered in other galaxies, and especially starburst galaxies, with ages below 1 Gyr \citep{schweizer_young_1999}. Their masses tend to be in the range of $10^{5-7}$ \Msun, which is in agreement with the age and the mass found for the ULX1 counterpart. However, the radii of young globular clusters are usually found to be around 3 pc, which may be too small for our object.

Another plausible explanation is that the stellar cluster associated with ULX1 is a nuclear star cluster (NSC; \citealt{neumayer_nuclear_2020}) that remains of its disrupted dwarf galaxy progenitor that used to be a satellite to NGC 2993. In the Milky Way, $\omega$ Cen is widely considered to be a NSC of its disrupted progenitor dwarf galaxy (\citealt{tsuchiya_capture_2003}, \citealt{souza_omegacat_2026}), due in part to the fact that it is the most massive globular cluster in the Milky Way with a mass of $3.55 \times 10^6$ \Msun\ \citep{baumgardt_catalogue_2018} and to the presence of a possible IMBH in its center \citep{haberle_fast-moving_2024}. Its half-mass radius is 10.04 pc \citep{baumgardt_catalogue_2018}. Assuming ULX1 is situated in such a tidally-formed NSC, the X-ray emission might come from the accreting massive central black hole of the disrupted dwarf galaxy. The ULX1 X-ray luminosity of 10$^{40}$ \ergs\ is in the range of luminosities for accreting massive black holes of dwarf galaxies \citep{mezcua_intermediate-mass_2018}. We also note that the photon index of ULX1 ($\Gamma \sim 1.5$) is consistent with values found in AGN. The observed gas in-falling into NGC 2993 at the position of ULX1 (see Sect \ref{subsec:na1}) might be seen as a remaining tidal tail. 
According to \cite{neumayer_nuclear_2020}, NSCs situated in late-type galaxies are made up of stars with different ages, usually dominated by stars older than a few Gyr but also often containing stars younger than 100 Myr \citep{kacharov_stellar_2018}, especially in NSCs located in earlier-type spirals \citep{rossa_hubble_2006}. The presence of younger stellar populations in NSCs may indicate star-formation periodically triggered by in-falling gas.


\subsubsection{MUSE counterpart}
\label{subsubsec:integ_spec}

While our spatially-resolved maps have already revealed the properties of the environment of ULX1, in particular its enhanced \nii/H$\alpha$ ratio (e.g. Figs~\ref{fig:bpt_zoom}), we extracted a spectrum of the ULX counterpart to understand the cause of such an enhancement. In particular, we wanted to distinguish the ionization that may be related to the ULX from the ionization due to the stellar UV radiation, which seems to be present everywhere in the galaxy (see Appendices \ref{app:camel} and \ref{app:bpt}). We extracted an average spectrum from an aperture with a diameter around 10\% smaller than the PSF FWHM at the energy of the \nii$\lambda$6583 line to make sure that we avoid contamination from regions external to the ULX. We also extracted an average spectrum from the background which is dominated by ionization from HII regions.

We modeled the emission lines for the source and background spectra using \texttt{CAMEL}. This was done both for spectra obtained from the original cube and from the continuum-subtracted cube (corrected for the stellar continuum and absorption lines, see Sect. \ref{subsubsec:stellar_kinem}). This allows us to have an upper limit for the flux ratios, since in the original cube the H$\alpha$ and H$\beta$ emission lines are underestimated. In the continuum-subtracted cube they may be overestimated, in that case the corresponding values should be seen as the lower limit. We inserted the resulting line ratios into the BPT diagrams created using the continuum-subtracted cube in Fig. \ref{fig:bpt_points} (see Appendix \ref{app:bpt} for the full galaxy diagrams and associated maps). 
As expected, H$\alpha$ and H$\beta$ fluxes rise in the continuum subtracted cube, therefore the line ratios become smaller. 
Subtracting the background does not have a significant effect on the \sii\ and \oi\ fluxes, however the \nii/H$\alpha$ ratio rises significantly. We also observe a rise in the \oiii/H$\beta$ ratio in the continuum and background subtracted spectrum. In the original background subtracted spectrum the H$\beta$ line is not detected, and we can only provide a lower limit on the corresponding \oiii/H$\beta$ ratio.\\

\begin{figure}[ht!]
    \centering
    \includegraphics[width=1\linewidth]{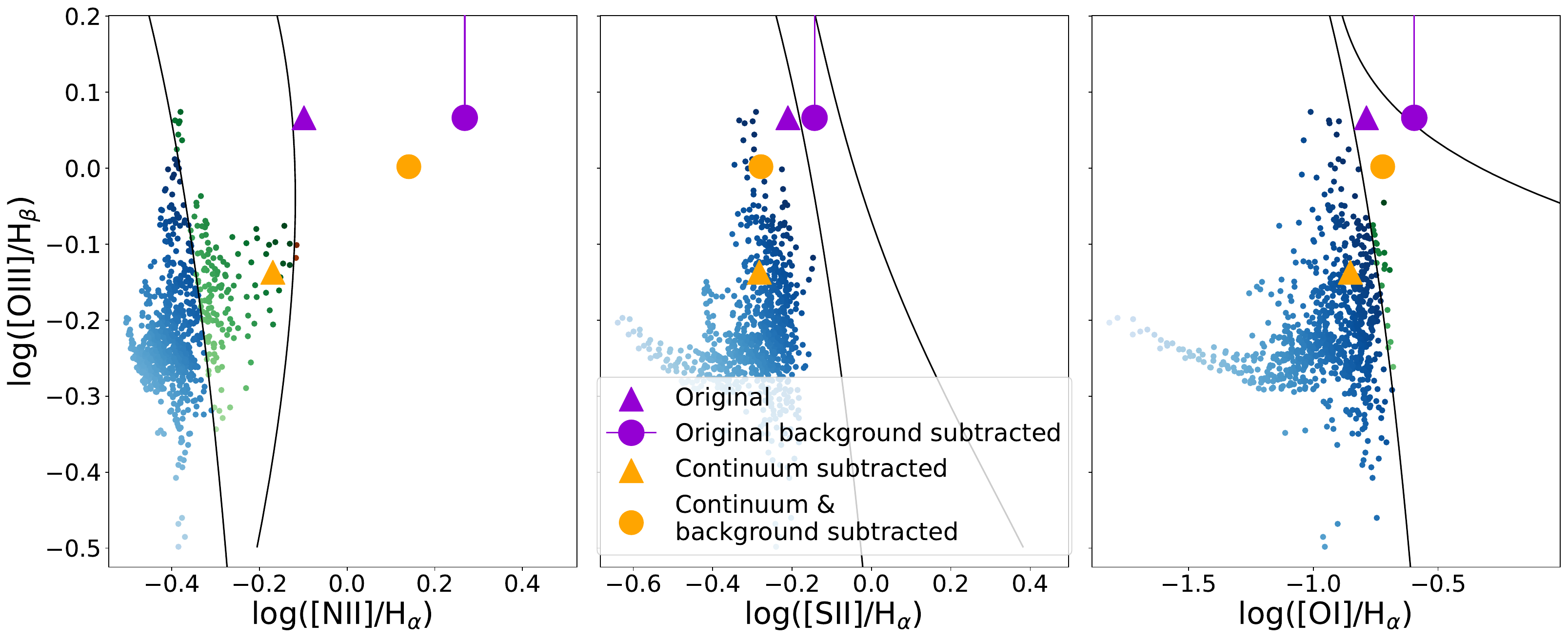}
    \caption{BPT diagrams (see Fig. \ref{fig:bpt_zoom}), where points correspond to the line ratios obtained from an average spectrum extracted in an aperture around the ULX1 counterpart. Purple points show results from the original cube, orange points show results from the cube from which the continuum and the absorption lines have been subtracted. Circles correspond to the background-subtracted line ratios, and triangles show ratios without subtracting the background. The y-axis position of the purple circle is a lower limit.}
    \label{fig:bpt_points}
\end{figure}

Hereafter, we outline some possible explanations for the \nii/H$\alpha$ enhancement. We note that the data quality does not allow us to fully establish its nature, but we can rule out some scenarios. These explanations can be divided into those invoking the ULX as the source of the enhancement (Sects. \ref{par:photoionization} and \ref{par:shocks}) and an overabundance of nitrogen (Sect. \ref{par:N_enrich}), although they are not necessarily mutually
exclusive.

\paragraph{Hard photoionization} 
\label{par:photoionization}
Perhaps the most obvious explanation is the influence of the ULX itself via its X-ray excitation. At face value, our BPT diagrams do not seem to support this scenario (Figs.~\ref{fig:bpt_zoom} and ~\ref{fig:bpt_points}). Particularly, the lack of enhanced \oiii/H$\beta$ and \oi/H$\alpha$ seen in other ULX ionized regions \citep[e.g.][]{gurpide_muse_2022} would argue against this scenario. However, the coarse spatial resolution with respect to the size of the source can easily dilute the line ratios. For instance, the enhanced \oi/H$\alpha$ produced in outer regions of X-ray photoionized nebulae by highly penetrating X-rays (\citealt{gurpide_muse_2022}, \citealt{gurpide_quasi-isotropic_2024}) will get diluted in integrated spectra. In our case, this dilution effect can be enhanced, given that stellar ionization of the surrounding ISM is strong (Sect. \ref{subsec:ion_field}), especially if the ULX is embedded in a stellar cluster (Sect.~\ref{subsubsec:stellar_counter}). Whether such a combination of factors (spatial integration coupled with strong stellar continuum contribution) can explain the absence of significantly enhanced \oiii/H$\beta$ while boosting \nii/H$\alpha$ cannot be answered without detailed photoionization modeling, which is beyond the scope of this work.


\paragraph{Shocks} 
\label{par:shocks}
To see if the \nii/H$\alpha$ ratio can be explained by shocks, we used the shock-only and shock+precursor models computed by \cite{alarie_extensive_2019} using the \texttt{MAPPINGS V} code \citep{sutherland_mappings_2018}. We tried two sets of grids: an update of the \cite{allen_mappings_2008} grid that was previously calculated using \texttt{MAPPINGS III} with solar abundances, and an extension of this grid computed for solar abundances from \cite{gutkin_modelling_2016}.
Only some shocks-only models with shock velocities $>250$ \kms\ can explain high \nii/H$\alpha$ ratios together with low \oiii/H$\beta$ ratios. However, when shock velocities are higher than 150 \kms, a precursor is expected \citep{dopita_spectral_1996}. No model with a precursor can explain the observed emission line ratios. For more information, see Appendix \ref{app:mappings}. Furthermore, these combinations of parameters cannot explain the observed \sii/H$\alpha$ and \oi/H$\alpha$ ratios. These ratios do not change at the ULX1 position compared to its surroundings.
In addition, a good indicator of the presence of shocks is the line broadening at the position where the shock is expected. However, no such broadening was observed at the position of ULX1 for any of the studied emission lines (see bottom panel of Fig. \ref{fig:camel}, Appendix \ref{app:camel}). Therefore, we conclude that the high \nii/H$\alpha$ ratio is unlikely to be explained by shocks, given that the only models that reach this ratio are not physically motivated.


\paragraph{Local enrichment in nitrogen}
\label{par:N_enrich}

A more likely explanation is a localized enrichment of nitrogen. In that case, BPT diagrams may classify the corresponding region as a LINER \citep{reyes-perez_consistent_2015}. 
Nitrogen-rich bubbles are observed around some Wolf-Rayet (WR) stars (\citealt{esteban_spatially_1992}, \citealt{stock_herschel-pacs_2014}, \citealt{reyes-perez_consistent_2015}). Evidence of a ULX having a WR companion has been found \citep{liu_puzzling_2013}, although this evidence is contested \citep{laycock_revisiting_2015}.

To quantify whether the observed nitrogen abundance can be produced by a single star, we calculated the nitrogen abundance at the ULX1 position. To do so, we needed the corresponding electron temperature and the electron density. To calculate these, we used the continuum and background subtracted spectrum summed over an aperture around the ULX1 counterpart (see beginning of Sect. \ref{subsubsec:integ_spec}). We corrected all the line fluxes for the galactic extinction using the extinction law from \cite{cardelli_relationship_1989} with R$_V$=3.1 and E(B-V)=0.0520 \citep{schlafly_measuring_2011}. We then also corrected the fluxes for the extinction within the galaxy using the Balmer decrement and the extinction law from \cite{calzetti_dust_2000} with R$_V$=4.05 (see Appendix \ref{app:extinction} for more details and the extinction map of the galaxy). 
As described in Sect. \ref{subsec:ne}, we calculated the electron density from the \sii\ lines ratio and obtained 130 cm$^{-3}$. An indicator of the electron temperature is the ($\nii\lambda6583+\nii\lambda6548) / \nii\lambda5755$ ratio. The \nii$\lambda$5755 emission line is not detected at the position of ULX1. Therefore, we calculated its 3$\sigma$ detection limit which we used as an upper limit for flux (< 5.0 $\times$ 10$^{-18}$ erg~s$^{-1}$~cm$^{-2}$). Thus, we found an upper limit of 23900 K on the electron temperature. This allowed us to calculate a lower limit on the nitrogen abundance ($N^+/H^+$)\footnote{We use $N^+$ and $H^+$ to talk about the singly ionized nitrogen \nii\ and hydrogen, respectively.} at the position of ULX1 using the function \texttt{getIonAbundance} from \texttt{PyNeb}. We found $12+\log{(N^+/H^+)} > $7.17. From this lower limit alone, we are not able to conclude whether the nitrogen abundance at the ULX1 position indicates the presence of a WR nebula or a simple \hii\ region.

The nitrogen mass $M(N)$ is obtained from the $\nii\lambda6583$ line intensity, the line emissivity $\epsilon(\nii\lambda6583, T_e)$, 
the electron density $n_e$, and the $N^+$ ionic fraction, using the following equations:

\begin{equation}
    M(N) = \frac{N}{N^+} \times M(N^+) = \frac{N}{N^+} \times n_{N^+} \times V \times m_N, 
\end{equation}

where $N/N^+$ is the inverse of the $N^+$ ionic fraction, $n_{N^+}$ is the $N^+$ density, $V$ is the volume of the emitting region, and $m_N$ is the mass of the N atom. The product $n_{N^+} \times V$ can be derived from:

\begin{equation}
    L(\nii\lambda6583) = \epsilon(\nii\lambda6583, T_e) \times n_e \times n_{N^+} \times V, 
\end{equation}

where $L(\nii\lambda6583) = F(\nii\lambda6583) \times (4\pi D^2)$. $F(\nii\lambda6583)$ is the total continuum and background subtracted flux in the aperture corrected from extinction. $D$ the distance to the galaxy. This leads to:

\begin{equation}
    M(N) = \frac{N}{N^+} \times \frac{L(\nii\lambda6583) \times m_N}{\epsilon(\nii\lambda6583, T_e) \times n_e}
\end{equation}

Following \cite{peimbert_densities_2013} and using \texttt{PyNeb}, the line emissivity (in erg~s$^{-1}$~cm$^{3}$) is approximated by a function of $T_e$ (in K) of the form:

\begin{equation}
    \epsilon(\nii\lambda6583, T_e) = 1.213\times 10^{-19} \times T_e^{-0.35} \times e^{-21000/T_e}
\end{equation}

This leads to the estimation of the nitrogen mass of:

\begin{equation}
\begin{split}
    M(N) = 0.202 M_\odot \times & \frac{N}{N^+} \times \frac{130}{n_e} \times \\
    & \times \frac{0.0122}{T_e^{-0.35} \times e^{-21000/T_e} } \times \left[\frac{D}{30.5 \mathrm{Mpc}}\right]^2
\end{split}
\end{equation}

where 0.202 \Msun\ is the lower limit on the nitrogen mass in the emitting gas computed using $T_e = 23900$ K and $N/N^+ = 1$.\\

In the following, we explore possible scenarios that might explain the observed apparent nitrogen enrichment. However, since we are only able to obtain a lower limit on the nitrogen mass, these scenarios will require additional data to be fully tested. 
Given that the usual mass of a star in a WR stage is between 10 and 25 \Msun\ \citep{crowther_physical_2007}, our lower limit on the nitrogen mass is around 1\% to 2\% of the total mass of the star, and we cannot exclude that this amount of mass may have been ejected from a WR star. However, if we take $T_e = 8000$ K and $N/N^+ = 1.5$, we estimate a nitrogen mass of 1.18 \Msun. According to \cite{brinchmann_galaxies_2008}, a WR star ejects a median value of nitrogen mass of 0.014 \Msun\ over a period of 2 Myr, therefore several WR stars may be needed to produce this amount of nitrogen mass. 
Sizes of WR bubbles are usually below 10 pc in radius. The \nii\ counterpart is not spatially resolved by MUSE, so we only know that it has a diameter below 24 pc.
In addition, the ULX seems to be situated in a star cluster resolved in the \textit{HST} image (Sect. \ref{subsubsec:stellar_counter}). Thus, it is plausible that several WR stars are responsible for the apparent nitrogen enrichment. WR stars of WC-type (those that produce carbon lines) are associated with a red bump at around 5808 \AA\ from the \civ\ emission lines, which we do not observe in our spectrum. To confirm the presence of WN-type WR stars (those that produce nitrogen lines), we would need to observe this source with a spectrograph with a wavelength range extending further into the blue to be able to look for the blue bump at around 4686 \AA. This feature is a blend of several emission lines (such as \heii, \niii, \nv, \ciii\ and \civ). 

From these observations we cannot confirm whether the \nii\ excess is related to ULX1, although it is surprising to observe an overabundance of nitrogen that solely appears in the part of the galaxy associated with an ULX source.
ULXs seem to be similar to high-mass X-ray binaries (HMXBs) accreting in the super-Eddington regime, since observational evidence shows that the ULX luminosity function appears to prolong the one from HMXBs up to $10^{40}$ \ergs \citep{mineo_x-ray_2012}. Companion stars in ULXs are typically intermediate (8-10 \Msun) or massive (> 10 \Msun) stars (\citealt{motch_mass_2014}, \citealt{heida_discovery_2019}, \citealt{rodriguez_castillo_discovery_2020}). During the binary massive star evolution, after the formation of the first compact object there is a short stage of common envelope (CE) during which the compact object is engulfed within the envelope of the companion star. At the end of this CE period, part of the companion star envelope is expelled. This may lead to the formation of a helium star. A helium companion star of an ULX has been observed in NGC 247 ULX-1 by \cite{zhou_identification_2023-1} using MUSE data. We may speculate that a system exiting its CE phase may strip a star of its envelope even further, expelling large amounts of nitrogen. In that case, the high levels of \nii\ that we observe may be explained by a massive companion star. 
An overabundance of nitrogen attributed to the companion star of a ULX has also been observed in high-resolution X-ray spectroscopy of Holmberg II X-1 \citep{barra_examining_2024}.


\subsection{NGC 2993 central X-ray sources}

There are two X-ray sources near the center of the galaxy (see Fig. \ref{fig:chandra}). The projected distance from the optical center of NGC 2993 to X1 is 1.35\arcsec\ (230 pc), and to X2 it is 2.57\arcsec\ (440 pc). Given that the apparent isophotal diameter of NGC 2993 is 1.35$\pm$0.12\arcmin\ (from HyperLeda\footnote{\url{http://atlas.obs-hp.fr/hyperleda/}}, \citealp{makarov_hyperleda_2014}), the galactocentric ratios are 0.03 and 0.06 for X1 and X2 respectively. Since NGC 2993 is a strongly interacting galaxy, an offset from the optical center is not unexpected. Therefore, we cannot exclude that X1 or X2 might host the massive black hole of NGC 2993, since AGN have been found to be spatially offset from the stellar core of their galaxies by several kpc \citep{barrows_spatially_2016}.

Both X1 and X2 exceed 10$^{39}$ \ergs\ in the 0.5$-$7.0 keV band. The X-ray luminosities would point toward a low luminosity AGN. The photon indexes of X1 ($\Gamma \sim 1.65$) and X2 ($\Gamma \sim 1.28$) are consistent with values expected for an AGN. 
No significant emission line broadening -- that could be attributed to a possible AGN activity -- is observed at the X1 and X2 positions. Assuming that X1 hosts the massive black hole (MBH) of the galaxy, we use the M-$\sigma$ relation between the MBH mass and the stellar velocity dispersion in its bulge \citep{kormendy_coevolution_2013} to compute an upper limit on the MBH mass. 
The median stellar velocity dispersion in an aperture of 0.6\arcsec\ radius around the X1 position is 61.1 \kms. So, the MBH mass would be $\lesssim1.72 \times 10^6$ \Msun. 
This mass estimate is in the range expected for a MBH in a galaxy with a $6 \times 10^9$ \Msun\ mass like NGC 2993 \citep{kormendy_coevolution_2013}. The corresponding Eddington luminosity would be $\lesssim1.99 \times 10^{44}$ erg s$^{-1}$, leading to the Eddington fraction of $\gtrsim4.08 \times 10^{-5}$. 
A similar estimate can be computed for X2. Such low Eddington fractions may explain the lack of influence of X1 and X2 on the surrounding gas. Indeed, low-luminosity AGN are often found to be accreting far below their Eddington limit \citep{gavignaud_eddington_2008}.
In addition, low-luminosity AGN may have radiatively inefficient accretion flows (\citealt{narayan_low-luminosity_2005}, \citealt{trump_nature_2009}), which leads to intrinsically weak optical emission lines. Therefore, these AGN may often be misclassified as BPT \hii\ regions in galaxies where ionization by the stellar UV radiation is strong (\citealt{agostino_crossing_2019}, \citealt{agostino_vlt-muse_2023}), which is the case for NGC 2993.  
\medskip

On the other hand, X1 and X2 may alternatively be considered as ULX candidates based on their X-ray luminosities larger than $10^{39}$ erg s$^{-1}$ and spectral parameters shown in Table~\ref{tab:xray}.
X2 does not seem to be associated with any variations in surface brightness or kinematics in the optical data. Based on the formalism outlined in Sect. \ref{subsec:hst_imaging}, 
we computed a 3$\sigma$ limiting Vega magnitude of 22.14 mag (corrected for the galactic extinction) at the position of X2. This gives us a lower limit on the X-ray-to-optical ratio log(f$_X$/f$_V$) > 0.96. This lower limit does not allow us to discriminate on the exact nature of X2.

X1 is associated with a high concentration in surface brightness in the \textit{HST} image (Fig. \ref{fig:ngc2993}). It is also associated with an increase in surface brightness from the emission lines. In addition, the electron density reaches its maximum ($\approx$ 330 cm$^{-3}$) at the position of X1 (Fig. \ref{fig:el_dens}). We conclude that X1 is likely associated with a star-forming region. 

As shown in the BPT diagrams of the galaxy (Appendix \ref{app:bpt}), the dominant ionization process around the positions of X1 and X2 seems to be photo-ionization by UV radiation coming from massive hot OB stars, with no significant indication of photo-ionization by extreme UV/X-ray photons coming from accreting compact object sources, nor shock ionization. Since X1 is embedded in a star forming region, the presence of a bubble produced by X1 may be outshined by the emission from the star forming region. For X2, we may argue that it is a too young ULX system to have produced a significant impact on the surrounding gas.

\section{Conclusion}
\label{sec:conclusion}

We used archival X-ray observations from \textit{Chandra} and \textit{XMM} telescopes to identify four X-ray sources in the NGC 2993 galaxy: X1 and X2 near its center and two ULX candidates ULX1 and ULX2. Using archival \textit{HST} images, we studied the optical counterparts of the two ULXs. 
We used an archival \textit{VLT}/MUSE datacube to study spatially resolved properties of NGC 2993. To do so, we computed stellar properties and kinematics by fitting stellar population templates to the continuum part of the spectrum; and we computed gas kinematics by fitting Gaussian lines to the observed emission lines. This was done to obtain information about the environment of ULX1, X1 and X2. The main results of the paper are given below.

\begin{itemize}
    \item[$\circ$] The gas movement in NGC 2993 is in a perturbed state, possibly due to its gravitational interaction with NGC 2992. The velocity of the absorption doublet \nai\ seems to indicate the presence of gas falling onto the plane of the galaxy. In some regions, there are clearly distinguishable double components in the emission lines, which also favors extra-planar gas and further suggests that the ionization field is strong enough to ionize gas outside of the disk of the galaxy. 
    \item[$\circ$] The X-ray to optical ratios of ULX2 computed using \textit{HST} observations in two filters seem consistent with ULXs, although more data is needed to confirm its ULX nature.
    ULX1 is situated in a stellar cluster of 5.1$\pm$1.1 pc in radius resolved by \textit{HST}. We find the age of 0.32 Gyr and an upper limit on the mass of $3.50\times10^6$ \Msun\ for the cluster using its stellar continuum. We confirm that the host of this source is associated with NGC 2993 using its optical emission lines, thus confirming the ULX nature of ULX1.
    \item[$\circ$] There is an excess in the \nii\ emission at the position of ULX1. 
    It is not resolved by MUSE, with an upper limit on the radius of 12 pc. 
    We find a lower limit on the nitrogen mass in the emitting gas of 0.202 \Msun. This possible nitrogen enrichment at the position of ULX1 might be explained by the presence of one or more WR stars in the associated stellar cluster. It might also be related to the ULX system, if the nitrogen layers were stripped from the ULX massive companion star in the past. 
    \item[$\circ$] X1 and X2 are located in the central part of the galaxy. The surroundings of X1 and X2 do not show any evidence of ionization from these X-ray sources. X1 seems to be situated in a highly star-forming region with elevated electron density. Assuming X1 harbors the massive black hole of the NGC 2993 galaxy, we find an upper limit of $1.72 \times 10^6$ \Msun\ on the black hole mass and a very low Eddington fraction ($\gtrsim4.08 \times 10^{-5}$) that might account for the absence of clear impact on its surroundings. 
    We could not relate X2 to anything of notice on the stellar and ionized gas maps. Its lower limit on log(f$_X$/f$_V$) > 0.96 does not allow us to conclude on its nature.
\end{itemize}

To confirm whether the apparent nitrogen enrichment does come from WR stars, we would need observations in the bluer end of the spectrum, using for example the extended mode on MUSE which starts at 4600 \AA. Indeed, it would be useful to check whether the WR blue bump at around 4686 \AA\ is present at the ULX1 position. In addition, observing the \heii\ emission line may have the potential to confirm the presence of a highly energetic source at the position of ULX1, since \heii\ is only ionized by high energy photons. 
Although part of this \heii\ emission (if detected) may be due to ionization by WR stars \citep{shirazi_strongly_2012}, if only the narrow \heii\ emission line is detected but there is no detection of the broad component associated with the blue WR bump (and given that the upper limit on this non-detection is low enough), photoionization by the ULX would be favored as the reason for the \heii\ emission. 
It would also be useful to get a better constraint on the electron temperature and on the ratio between the singly and doubly ionized ions at the ULX1 position by observing the \oii($\lambda\lambda$3727,3729) doublet, for example with the upcoming \textit{VLT}/BlueMUSE \citep{richard_blue_2024} instrument.
\begin{acknowledgements}
We thank the anonymous referee for providing constructive comments. This study has been partially supported through the grant EUR TESS N°ANR-18-EURE-0018 in the framework of the Programme des Investissements d'Avenir. This work was supported by the CNRS-INSU through the International Mobility Support program and the French Chilean Laboratory for Astronomy (FCLA, IRL 3386). We would like to thank Simon Dupourqué for his help and useful advice on the analysis of X-ray spectra. 
C.M. acknowledges the support of grants UNAM/DGAPA/PAPIIT IN113226.
\end{acknowledgements}

\bibliographystyle{bibtex/aa}
\bibliography{bibtex/articleNGC2993}

\appendix
\onecolumn 

\section{Astrometric corrections and source positions}
\label{app:astrometry}

To ensure the correspondence between the X-ray sources and their positions in the optical data, we astrometrically corrected all the observations used in this work. The corresponding coordinate shifts are given in Table \ref{tab:corrections}. First, we used \texttt{DAOStarFinder} from the python package \texttt{photutils} 1.11.0 to detect sources in the \textit{HST} ACS/WFC/F606W filter. We then matched these sources to the Gaia DR3 catalog \citep{vallenari_gaia_2023} and found 13 matches. We then aligned the observation in the ACS/WFC/F435W filter to the astrometrically corrected observation in the ACS/WFC/F606W filter. We then applied the same method to the whitelight MUSE image and found 10 matches with the Gaia DR3 catalog. The transformation found for the whitelight image was then applied to the whole cube. 

\begin{sidewaysfigure}[h!]
    \centering
    \includegraphics[width=\columnwidth]{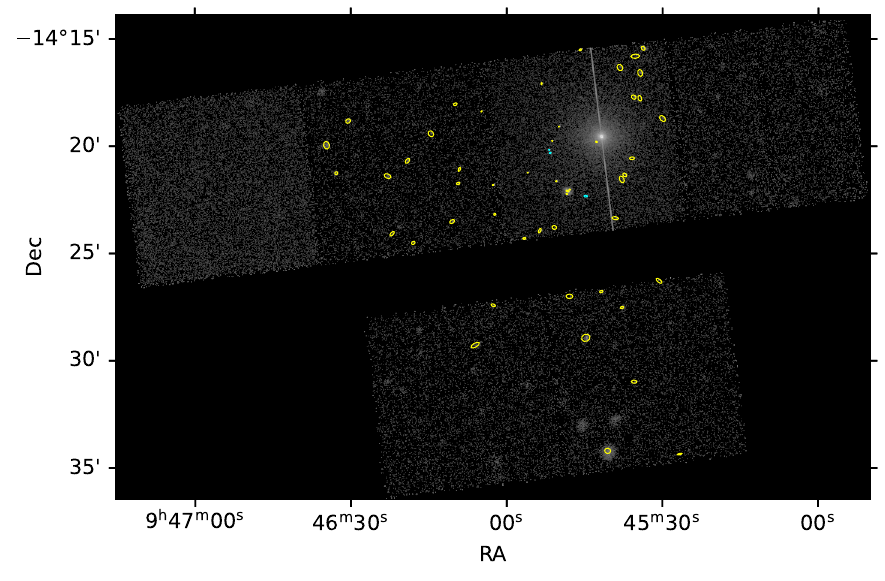}
    \caption{Full \textit{Chandra} \textit{3956} broad band (0.5$-$7.0 keV) image showing the 53 sources detected using \texttt{wavdetect} with scales=(1, 2, 2.8, 4, 6). The three sources used for the astrometric correction are shown in cyan, other detected sources are shown in yellow.}
    \label{fig:chandra_wavdetect}
\end{sidewaysfigure}

To derive the positions of the X-ray sources, we used the \textit{3956} \textit{Chandra} observation which has a better quality than the \textit{11858} one. Indeed, although the exposure time for the \textit{11858} observation is larger than for the \textit{3956} observation, their exposure maps show that there are more total counts in the \textit{3956} observation. This is not surprising since there is a seven-year gap between the two observations. As the ACIS instrument deteriorates over time, we get less counts for the same amount of photons \citep{plucinsky_complicated_2018}. In addition, since the \textit{11858} observation was made with HETG, its effective collecting area is smaller. To astrometrically correct the \textit{3956} \textit{Chandra} observation, we used \texttt{wavdetect} to find sources in the \textit{Chandra} image. We set the scale parameter to (1, 2, 2.8, 4, 6), which means that the \textit{Chandra} image was correlated with Mexican hat wavelets of radii from 1 to 6 pixels. We chose these scales so that \texttt{wavdetect} detects all four X-ray sources in NGC 2993 and does not detect extended sources which would need higher scale values.
We found 53 X-ray sources, excluding the saturated center of NGC 2992 and detections due to the corresponding readout streak. Figure \ref{fig:chandra_wavdetect} shows the detected sources on the \textit{Chandra} broad band (0.5$-$7.0 keV) image. Originally, we tried to match them to the Gaia DR3 catalog using \texttt{wcs\_match}. Excluding the sources studied in the present analysis (X1, X2, ULX1 and ULX2), there were only two matches between the \textit{Chandra} observation and the catalog, and the resulting astrometric correction was rejected upon visual inspection. We decided to match the \textit{Chandra} sources to the astrometrically corrected \textit{HST} ACS/WFC/F606W image and found three matches shown in cyan ellipses in Fig. \ref{fig:chandra_wavdetect}. Astrometric correction using these three matches looked visually correct.

\begin{table}[h!]
    \centering
    \caption{Astrometric corrections of \textit{Chandra}, \textit{HST} and \textit{VLT}/MUSE data that were applied to each observation.}
    \renewcommand{\arraystretch}{1.3}
    \begin{tabular}{ccccc}
    \hline\hline
         & \textit{Chandra} & \makecell{\textit{HST} \\ ACS/WFC/ \\ F606W} & \makecell{\textit{HST} \\ ACS/WFC/ \\ F435W}& \textit{VLT}/MUSE \\ \hline
        RA & $-$0.0784 & $-$0.0551 & $-$0.0774 & 0.2796
 \\ 
        Dec & $-$0.1162 & 0.0223 & $-$0.0600 & 0.0570 \\ \hline
    \end{tabular}
    \tablefoot{The shift values are given in arcseconds.}
    \label{tab:corrections}
\end{table}

The positions of the X-ray sources were found by applying the \texttt{wavdetect} routine to the astrometrically corrected 3956 \textit{Chandra} observation. However, \texttt{wavdetect} underestimates the size of the error circle \citep{hong_x-ray_2005}. We calculated the 95\% confidence interval error circles using Eq. 5 from \cite{hong_x-ray_2005}. This equation uses net counts from the sources that we obtained from \texttt{wavdetect} (405 for ULX1, 311 for ULX2) and the aim point offset at the position of the sources that we obtained with the \texttt{dmcoords} function from \texttt{CIAO} (2.5\arcsec\ for ULX1, 1.8\arcsec\ for ULX2). The resulting error circle radii and the source coordinates are given in Table \ref{tab:positions}. For precise \textit{HST} counterpart identification, we need to consider the residual uncertainty between the \textit{Chandra} and \textit{HST} observations which is left after the astrometric correction. To do so, we computed the mean residual value between the \textit{Chandra} and \textit{HST} positions of the three matched sources, which was found to be  0.217\arcsec. Then, we quadratically summed this uncertainty with the X-ray source position uncertainty to obtain a better estimate of the error circle radius that should be used to look for \textit{HST} counterparts of the X-ray sources. We provide this updated uncertainty value in Table~\ref{tab:positions}. We used these updated error circle radii for ULX1 in Figures throughout the article, and for all of the X-ray sources in Figure \ref{fig:opt_counterparts}.

\begin{table}[h!]
    \centering
    \caption{Positions and the error circle radii of the X-ray sources in NGC 2993.}
    \renewcommand{\arraystretch}{1.3}
    \begin{tabular}{ccccc}
    \hline\hline
        Source & \makecell{RA \\(J2000)} & \makecell{Dec \\(J2000)} & \makecell{\textit{Chandra}\\ error radius (\arcsec )} & \makecell{\textit{Chandra}+\textit{HST}\\ error radius (\arcsec)} \\
        \hline
        ULX1 & 09h45m47.90s & $-$14d22m01.95s & 0.331 & 0.396 \\
        ULX2 & 09h45m50.47s & $-$14d21m38.69s & 0.322 & 0.388 \\
        X1 & 09h45m48.42 & $-$14d22m05.34 & 0.316 & 0.383 \\
        X2 & 09h45m48.17 & $-$14d22m05.341 & 0.329 & 0.394 \\
        \hline
    \end{tabular}
    \tablefoot{The X-ray source positions are from \textit{Chandra}. The \textit{Chandra} error circle radius is given at the 95\% confidence interval. The \textit{Chandra} + \textit{HST} error circle radius additionally takes into account the residual uncertainty from the \textit{Chandra} to \textit{HST} astrometric correction.}
    \label{tab:positions}
\end{table}

\clearpage

\section{ULX2: X-ray spectral fitting}
\label{app:x_spectra}

\begin{figure}[h!]
    \centering
    \begin{subfigure}[c]{0.24\linewidth}
        \centering
        \includegraphics[width=\linewidth]{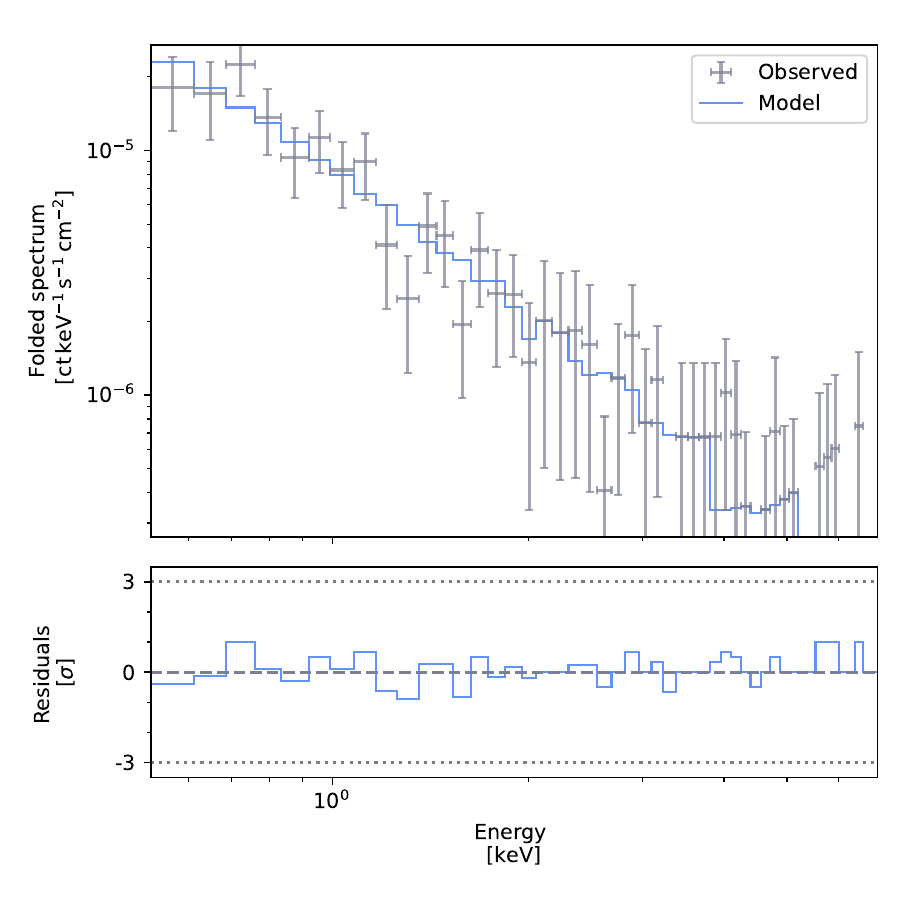}
        \phantomcaption        
    \end{subfigure}
    \begin{subfigure}[c]{0.24\linewidth}
        \centering
        \includegraphics[width=\linewidth]{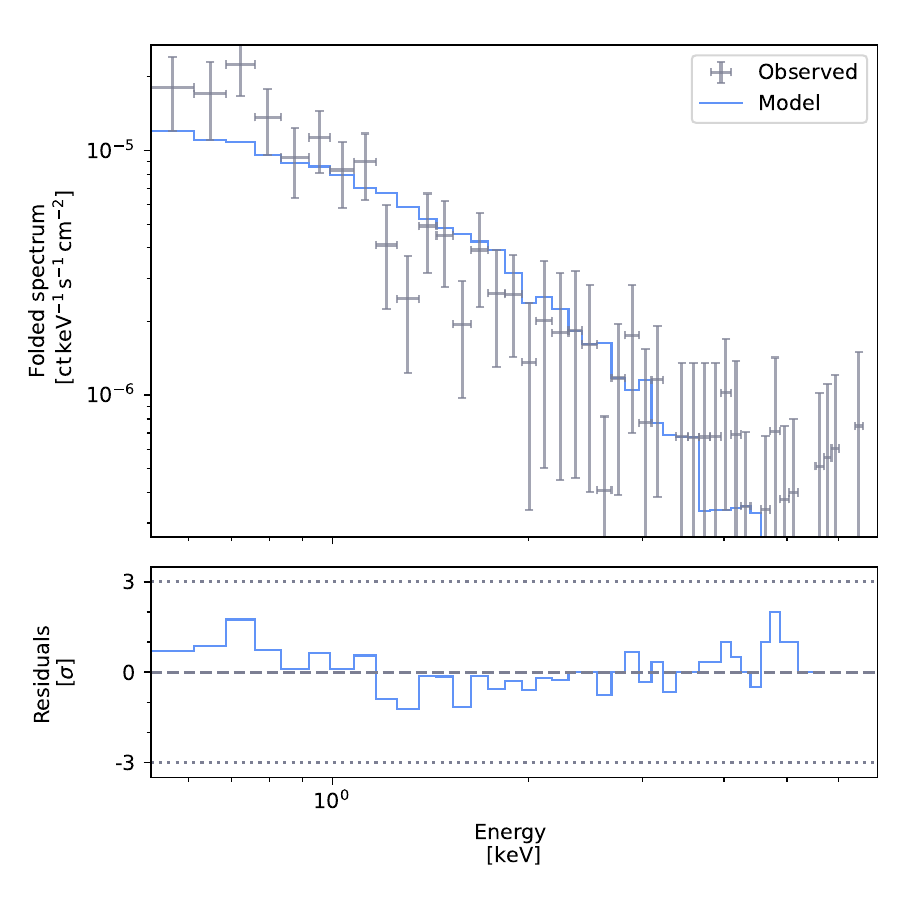}
        \phantomcaption        
    \end{subfigure}
        \begin{subfigure}[c]{0.24\linewidth}
        \centering
        \includegraphics[width=\linewidth]{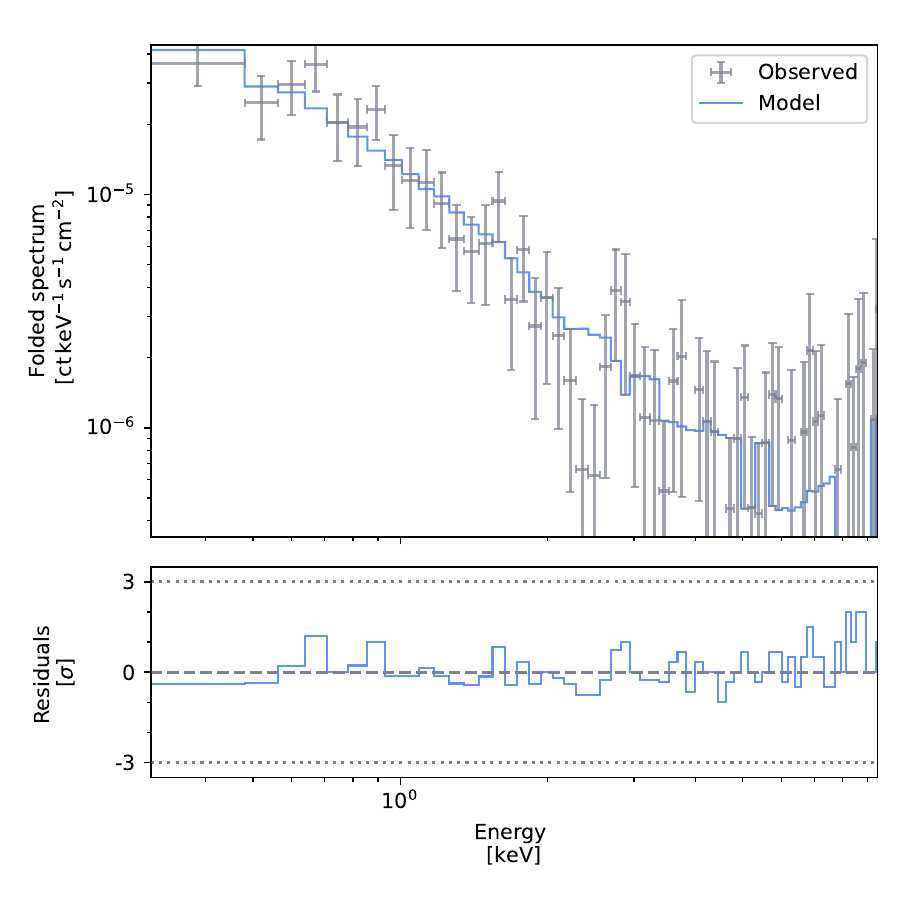}
        \phantomcaption        
    \end{subfigure}
    \begin{subfigure}[c]{0.24\linewidth}
        \centering
        \includegraphics[width=\linewidth]{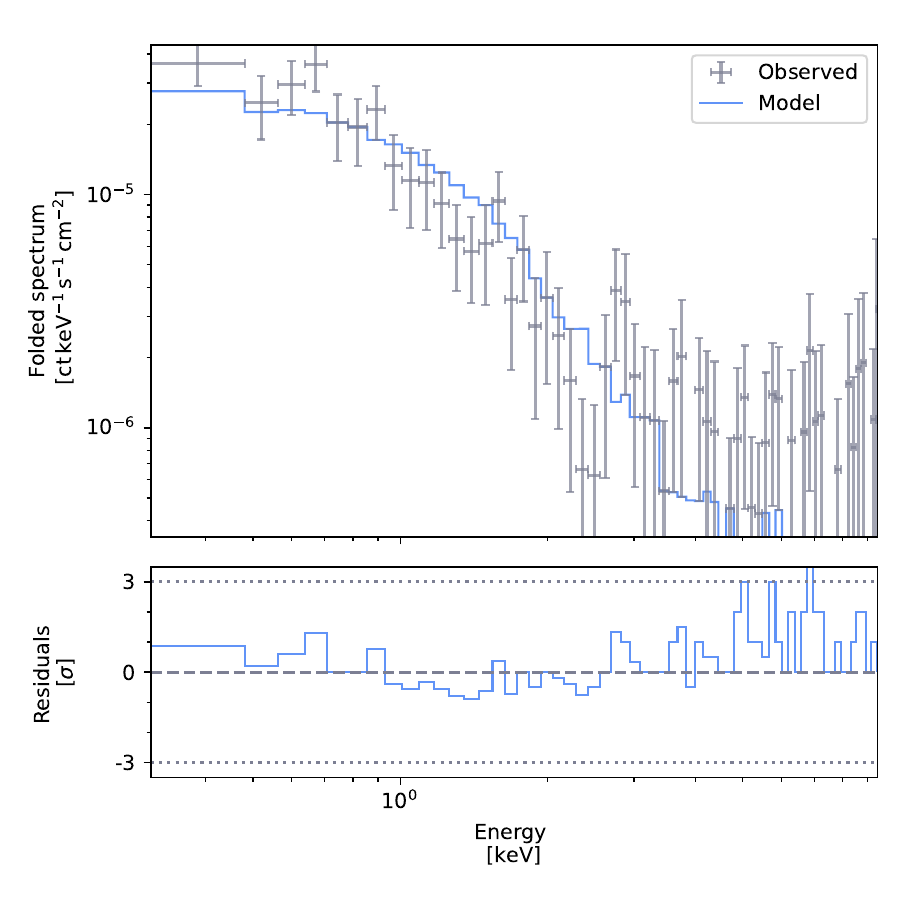}
        \phantomcaption        
    \end{subfigure}\quad
    \begin{subfigure}[c]{0.24\linewidth}
        \centering
        \includegraphics[width=\linewidth]{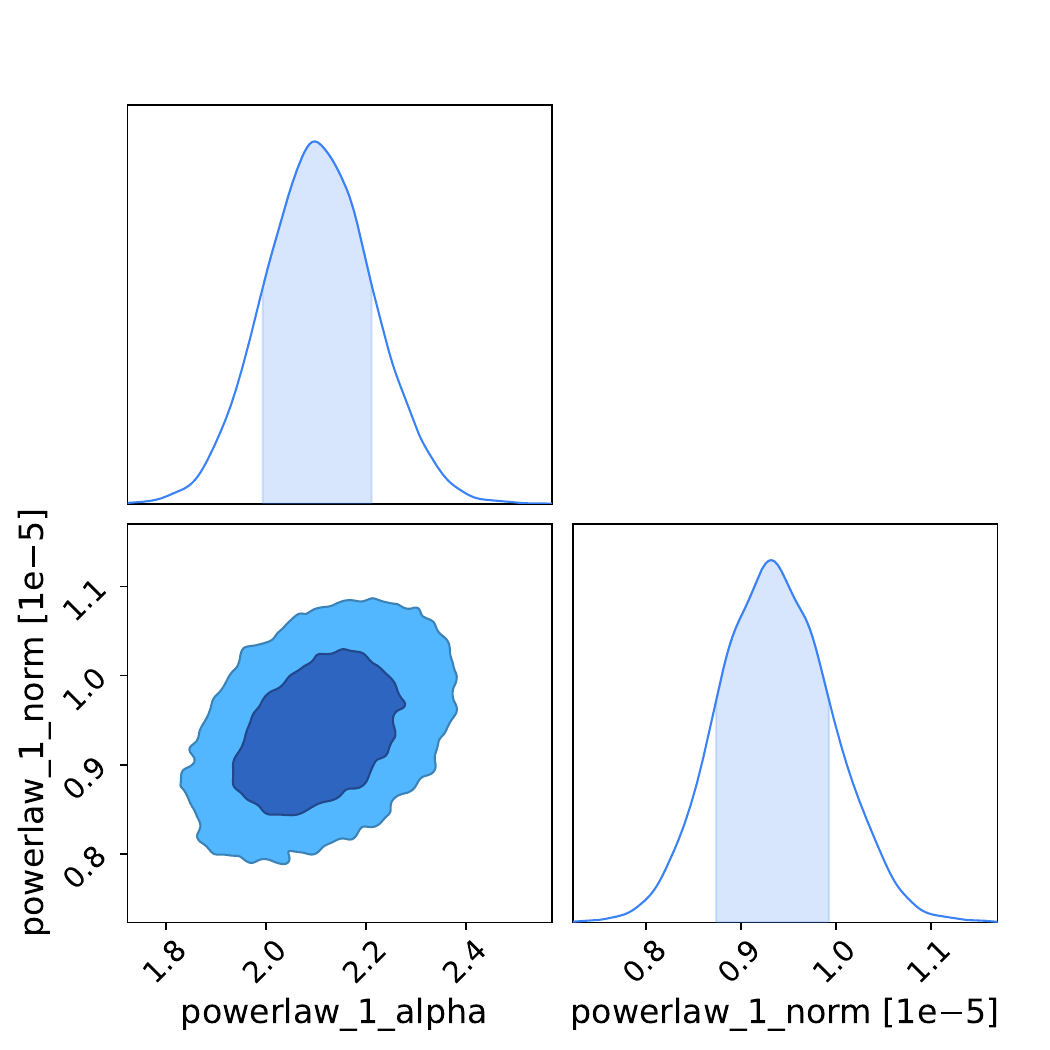}
        \phantomcaption        
    \end{subfigure}
    \begin{subfigure}[c]{0.24\linewidth}
        \centering
        \includegraphics[width=\linewidth]{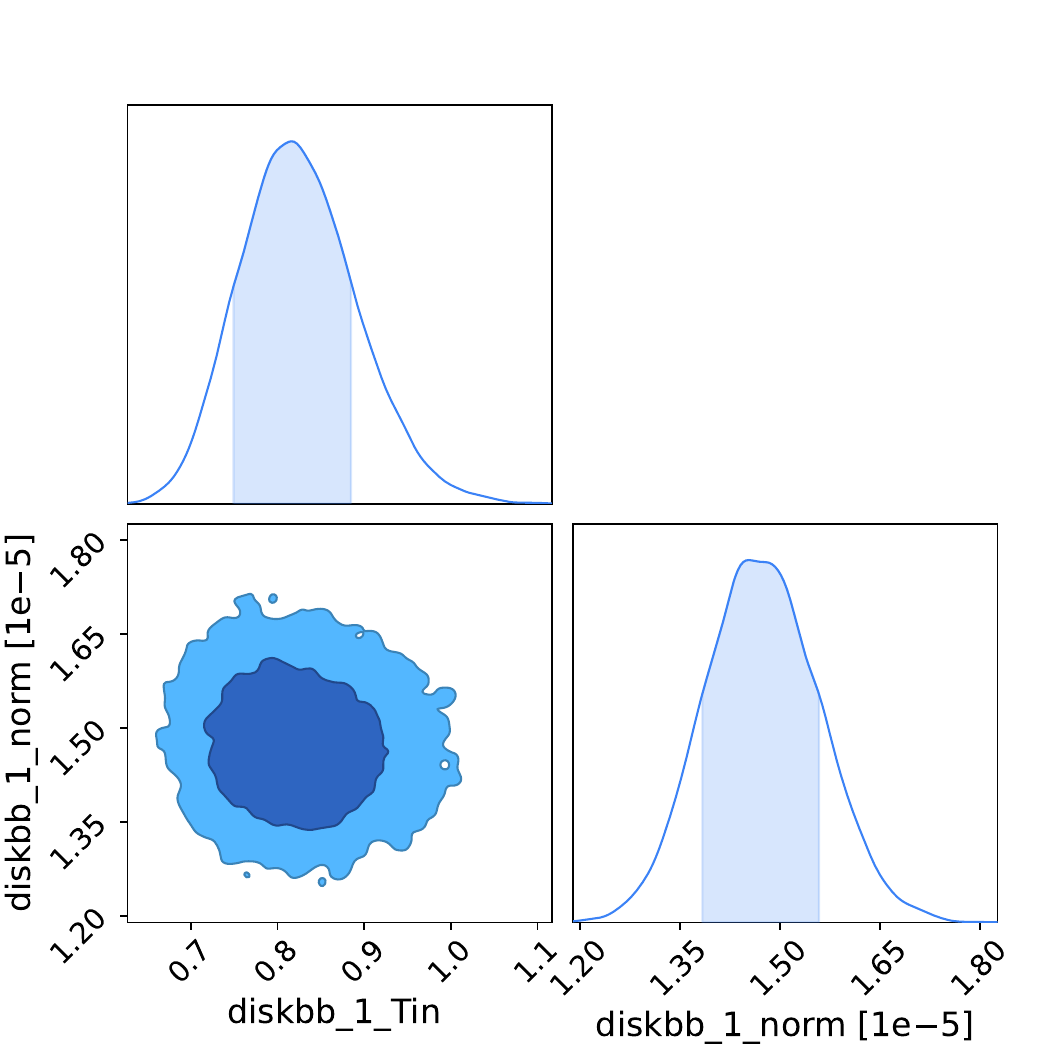}
        \phantomcaption        
    \end{subfigure}
    \begin{subfigure}[c]{0.24\linewidth}
        \centering
        \includegraphics[width=\linewidth]{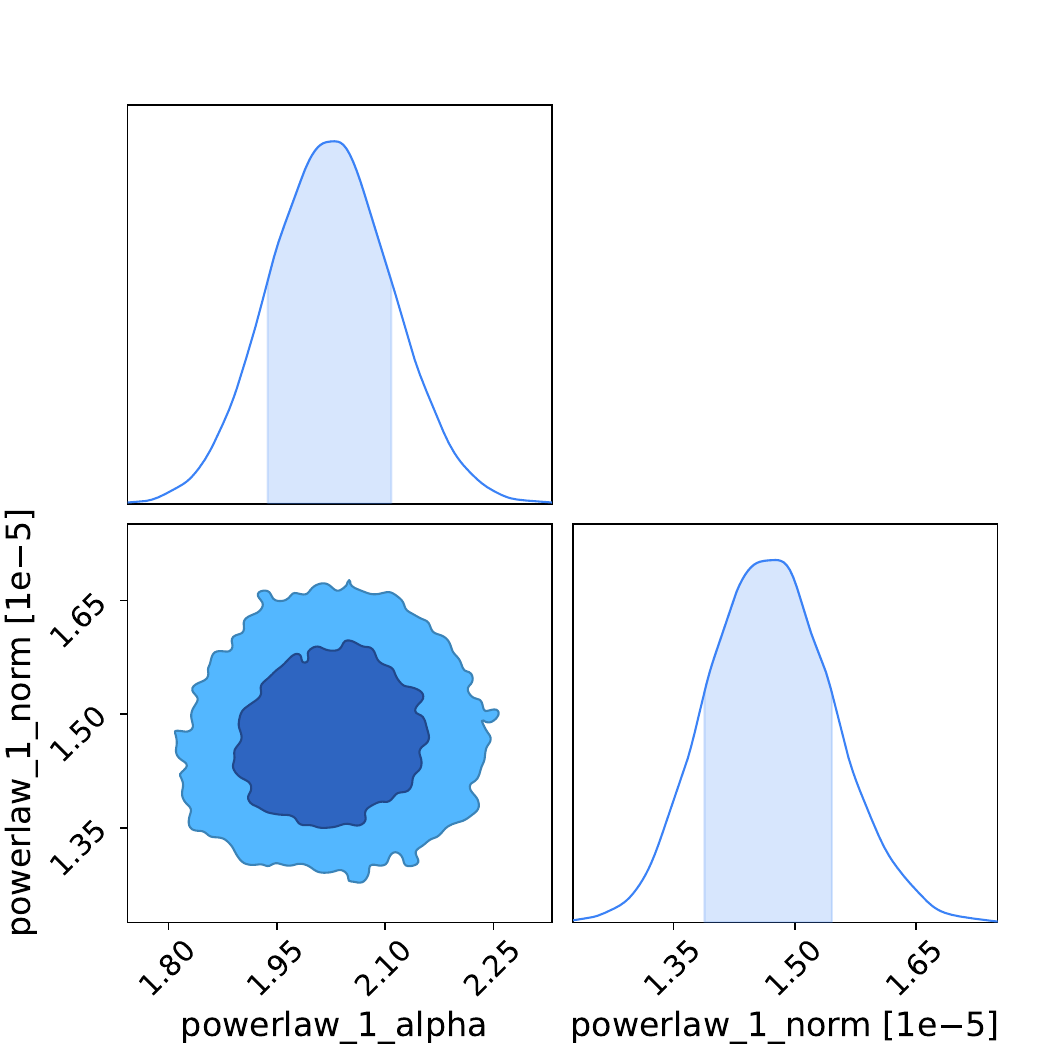}
        \phantomcaption        
    \end{subfigure}
    \begin{subfigure}[c]{0.24\linewidth}
        \centering
        \includegraphics[width=\linewidth]{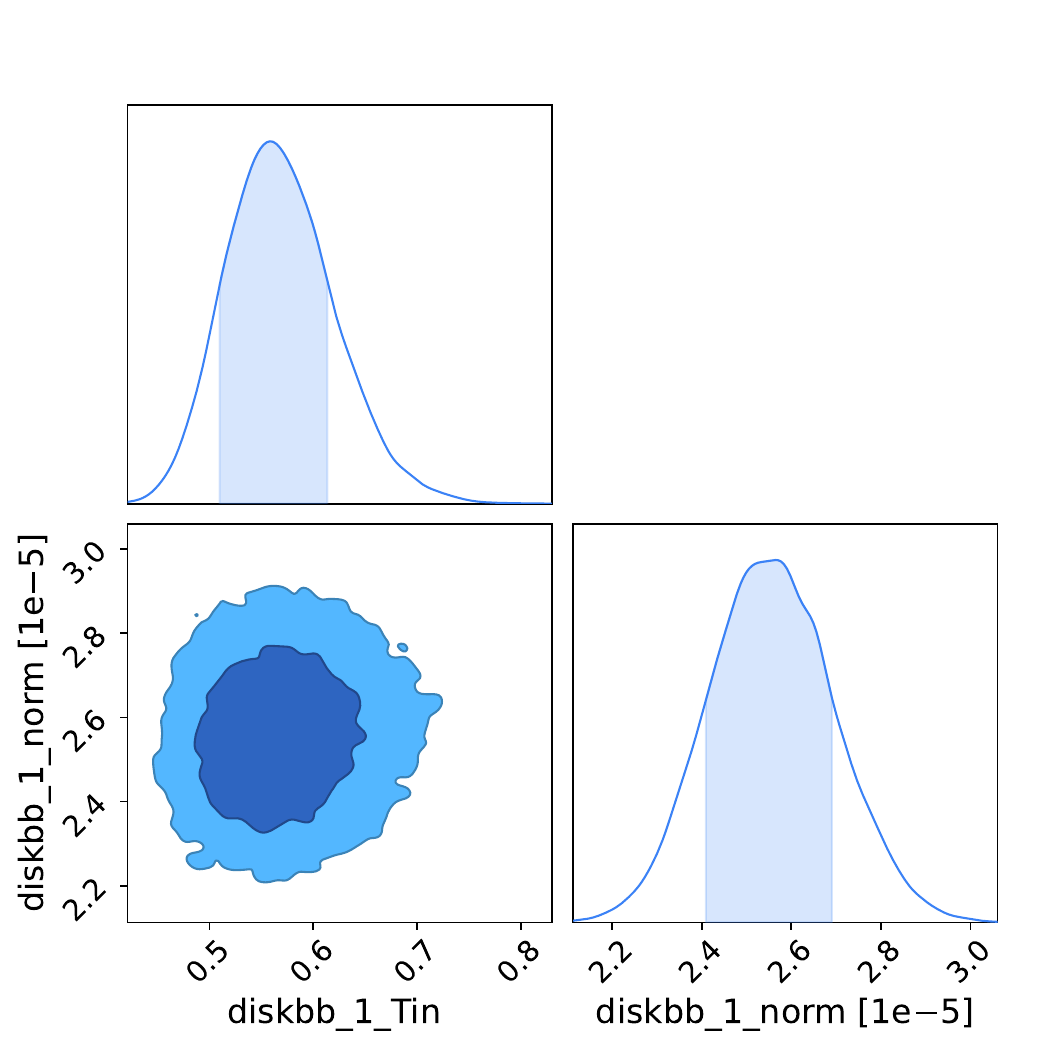}
        \phantomcaption        
    \end{subfigure}
    \caption{Top panels: Best-fit  models and the corresponding residuals from \texttt{jaxspec} for ULX2. From left to right: \textit{Chandra} 3956 data fit using \texttt{tbabs*powerlaw}, \textit{Chandra} 3956 data fit using \texttt{tbabs*diskbb}, \textit{XMM-Newton} pn data fit using \texttt{tbabs*powerlaw}, \textit{XMM-Newton} pn data fit using \texttt{tbabs*diskbb}. Bottom panels: corresponding corner plots showing the 1D (with the shaded region showing the 1$\sigma$ value range) and the 2D (with the shaded regions showing the 1$\sigma$ and 2$\sigma$ value ranges) projections of the posterior distribution of parameters. $T_{in}$ is in units of keV and $n_{\mathrm{H}}$ is in units of 10$^{22}$ cm$^{-2}$.}
    \label{fig:app_ulx2_spectrum}
\end{figure}

Figure \ref{fig:app_ulx2_spectrum} shows the ULX2 X-ray spectra and the corresponding fits using absorbed powerlaw and absorbed multi-blackbody accretion disk models. The fit is made using the \texttt{jaxspec} code. It is clear from the residuals that ULX2 is better modeled by \texttt{tbabs*powerlaw} than by \texttt{tbabs*diskbb}.

\section{Radial profile of ULX2}
\label{app:ulx2_profile}

\begin{figure}[h!]
    \centering
    \includegraphics[width=0.5\linewidth]{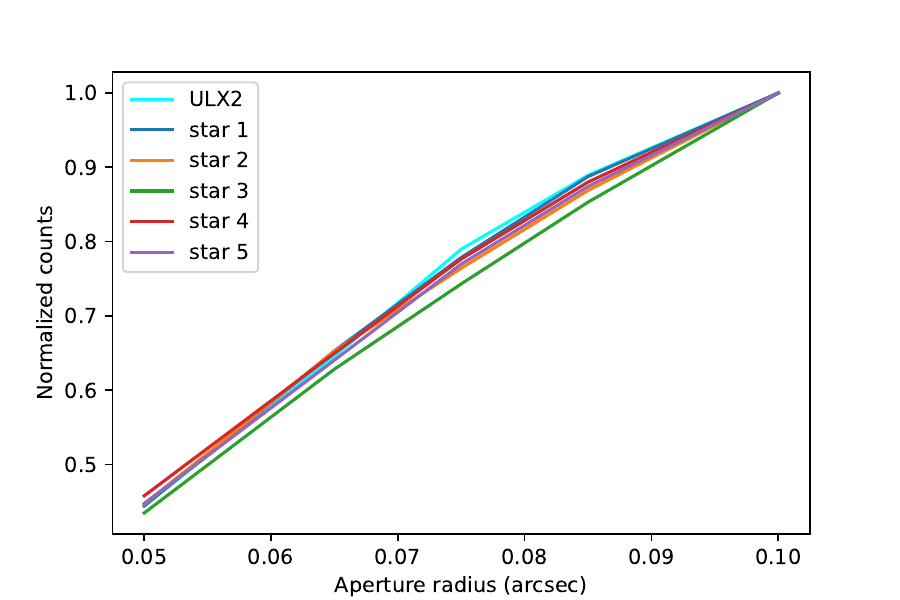}
    \caption{The radial profiles of the stars in the \textit{HST} ACS/WFC/F435W observation compared to the radial profile of the counterpart of ULX2.}
    \label{fig:ulx2_profiles}
\end{figure}

Figure \ref{fig:ulx2_profiles} is like Fig. \ref{fig:profiles} but for ULX2 in the \textit{HST} ACS/WFC/F435W filter. The stars used to create these profiles are different from those used to create the profiles in Fig. \ref{fig:profiles}, since the two \textit{HST} observations show different parts of the galaxy. The ULX2 counterpart seems unresolved.

\clearpage
\section{Zoom-ins on the ULX1 position}
\label{app:zooms}

\begin{figure*}[h!]
    \centering
    \begin{subfigure}[t]{0.28\linewidth}
        \centering
        \includegraphics[width=\linewidth]{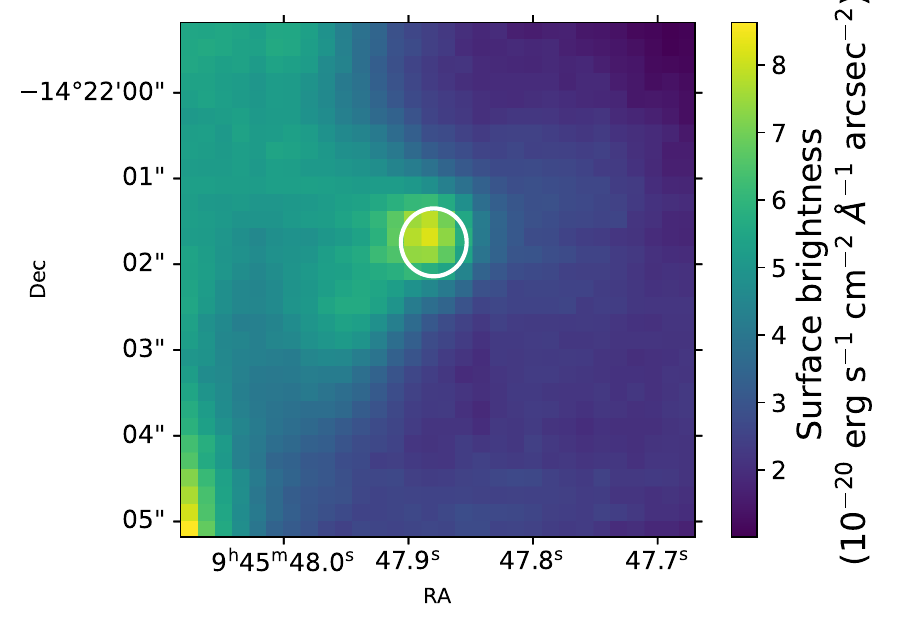}
    \end{subfigure}
    \begin{subfigure}[t]{0.28\linewidth}
        \centering
        \includegraphics[width=\linewidth]{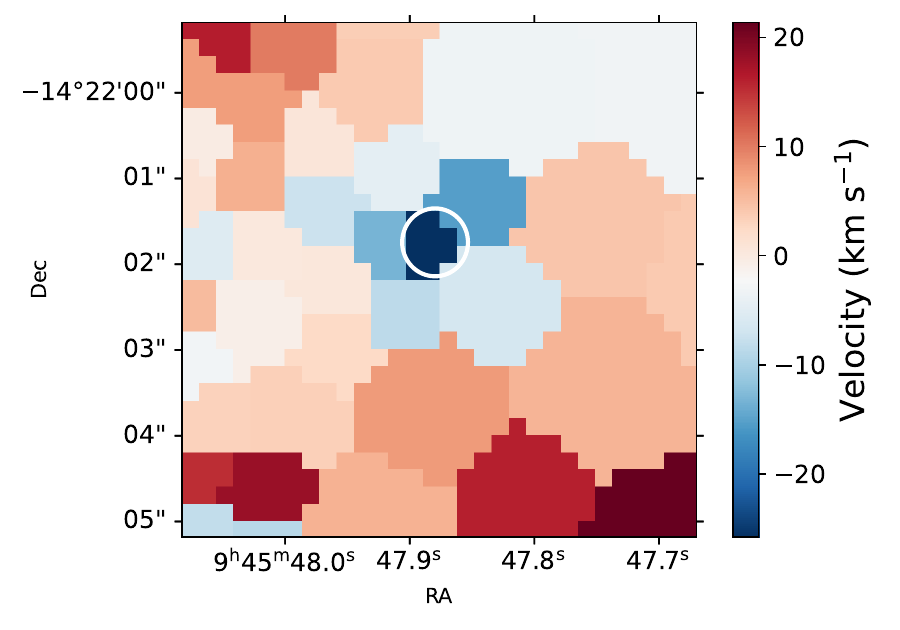}
    \end{subfigure}
    \begin{subfigure}[t]{0.28\linewidth}
        \centering
        \includegraphics[width=\linewidth]{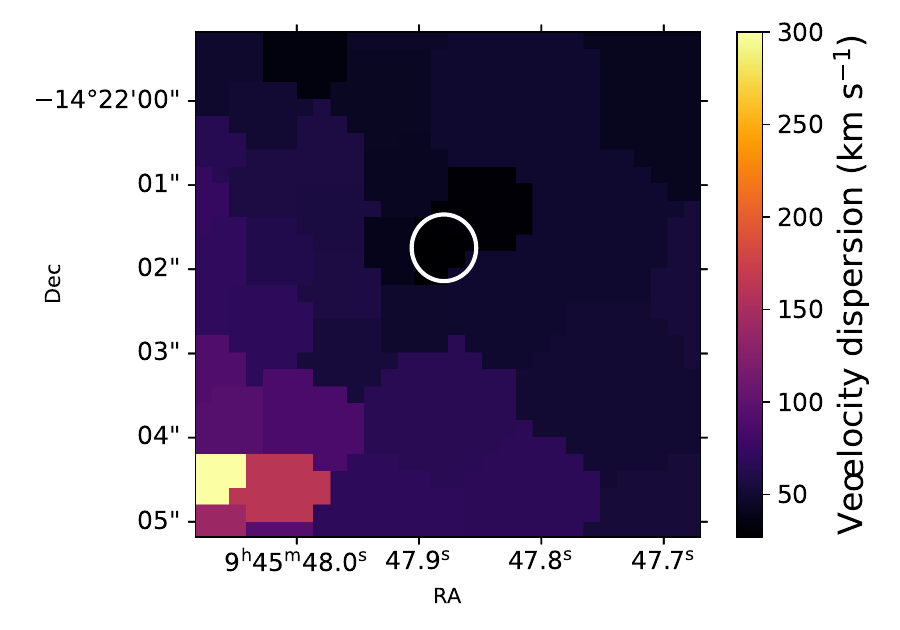}
    \end{subfigure}

    \begin{subfigure}[t]{0.28\linewidth}
    \centering
        \includegraphics[width=\linewidth]{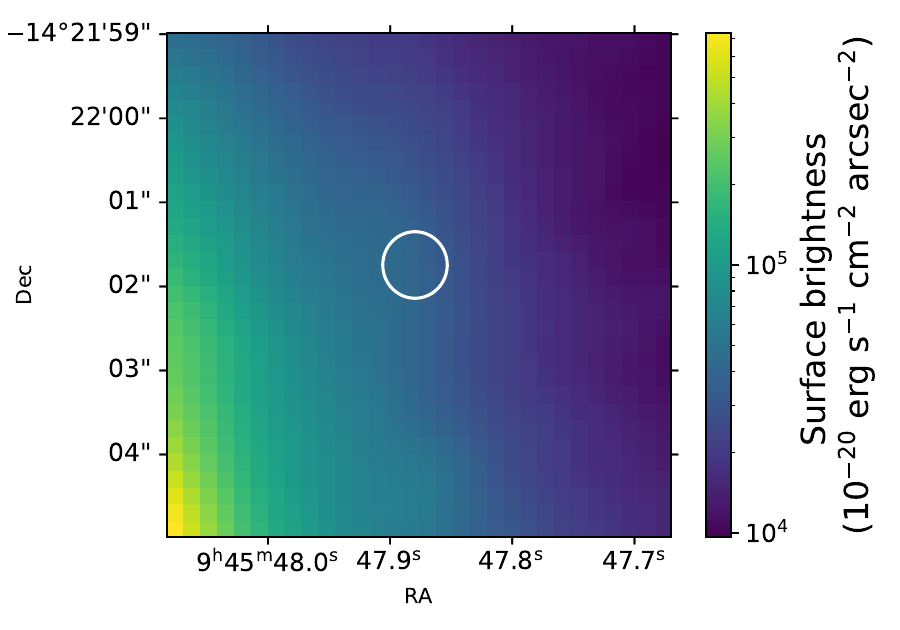}
    \end{subfigure}
    \begin{subfigure}[t]{0.28\linewidth}
        \centering
        \includegraphics[width=\linewidth]{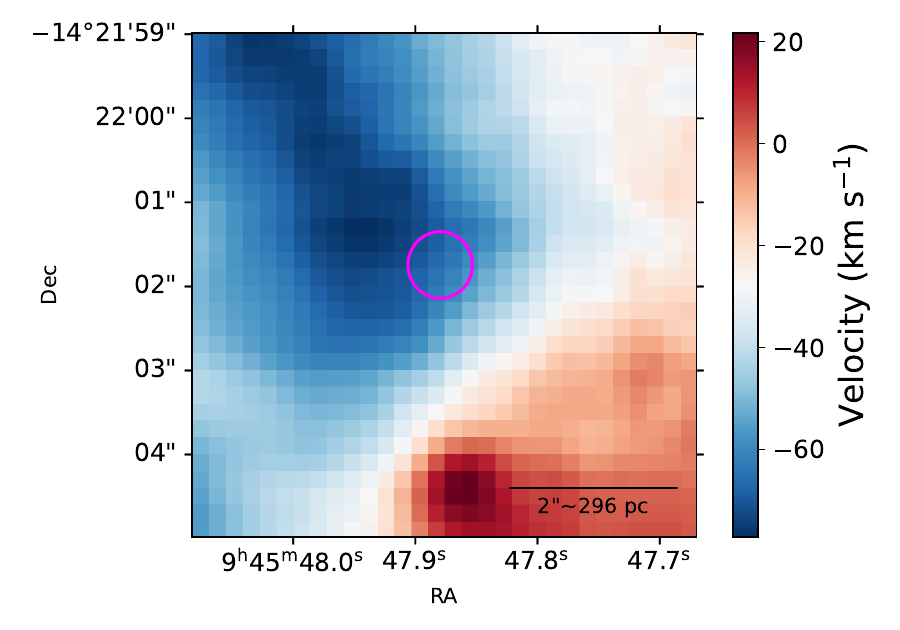}
    \end{subfigure}
    \begin{subfigure}[t]{0.28\linewidth}
        \centering
        \includegraphics[width=\linewidth]{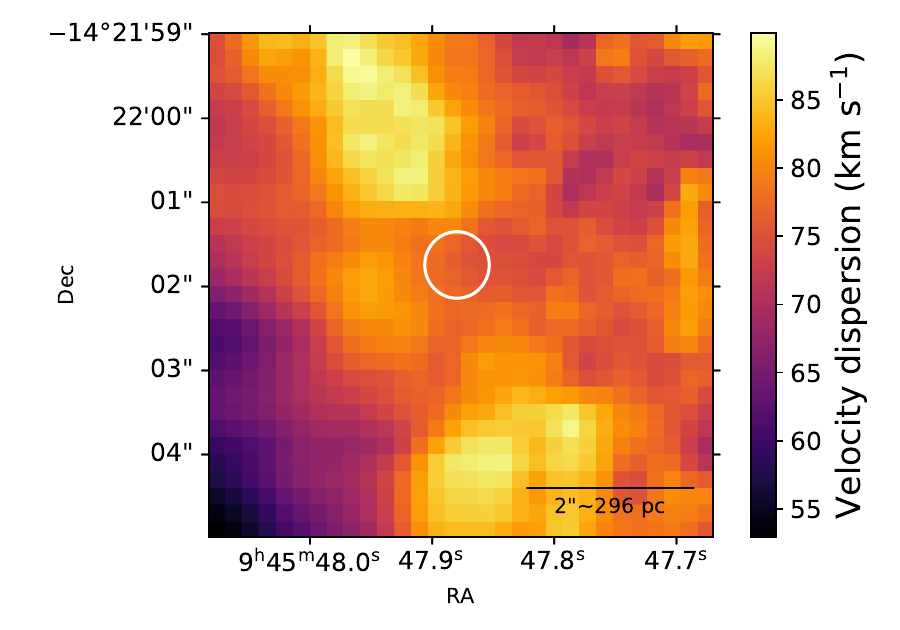}
    \end{subfigure}
    
    \caption{Comparison between stellar and gas properties. From left to right, first row: continuum surface brightness map from a wavelength range free of emission lines (5099$-$5487 \AA), stellar velocity map, stellar LoS velocity dispersion map; second row: H$\alpha$ surface brightness map, H$\alpha$ velocity map, H$\alpha$ velocity dispersion map.
    The circle marks the position of the ULX1 source.}
    \label{fig:zoom}
\end{figure*}

\begin{figure*}[h!]
    \centering
    
    \begin{subfigure}[t]{0.28\linewidth}
        \centering
        \includegraphics[width=\linewidth]{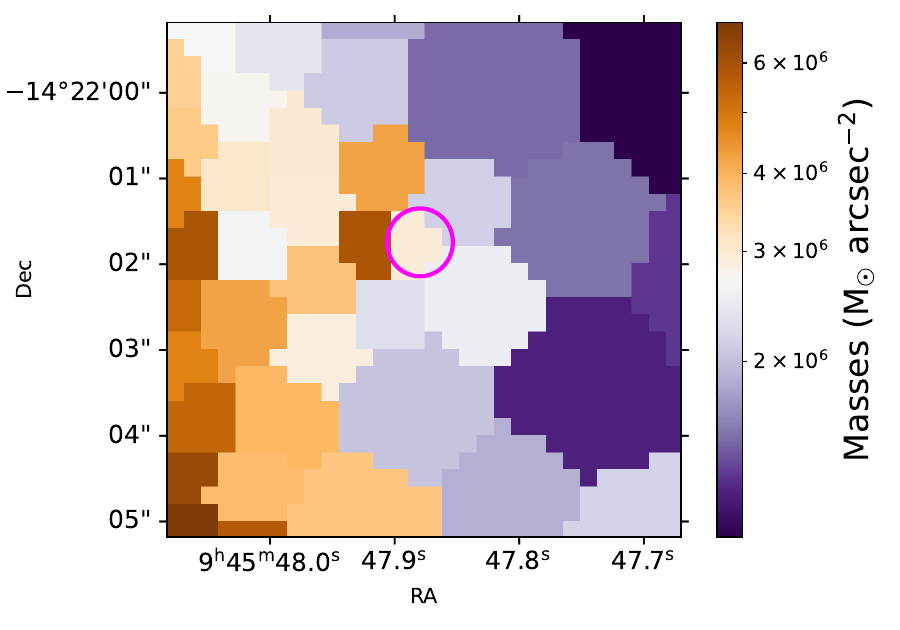}
    \end{subfigure}
    \begin{subfigure}[t]{0.28\linewidth}
        \centering
        \includegraphics[width=\linewidth]{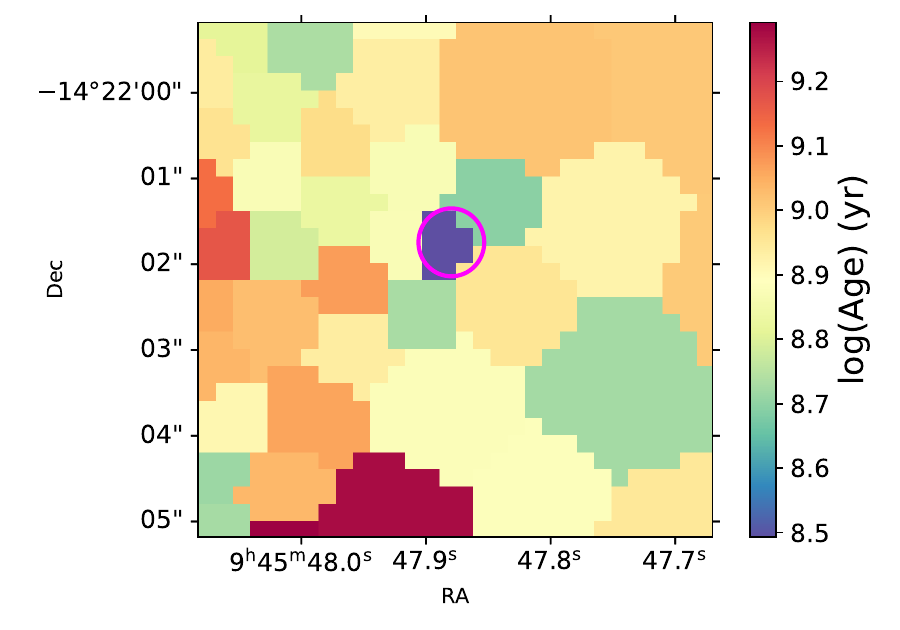}
    \end{subfigure}
    \begin{subfigure}[t]{0.28\linewidth}
        \centering
        \includegraphics[width=\linewidth]{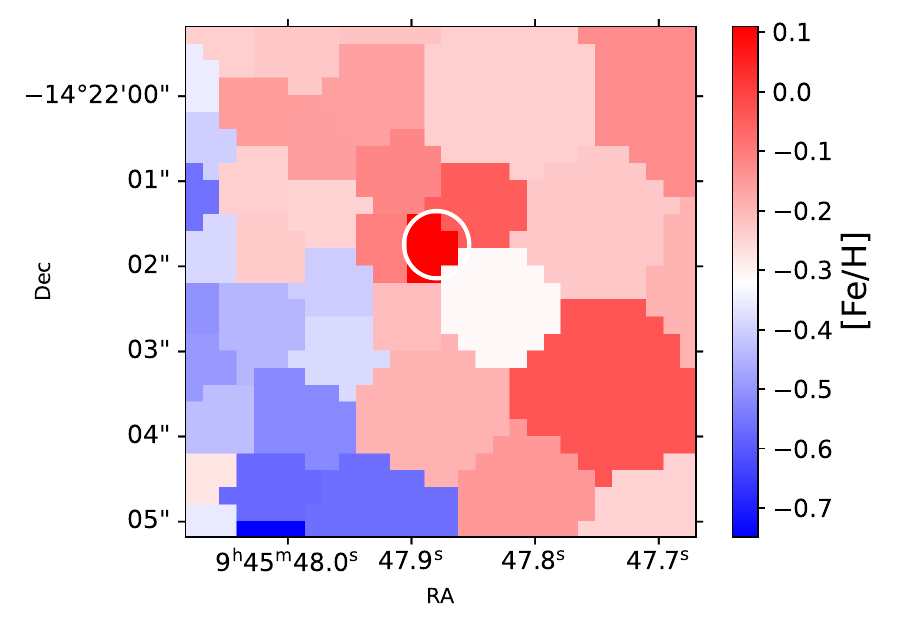}
    \end{subfigure}
    \begin{subfigure}[t]{0.28\linewidth}
        \centering
        \includegraphics[width=\linewidth]{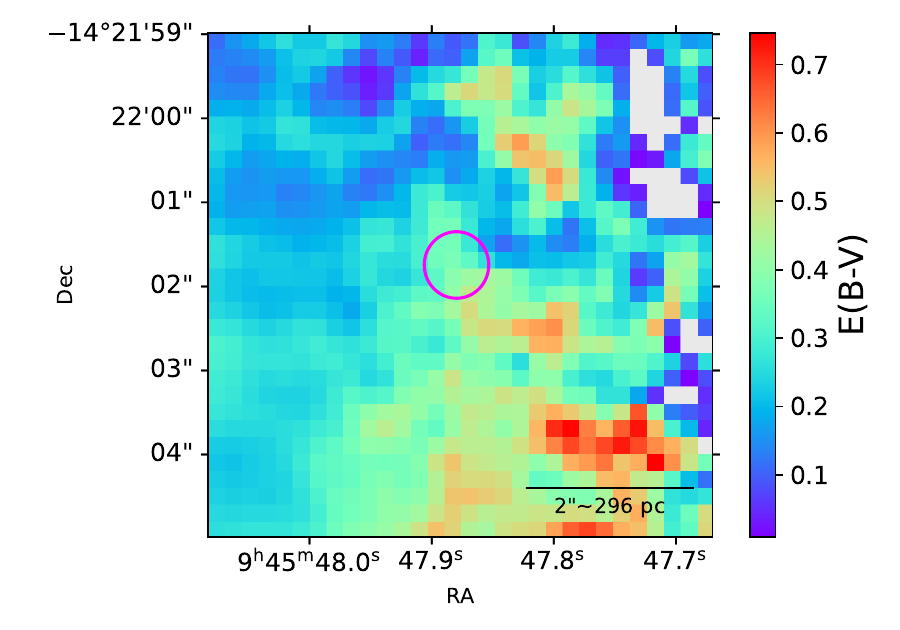}
    \end{subfigure}
    \begin{subfigure}[t]{0.28\linewidth}
        \centering
        \includegraphics[width=\linewidth]{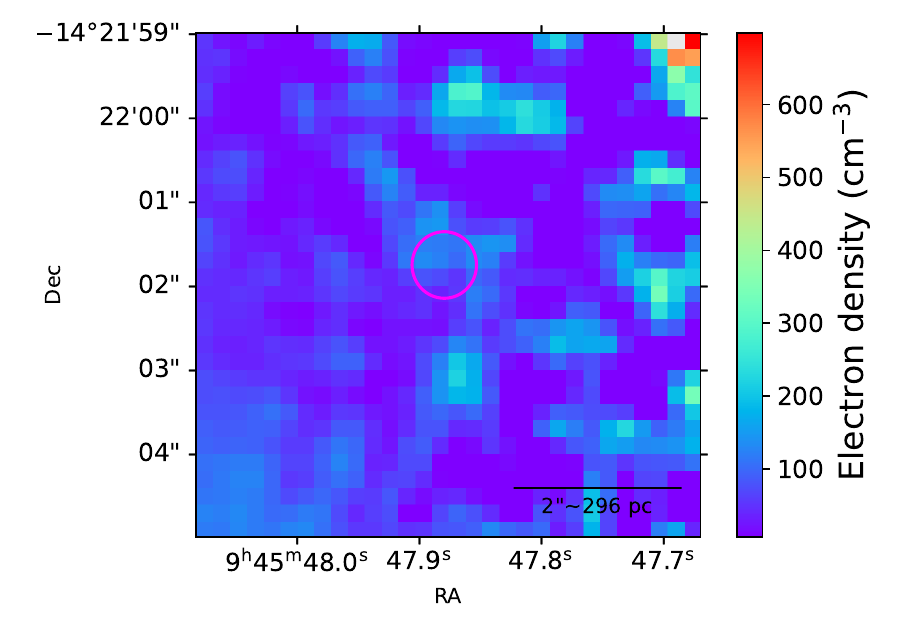}
    \end{subfigure}
    \begin{subfigure}[t]{0.28\linewidth}
        \centering
        \includegraphics[width=\linewidth]{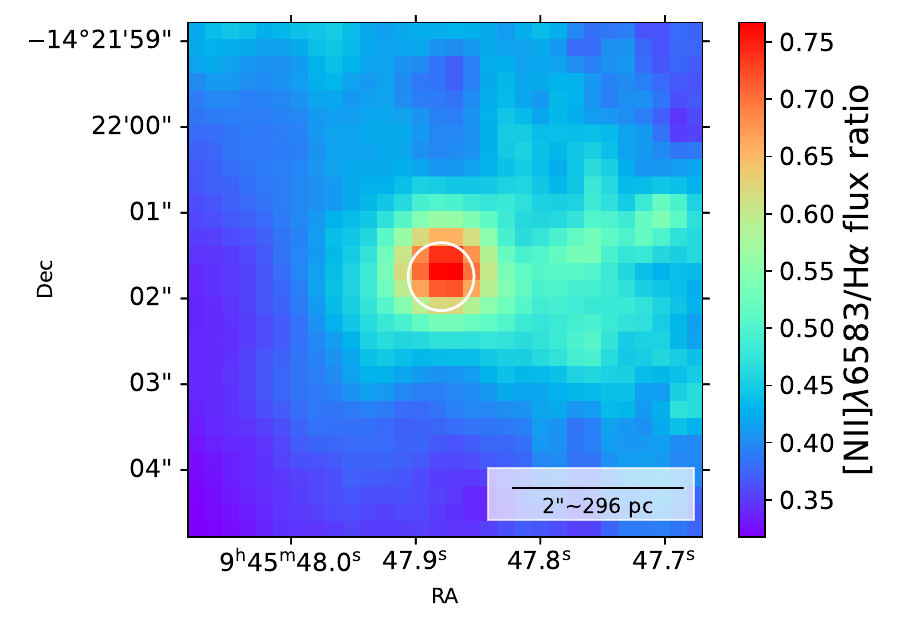}
        \phantomcaption
        \label{fig:n2ha_zoom}
    \end{subfigure}
    \begin{subfigure}[t]{0.28\linewidth}
        \centering
        \includegraphics[width=\linewidth]{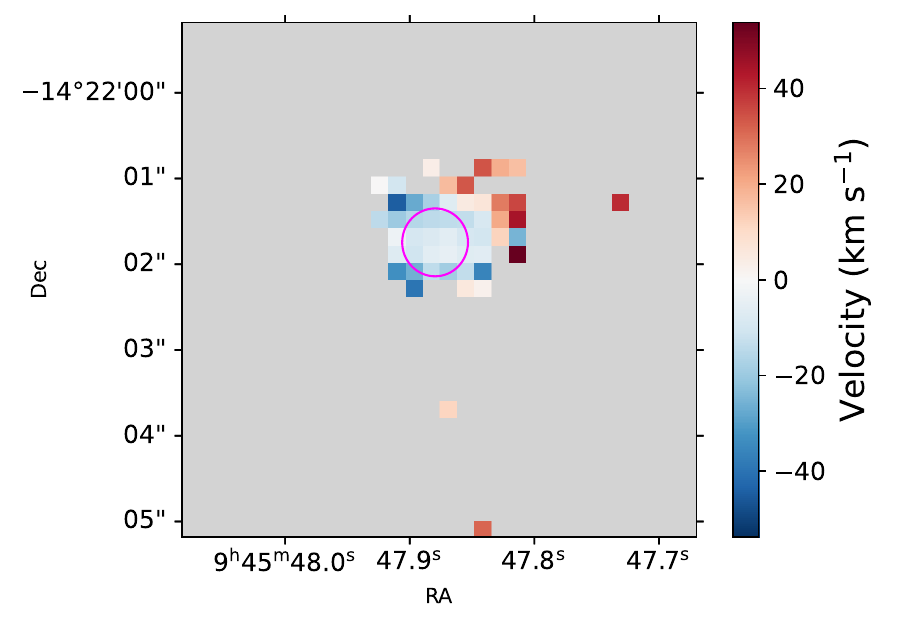}
        \phantomcaption
        \label{fig:na1_zoom}
    \end{subfigure}
        
    \caption{Additional maps. From left to right, first row: stellar mass, log of stellar ages, stellar metallicity; second row: extinction (see Fig.~\ref{fig:extinction}), electron density, \nii/H$\alpha$ ratio (see Fig.~\ref{fig:n2ha}); third row: \nai\ absorption lines velocity.
    The circle marks the position of the ULX1 source.}
    \label{fig:zoom2}
\end{figure*}

\clearpage

\section{Additional CAMEL maps}
\label{app:add_camel}

\subsection{CAMEL maps for other emission lines}
\label{app:camel}

\begin{figure}[h!]
    \centering
     \begin{subfigure}[b]{0.33\textwidth}
        \centering
        \includegraphics[width=0.9\linewidth]{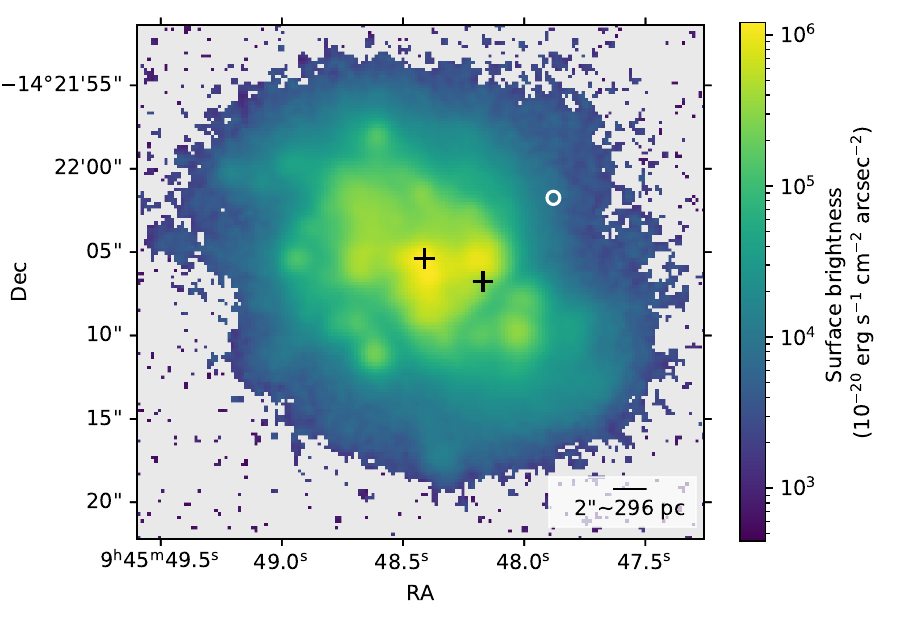}
    \end{subfigure}
    \begin{subfigure}[b]{0.33\textwidth}
        \centering
        \includegraphics[width=0.9\linewidth]{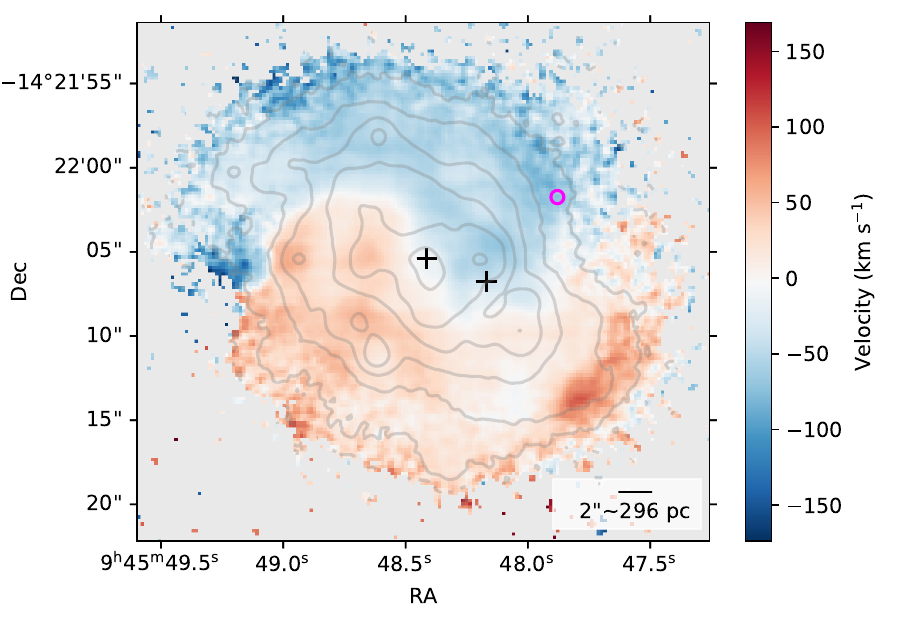}
    \end{subfigure}
    \begin{subfigure}[b]{0.33\textwidth}
        \centering
        \includegraphics[width=0.9\linewidth]{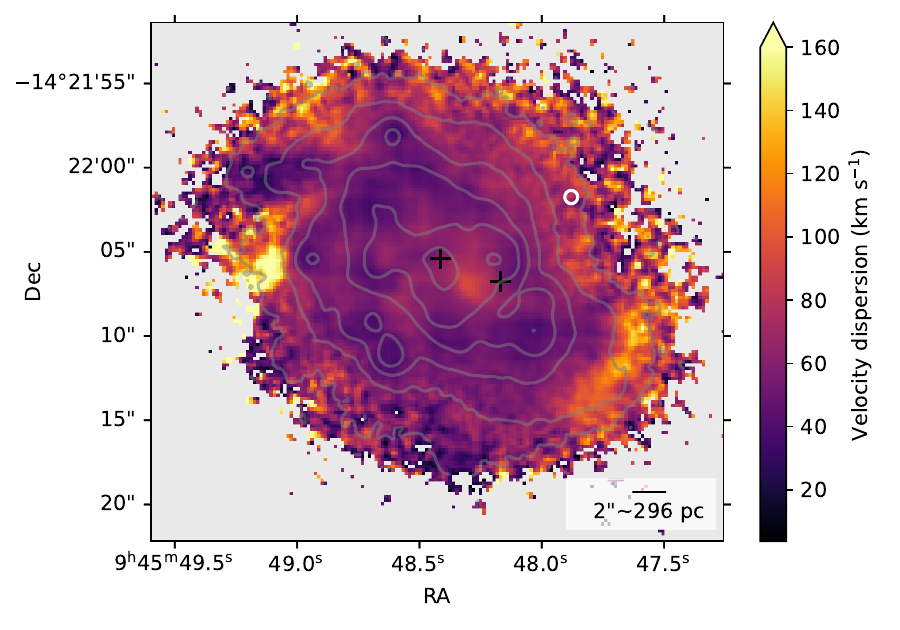}
    \end{subfigure}
    \begin{subfigure}[b]{0.33\textwidth}
        \centering
        \includegraphics[width=0.9\linewidth]{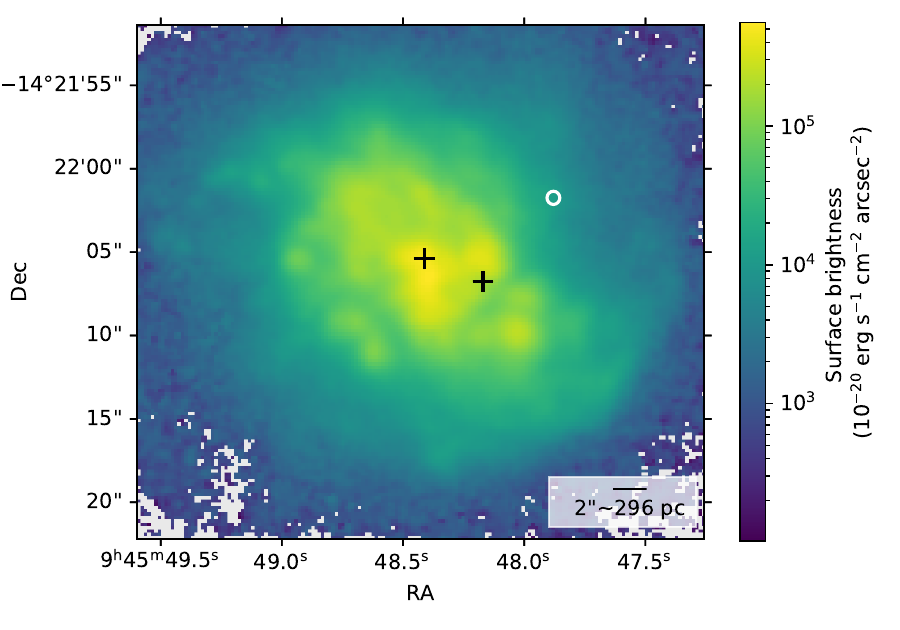}
    \end{subfigure}
    \begin{subfigure}[b]{0.33\textwidth}
        \centering
        \includegraphics[width=0.9\linewidth]{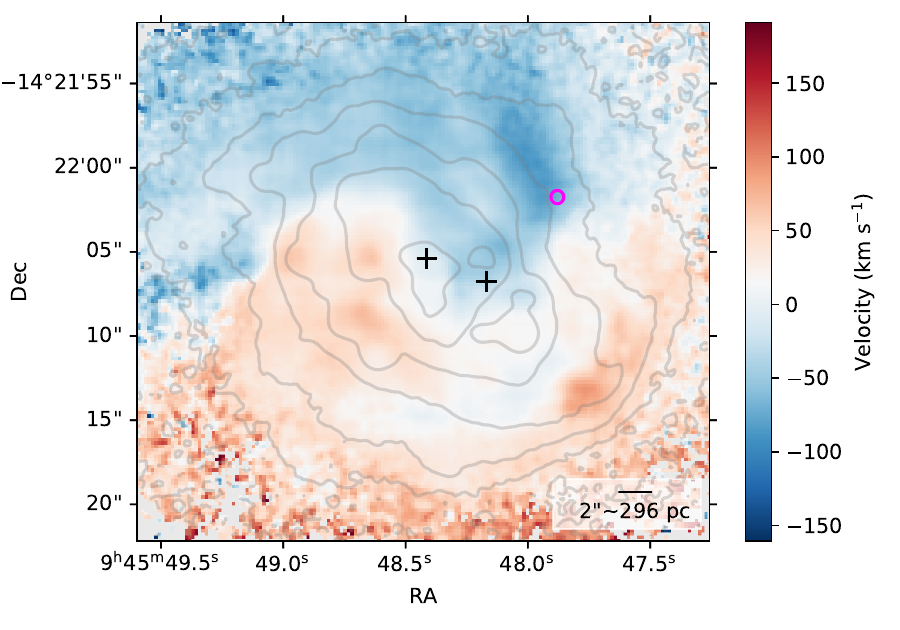}
    \end{subfigure}
    \begin{subfigure}[b]{0.33\textwidth}
        \centering
        \includegraphics[width=0.9\linewidth]{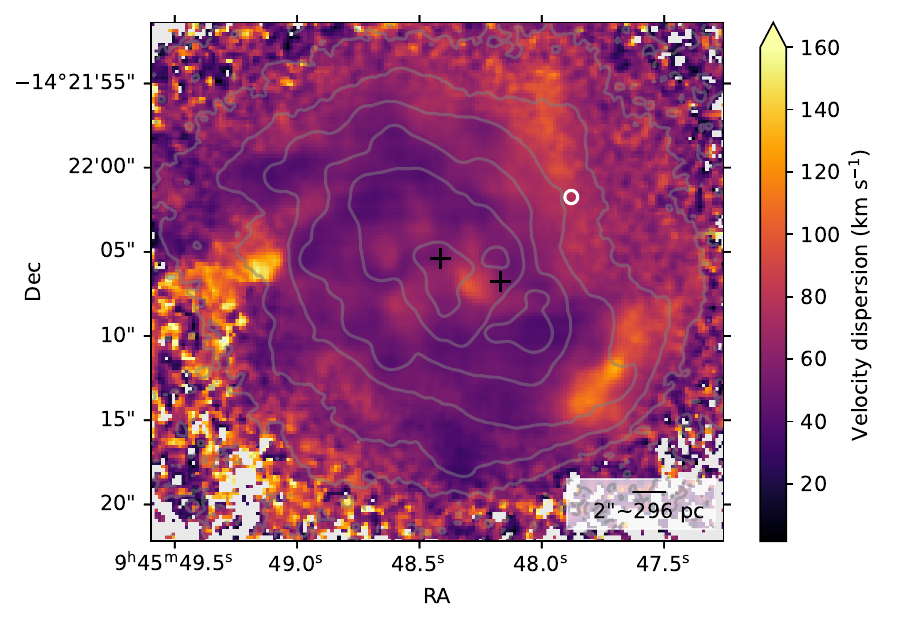}
    \end{subfigure}
     \begin{subfigure}[b]{0.33\textwidth}
        \centering
        \includegraphics[width=0.9\linewidth]{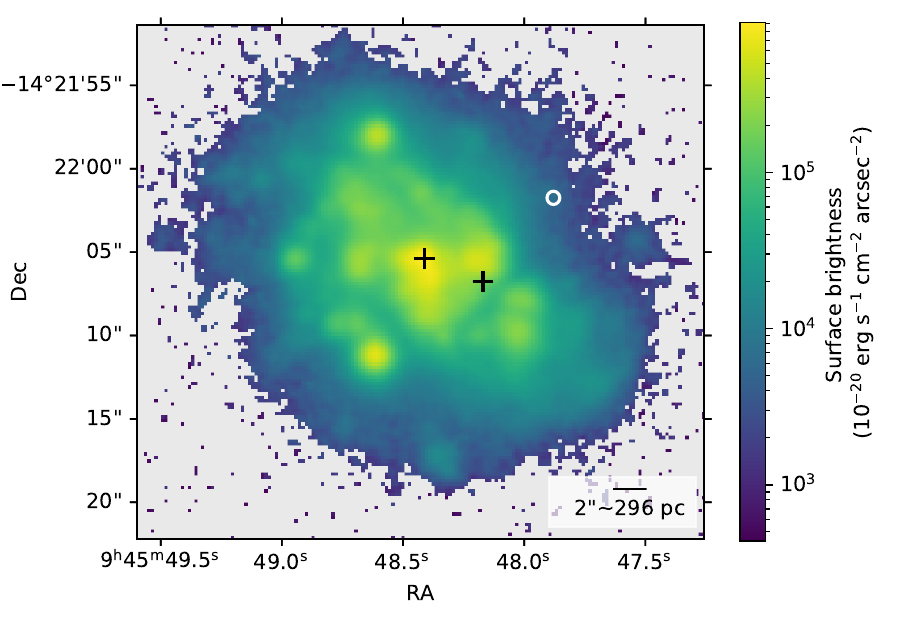}
    \end{subfigure}
    \begin{subfigure}[b]{0.33\textwidth}
        \centering
        \includegraphics[width=0.9\linewidth]{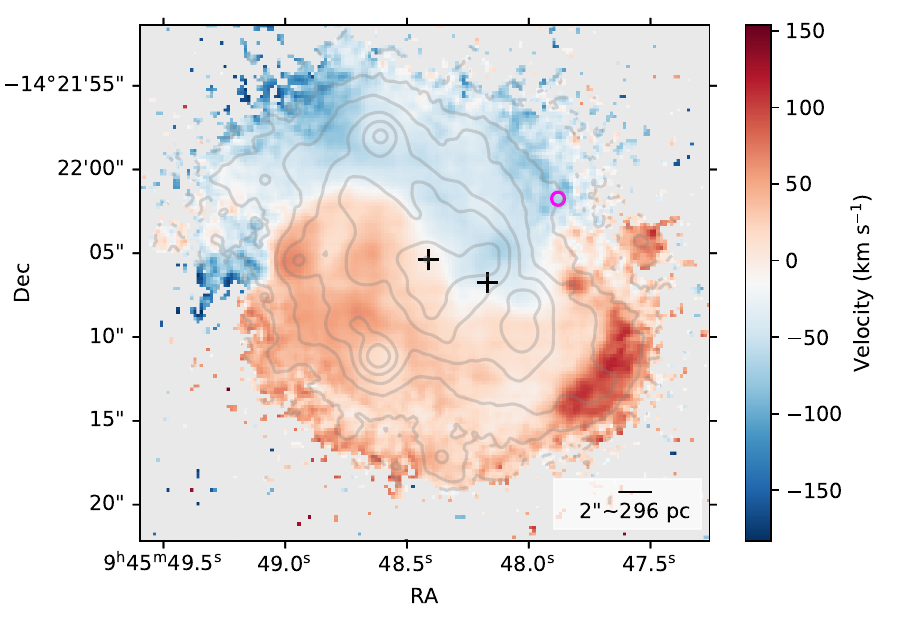}
    \end{subfigure}
    \begin{subfigure}[b]{0.33\textwidth}
        \centering
        \includegraphics[width=0.9\linewidth]{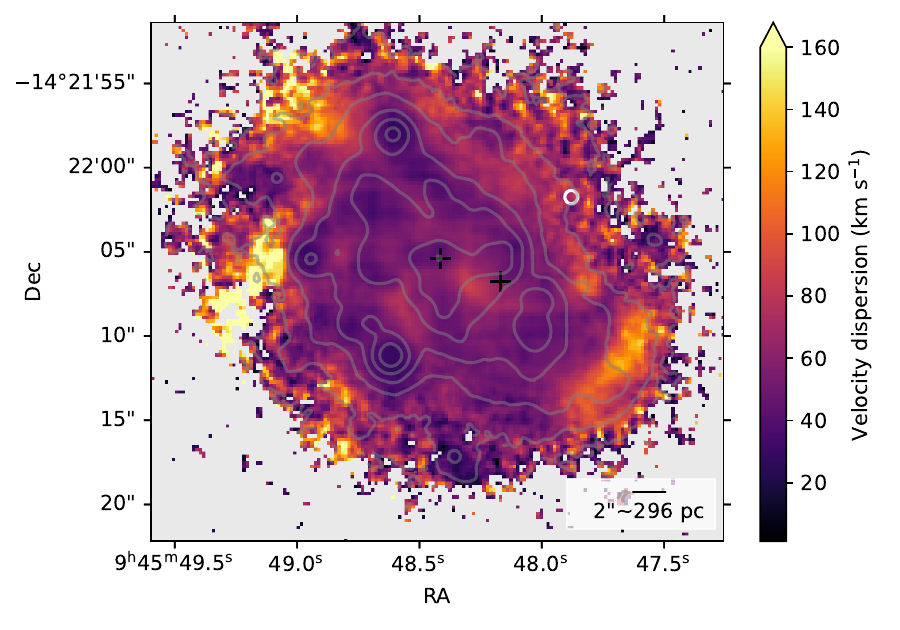}
    \end{subfigure}
    \begin{subfigure}[b]{0.33\textwidth}
        \centering
        \includegraphics[width=0.9\linewidth]{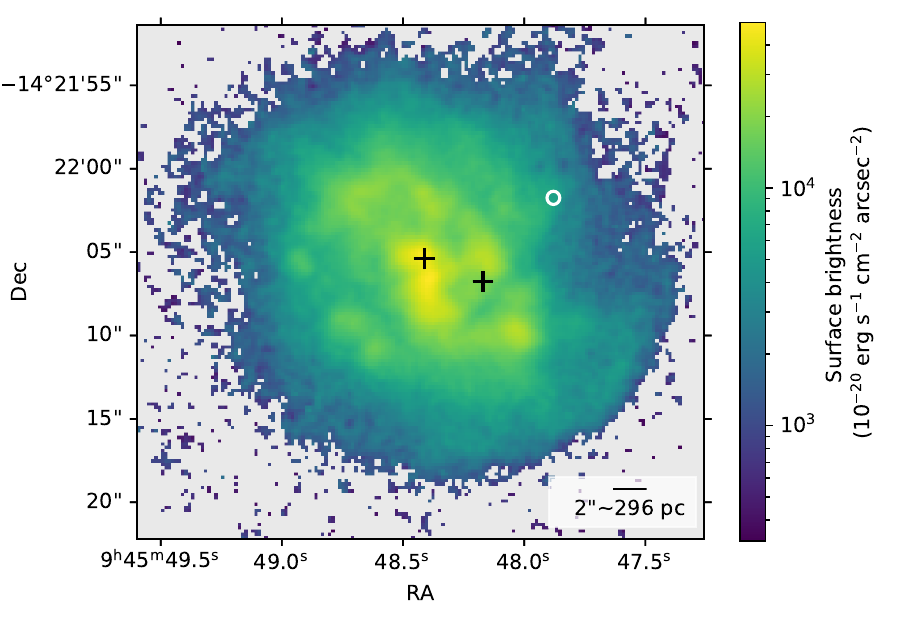}
    \end{subfigure}
    \begin{subfigure}[b]{0.33\textwidth}
        \centering
        \includegraphics[width=0.9\linewidth]{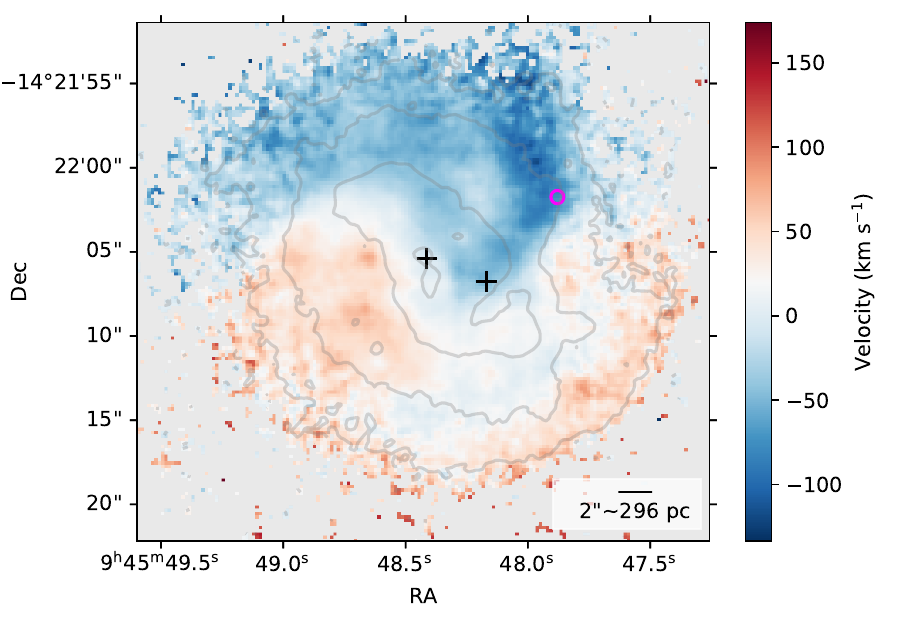}
    \end{subfigure}
    \begin{subfigure}[b]{0.33\textwidth}
        \centering
        \includegraphics[width=0.9\linewidth]{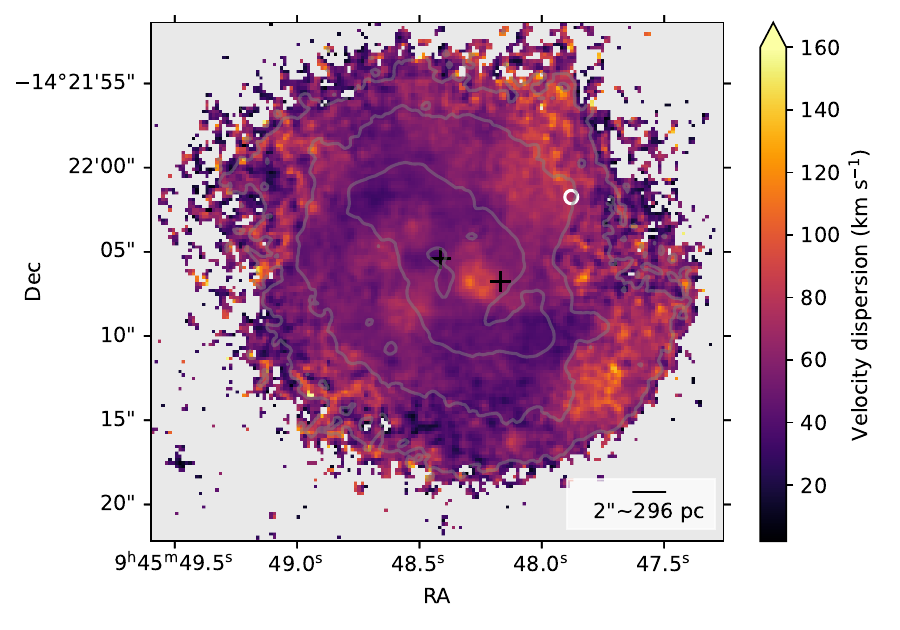}
    \end{subfigure}
    \begin{subfigure}[b]{0.33\textwidth}
        \centering
        \includegraphics[width=0.9\linewidth]{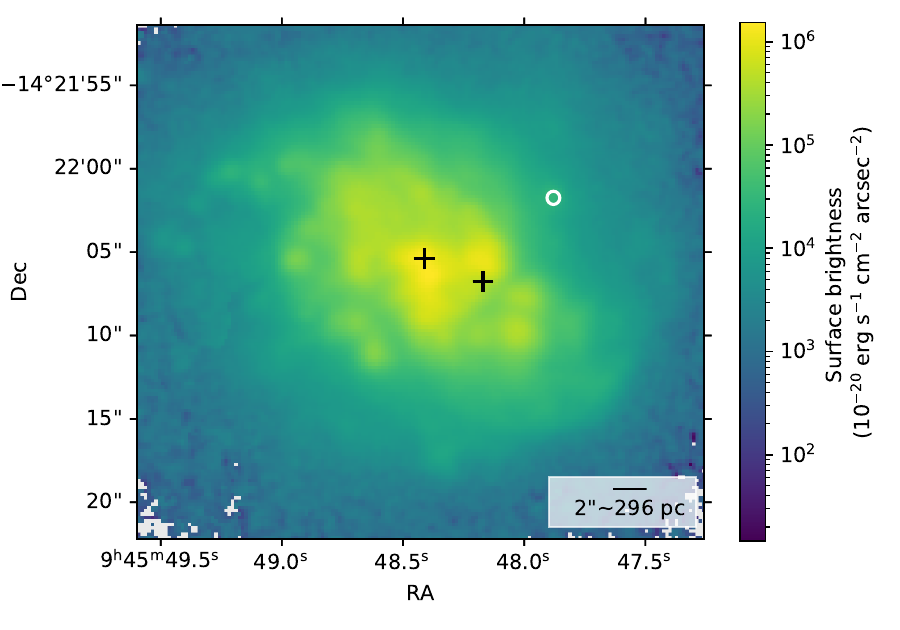}
    \end{subfigure}
    \begin{subfigure}[b]{0.33\textwidth}
        \centering
        \includegraphics[width=0.9\linewidth]{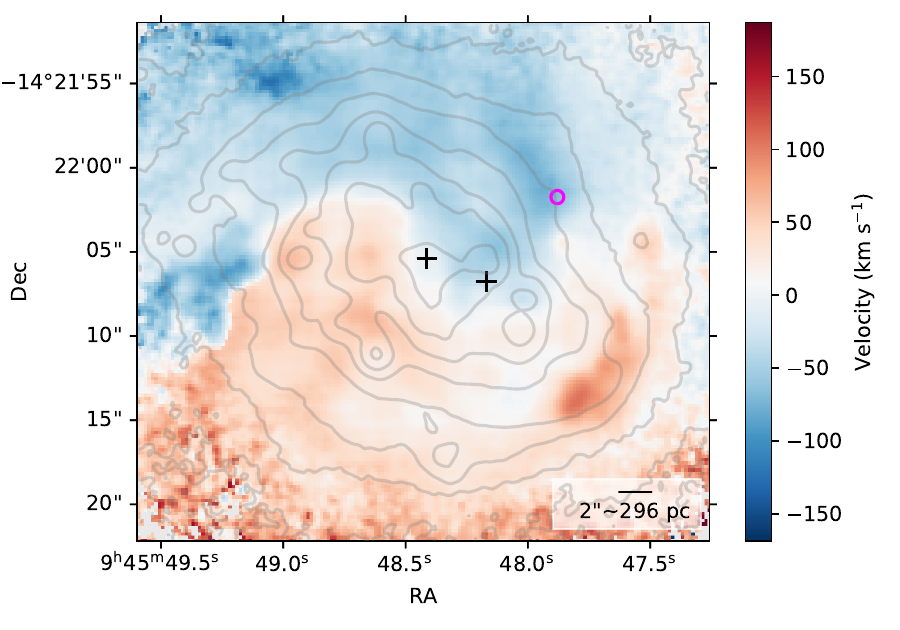}
    \end{subfigure}
    \begin{subfigure}[b]{0.33\textwidth}
        \centering
        \includegraphics[width=0.9\linewidth]{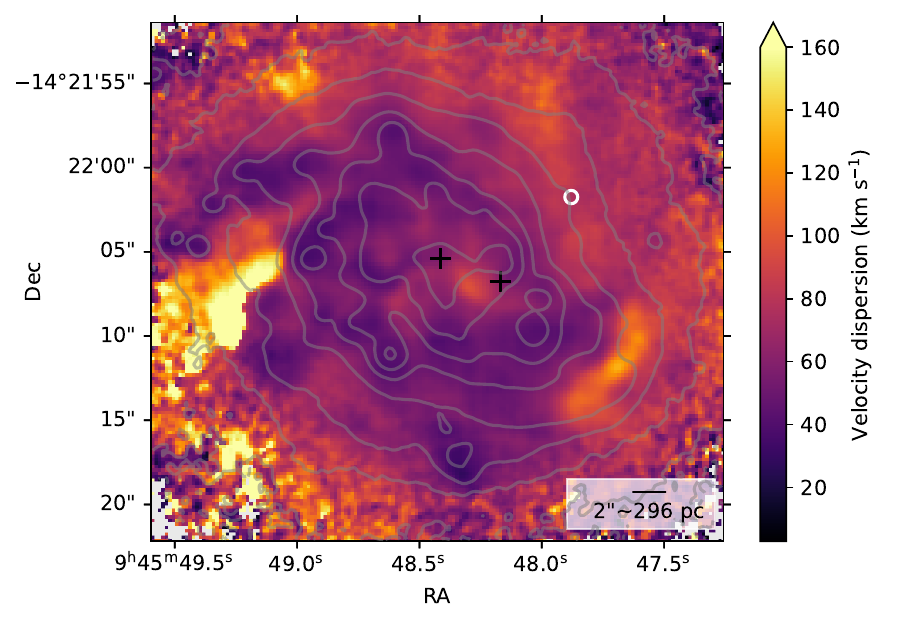}
    \end{subfigure}
    \hfill
    \caption{From top to bottom: H$\beta$, \sii$\lambda$6716, \oiii$\lambda$5007, \oi$\lambda$6300, \nii$\lambda$6583. Left to right: flux, velocity, velocity dispersion maps. Crosses and circle are as in Fig. \ref{fig:camel}. Contours on the velocity and velocity dispersion maps are from the surface brightness maps. Unlike in Fig. \ref{fig:camel} for the H$\alpha$ emission line, the flux and velocity maps are given as were fitted using a single component in all the regions of the galaxy. 
    The velocity and the velocity dispersion for \nii\ are the same as for H$\alpha$ since these lines were fit together. }
    \label{fig:camel_add}
\end{figure}

\subsection{Extinction}
\label{app:extinction}

To calculate the extinction map of the galaxy shown in Fig. \ref{fig:extinction}, we use the extinction law from \cite{calzetti_dust_2000} with R$_V = 4.05$. Given the mean E(B-V) value in the direction of the galaxy of 0.0520 \citep{schlafly_measuring_2011}, we calculate $\rm k(H\alpha) = 3.33$ and $\rm k(H\beta) = 4.60$. Assuming the electron temperature of 10000 K and electron density of 100~cm$^{-3}$ (which is close to the true values in the galaxy, see Sect. \ref{subsec:ne}), the intrinsic $\frac{\mathrm{H}\alpha}{\mathrm{H}\beta}$ ratio is 2.86 (Case B recombination from \citealt{osterbrock_astrophysics_2006}). We then use the Balmer decrement equation from \cite{momcheva_nebular_2013} to calculate the E(B-V) of the galaxy itself.

\subsection{\texorpdfstring{\nii/H$\alpha$}{[NII]/Ha} ratio}
\label{app:n2_ha}

\begin{figure}[h!]
    \centering
    \begin{minipage}[t]{0.45\linewidth}
        \centering
        \includegraphics[width=\linewidth]{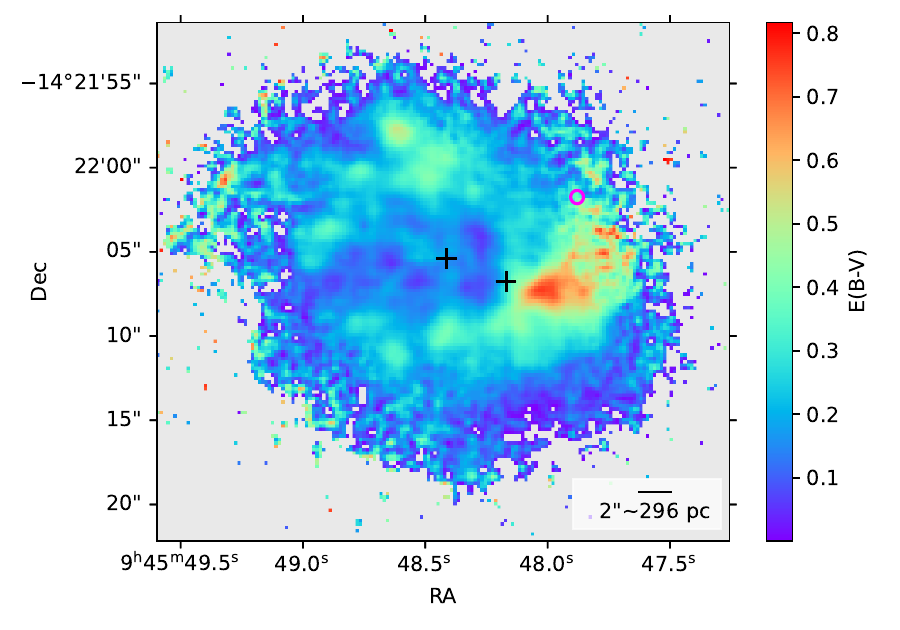}
        \caption{Extinction map from the Balmer decrement. Crosses and circle as in Fig. \ref{fig:camel}.}
        \label{fig:extinction}
    \end{minipage}\quad
    \begin{minipage}[t]{0.45\linewidth}
        \centering
        \includegraphics[width=\linewidth]{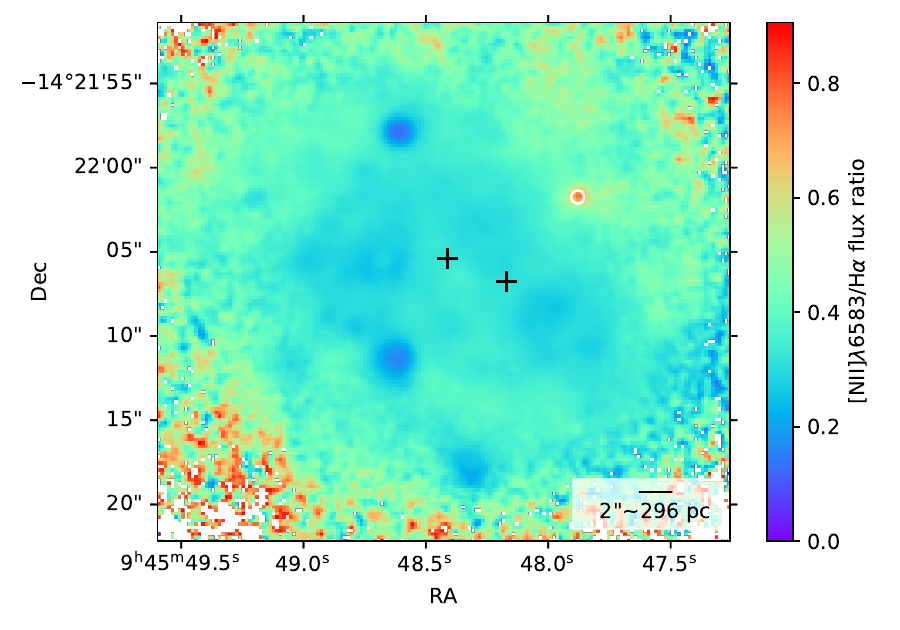}
        \caption{\nii$\lambda$6583/H$\alpha$ flux ratio. Note at the position of ULX1 (S/N = 78) an excess of \nii/H$\alpha$. Other areas with high values are due to S/N < 15. The white areas correspond to pixels for which S/N < 4. Crosses and circle are like in Fig. \ref{fig:camel}.}
        \label{fig:n2ha}
    \end{minipage}
\end{figure}

Figure \ref{fig:n2ha} shows the \nii$\lambda$6583/H$\alpha$ flux ratio map for the galaxy. We see that the value of 0.8 is only obtained at the position of ULX1 (S/N=78), except for the outer parts of the galaxy where S/N is below 15.

\section{BPT diagrams}
\label{app:bpt}

\begin{figure}[h!]
    \centering
     \begin{subfigure}[b]{\textwidth}
        \centering
        \includegraphics[width=0.85\linewidth]{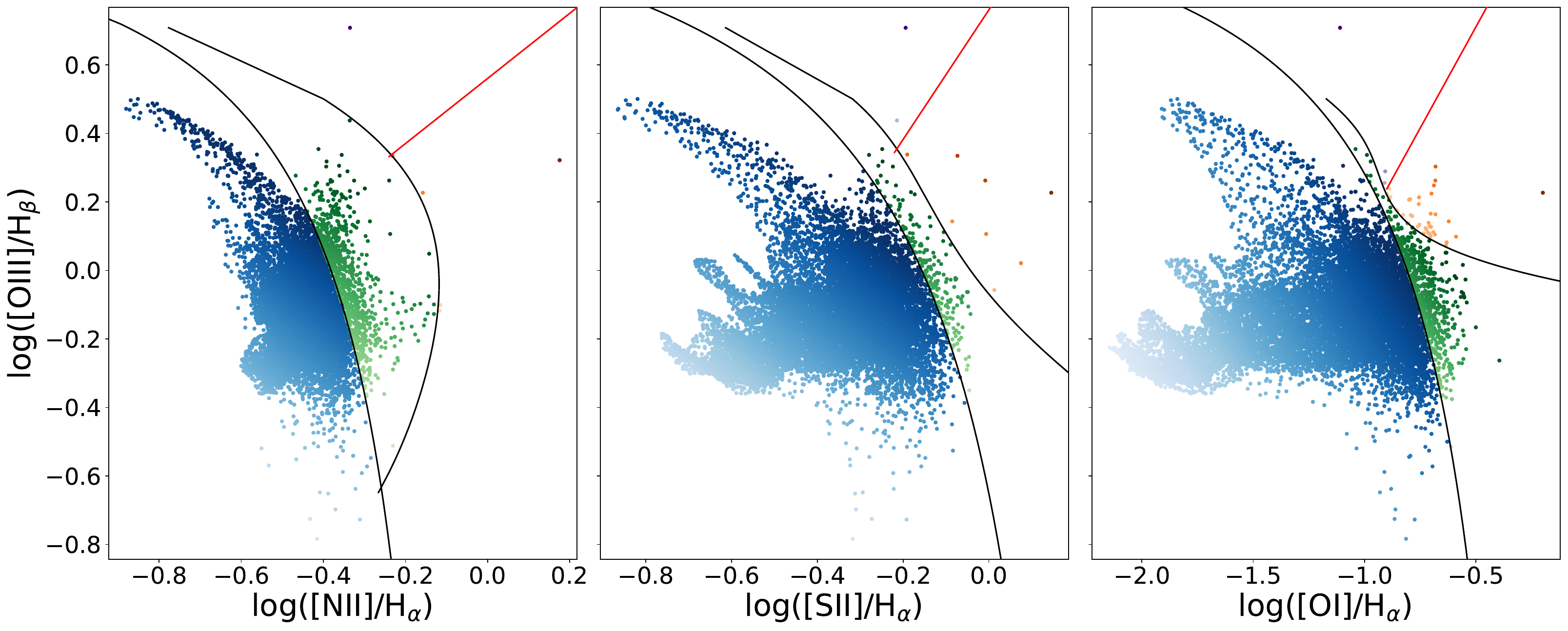}
        \phantomcaption
        \label{fig:bpt_full_diag}
    \end{subfigure}
    \begin{subfigure}[b]{\textwidth}
        \centering
        \includegraphics[width=\linewidth]{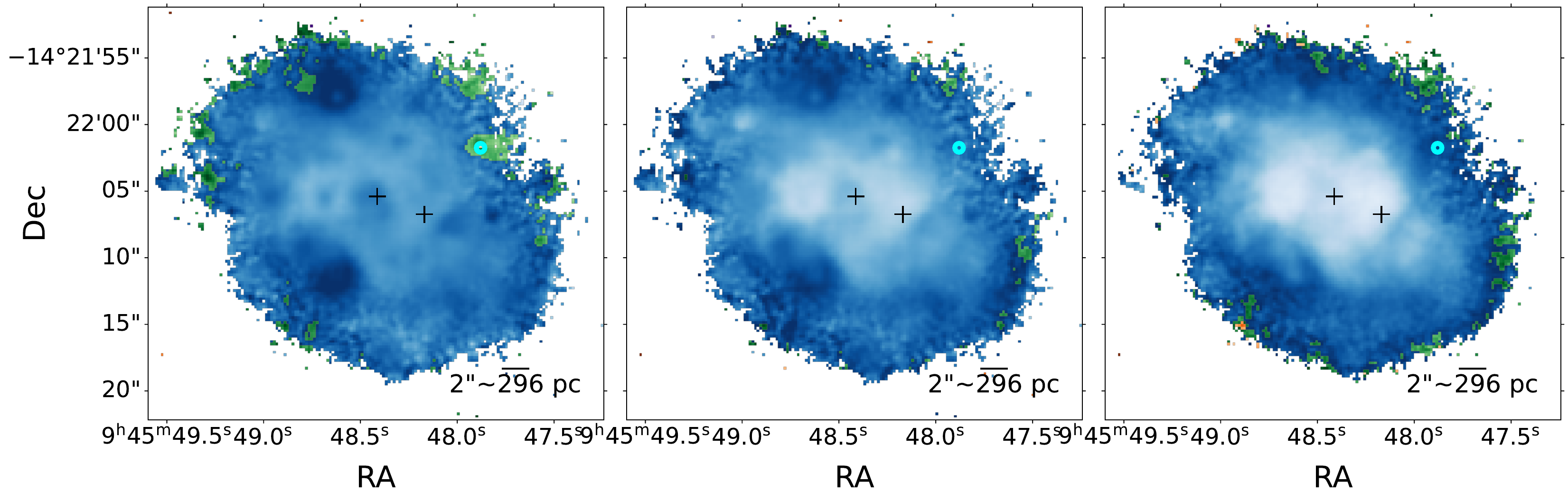}
        \phantomcaption
        \label{fig:bpt_full_map}
    \end{subfigure}
    \caption{BPT diagrams and their corresponding maps for the NGC 2993 galaxy. Crosses and circle are as in the Fig. \ref{fig:camel}. White pixels have S/N < 4.}
    \label{fig:bpt}
\end{figure}

Figure \ref{fig:bpt} shows the BPT diagrams (top panel), where to each point corresponds a pixel in the associated maps (bottom panels). The y-axis is always the \oiii$\lambda$5007/H$\beta$ (R3) ratio, and on the x-axis there are \nii$\lambda$6583/H$\alpha$ (N2), \sii$\lambda$6731/H$\alpha$ (S2) and \oi$\lambda$6300/H$\alpha$ (O1) ratios. The diagrams are separated into 4 regions according to \cite{law_sdss-iv_2021} which improves on the delimiters from \cite{kewley_host_2006} by applying them to spatially resolved galaxy data. 
Blue datapoints correspond to the \hii\ regions, where the gas is mainly ionized by the stellar UV radiation. The top region in the diagrams that would show purple points corresponds to AGN regions that are mostly photoionized by X-ray photons. The rightmost region in the diagrams (in red) corresponds to LINERs and usually indicates that the main source of ionization is shocks. The intermediate regions are in green.
As seen in the figures, gas in the whole galaxy seems to be mostly ionized by the stellar UV radiation, which is expected in a starburst galaxy. The only exception is in the N2-R3 diagram at the position of ULX1, which seems to be in the LINER region due to its elevated N2 ratio values. 

\clearpage

\section{Mappings}
\label{app:mappings}

\begin{figure*}[h!]
    \centering
    \begin{subfigure}[t]{0.47\textwidth}
        \centering
        \includegraphics[width=0.6\linewidth]{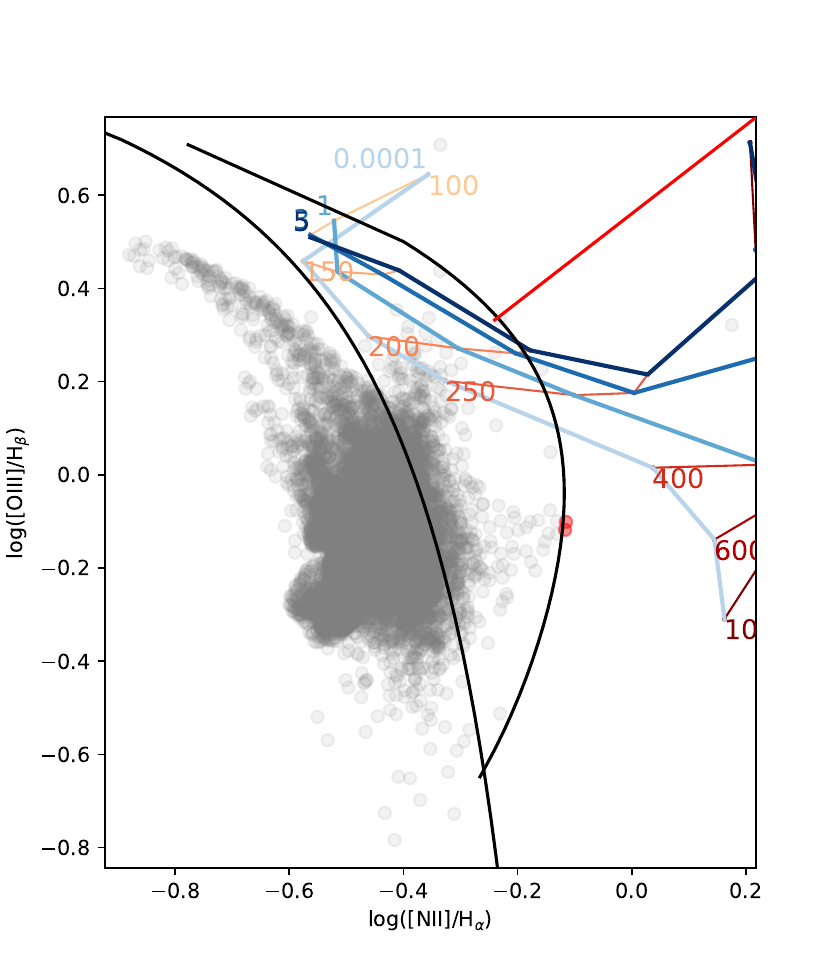}
        \caption{Shock only. Model from \cite{allen_mappings_2008} recomputed using \texttt{MAPPINGS V} with solar abundances from \cite{dopita_spectral_1996}.}
    \end{subfigure}\quad
    \begin{subfigure}[t]{0.47\textwidth}
        \centering
        \includegraphics[width=0.6\linewidth]{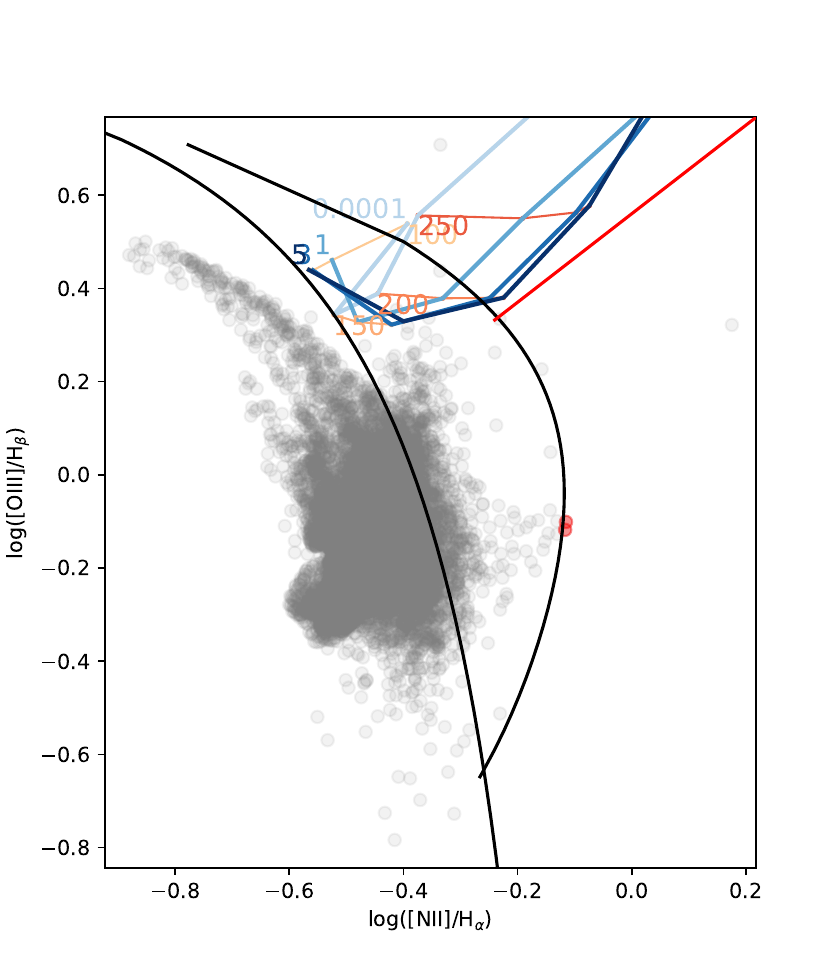}
        \caption{Shock + precursor. Model from \cite{allen_mappings_2008} recomputed using \texttt{MAPPINGS V} with solar abundances from \cite{dopita_spectral_1996}.}
    \end{subfigure}
    \begin{subfigure}[t]{0.47\textwidth}
        \centering
        \includegraphics[width=0.6\linewidth]{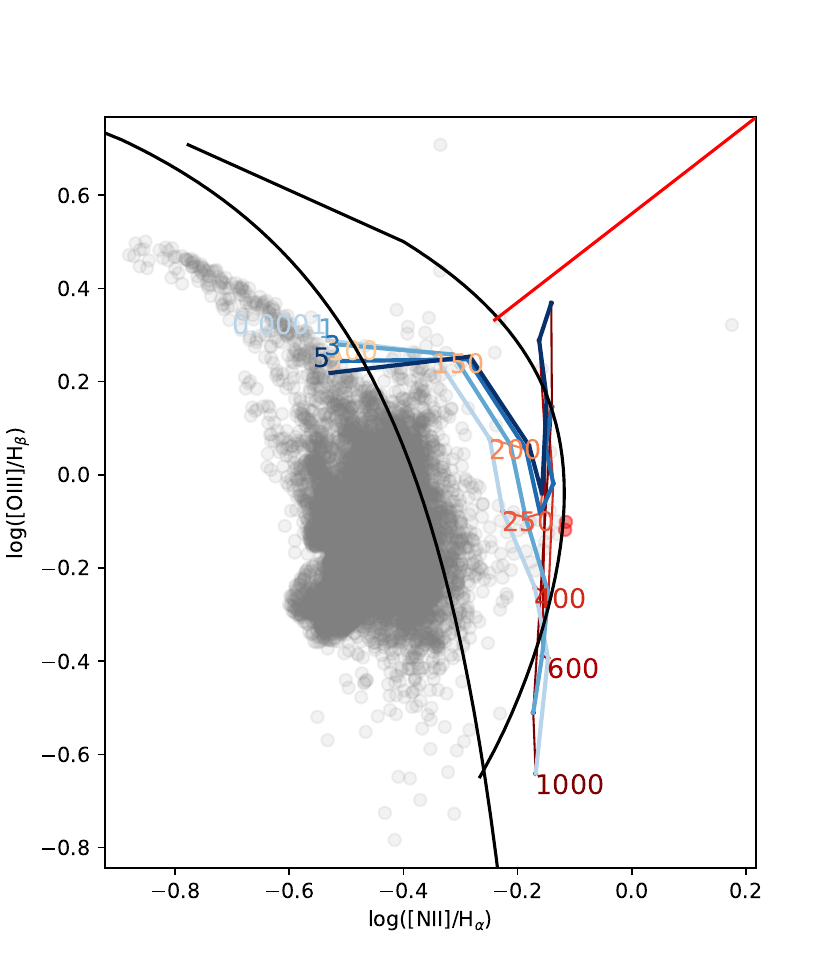}
        \caption{Shock only. Model from \cite{allen_mappings_2008} recomputed using \texttt{MAPPINGS V} with depleted solar abundances from \cite{dopita_modeling_2005}.}
    \end{subfigure}\quad
    \begin{subfigure}[t]{0.47\textwidth}
        \centering
        \includegraphics[width=0.6\linewidth]{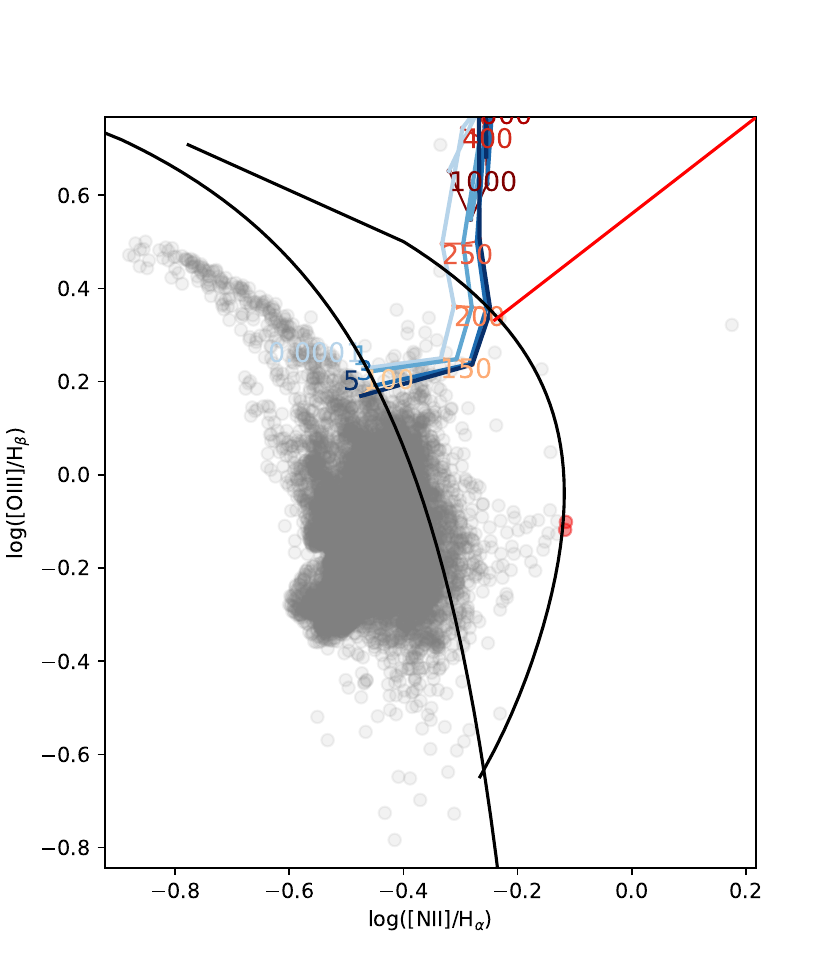}
        \caption{Shock + precursor. Model from \cite{allen_mappings_2008} recomputed using \texttt{MAPPINGS V} with depleted solar abundances from \cite{dopita_modeling_2005}.}
    \end{subfigure}
    \begin{subfigure}[t]{0.47\textwidth}
        \centering
        \includegraphics[width=0.6\linewidth]{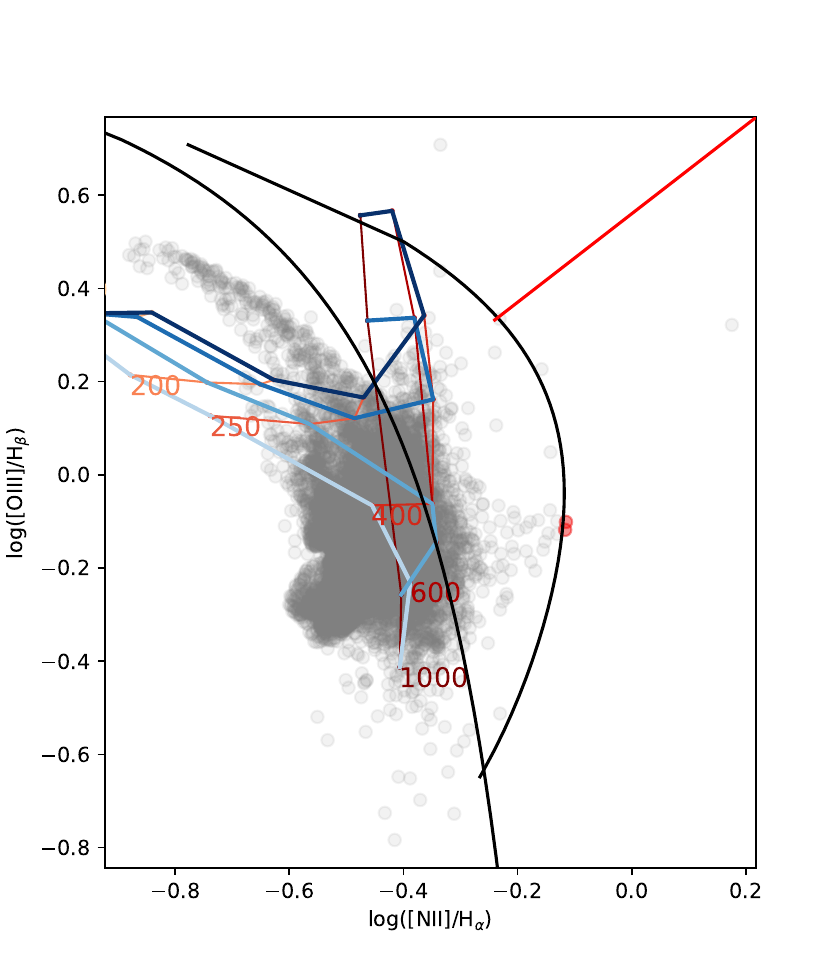}
        \caption{Shock only. Model extended for the abundances from \cite{gutkin_modelling_2016} with \texttt{MAPPINGS V}.}
    \end{subfigure}\quad
    \begin{subfigure}[t]{0.47\textwidth}
        \centering
        \includegraphics[width=0.6\linewidth]{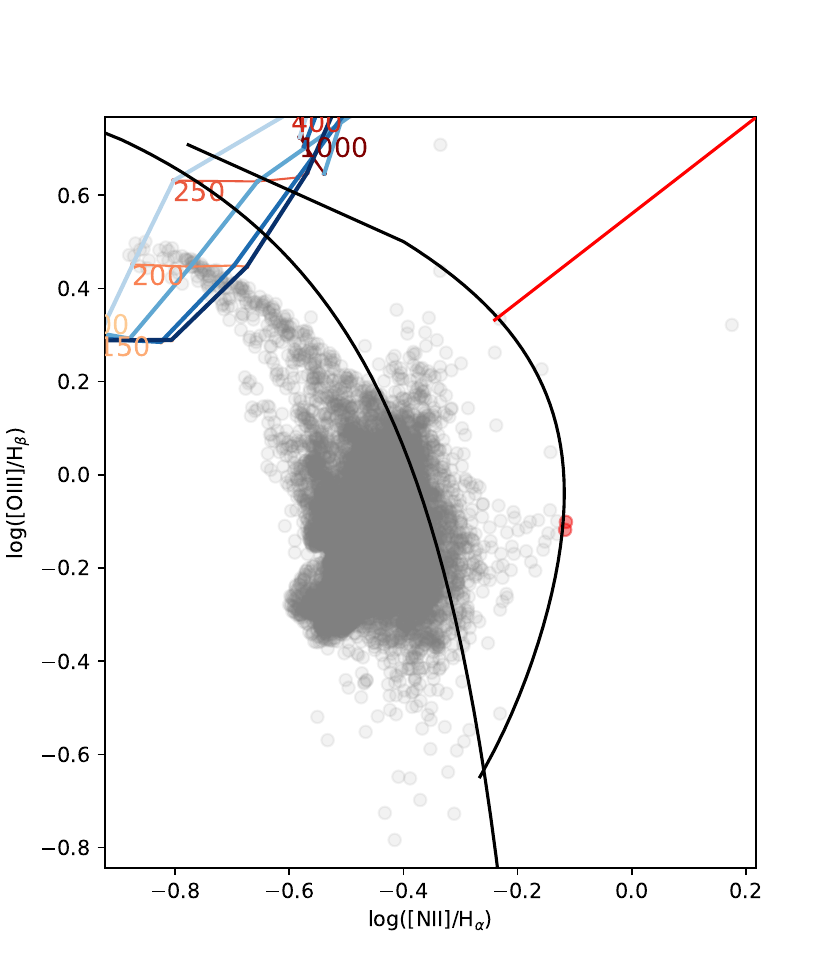}
        \caption{Shock + precursor. Model extended for the abundances from \cite{gutkin_modelling_2016} with \texttt{MAPPINGS V}.}
    \end{subfigure}
    \caption{\texttt{MAPPINGS V} shock models computed by \cite{alarie_extensive_2019}. The results are overlayed on the N2 BPT diagram. The two red points correspond to the two points in the LINER region at the position of ULX1. The blue lines connect points with the same pre-shock magnetic values (10$^{-4}$, 1, 3.23 and 5 $\mu$ G cm$^{3/2}$), with higher values having a darker color. The red lines connect points with the same shock velocities, their values are given in \kms.}
    \label{fig:mappings}
\end{figure*}

We used precursor+shock models computed by \cite{alarie_extensive_2019} using MAPPINGS V \citep{sutherland_mappings_2018}. We tested two models on our data: grids from \cite{allen_mappings_2008} recalculated with MAPPINGS V (top and middle panels in Fig. \ref{fig:mappings}) and an extension of those grids for low metallicity abundances using abundance sets from \cite{gutkin_modelling_2016} (bottom panels in Fig. \ref{fig:mappings}). We searched for the shock velocity values that may be able to explain the observed line flux ratios near the position of ULX1. 
In our case, we are looking for specific \nii/H$\alpha$ vs \oiii/H$\beta$ ratios, so the shock velocity values are overlayed on the N2 BPT diagram. The two red points correspond to the pixels at the position of ULX1 that are in the LINER region (see Fig. \ref{fig:bpt_zoom}).
In this Figure, we set the abundances to solar values, since in Sect. \ref{subsec:metal} we determined that in the \hii\ regions of NGC 2993 the abundances are nearly solar (i.e. almost everywhere in the galaxy except around the ULX1 position, where this calibration does not apply). In the top panel, we used solar abundances from \cite{dopita_spectral_1996}, in the middle panel depleted solar abundances from \cite{asplund_solar_2005} that were used in \cite{dopita_modeling_2005} and in the bottom panels $Z_{ISM}=0.010$ from \cite{gutkin_modelling_2016}. 
We also tried other abundance sets and none allowed us to explain the observed line ratios. Here, the pre-shock ISM density was set to 1 cm$^{-3}$, since the pre-shock densities are lower than post-shock densities \citep{dopita_spectral_1996} (see Sect. \ref{subsec:ne}). Changing the pre-shock density does not change our results. 

We show predictions for the shock only and for the shock + precursor models. Top and bottom panels show that the log(\nii/H$\alpha$) and log(\oiii/H$\beta$) ratio values around $-0.1$ cannot be explained using the corresponding shock models from \texttt{MAPPINGS}. 
The middle left panel seems to indicate that the observed \nii/H$\alpha$ vs \oiii/H$\beta$ ratios might be explained by shocks with velocities between 250 and 300 \kms. 
Although the bottom panel only shows the results for $Z_{ISM}=0.010$, \cite{gutkin_modelling_2016} provides a wide range of metallicities, and we find that for metallicities $Z_{ISM}>0.014$ the observed combination of \nii/H$\alpha$ and \oiii/H$\beta$ can be explained by a shocks only model with shock velocities $\sim600$ \kms.
However, the shock only model should be used in gas-poor environments, and shock + precursor model in gas-rich environments, and the latter seems to better correspond to our case. In addition, when the shock velocity is higher than 150~\kms , there is usually a precursor \citep*{dopita_spectral_1996}. 
Therefore, it seems unlikely that the observed line ratios could be explained by shocks.

\clearpage

\end{document}